\documentclass[a4paper,prd,superscriptaddress,twocolumn,aps,reprint,notitlepage,nofootinbib,floatfix,longbibliography]{revtex4}
\usepackage[usenames]{color}
\usepackage[pdftex, unicode, colorlinks, bookmarks=true, citecolor=blue, link color=blue, urlcolor=blue]{hyperref}
\usepackage{amsmath, amssymb, amsfonts}
\usepackage[utf8]{inputenc}
\usepackage{graphicx}
\usepackage{multirow}

\newcommand{\bra}[1]{\langle#1|}
\newcommand{\ket}[1]{|#1\rangle}
\newcommand{\vb}[1]{\pmb{#1}}

\newcommand{\abs}[1]{\left|\kern0.1em#1\kern0.1em\right|}
\newcommand{\atan}[1]{{\rm arctan}\left(#1\right)}
\newcommand{\acot}[1]{{\rm arccot}\left(#1\right)}
\newcommand{\til}[1]{\widetilde{#1}}

\newcommand{\arccot}{\mathrm{arccot}}

\newcommand{\RITM}{R_{\rm ITM}}
\newcommand{\RETM}{R_{\rm ETM}}
\newcommand{\TITM}{T_{\rm ITM}}
\newcommand{\TETM}{T_{\rm ETM}}
\newcommand{\tauArm}{\tau_{\rm arm}}
\newcommand{\phiSR}{\phi_{\rm SR}}
\newcommand{\hSQL}{h_{\rm SQL}}

\newcommand{\intinfty}{\int_{-\infty}^{\infty}\!}
\usepackage{xifthen}
\newcommand{\vq}[2][]{                
  \ifthenelse{\isempty{#1}}           %
    { \hat{\pmb{#2}} }                
    { \hat{\pmb{#2}}_\mathrm{#1} }    
}
\newcommand{\vqtil}[2][]{             
  \ifthenelse{\isempty{#1}}           %
    { \hat{\til{\pmb{#2}}} }                
    { \hat{\til{\pmb{#2}}}_\mathrm{#1} }    
}
\newcommand{\tq}[2][]{                
  \ifthenelse{\isempty{#1}}           %
    { \mathbb{#2} }                   
    { \mathbb{#2}_\mathrm{#1} }       
}
\newcommand{\vs}[2][]{                
  \ifthenelse{\isempty{#1}}           %
    { \mathbf{#2} }                   
    { \mathbf{#2}_\mathrm{#1} }       
}

\newcommand{\tr}{ {\mathsf{T}} }

\newcommand{\matrY}[1][]{
  \ifthenelse{\isempty{#1}}           %
    {                    
    \begin{pmatrix}
       y_1 & y_2 \\
      -y_2 & y_1
    \end{pmatrix}
    }
  {                  
    \begin{pmatrix}
       y_1^\mathrm{#1} & y_2^\mathrm{#1} \\
      -y_2^\mathrm{#1} & y_1^\mathrm{#1}
    \end{pmatrix}
  }
}
\newcommand{\matrYOm}[1][]{
  \ifthenelse{\isempty{#1}}           %
    {                    
    \begin{pmatrix}
       y_1(\Omega) & y_2(\Omega) \\
      -y_2(\Omega) & y_1(\Omega)
    \end{pmatrix}
    }
  {                  
    \begin{pmatrix}
       y_1^\mathrm{#1}(\Omega) & y_2^\mathrm{#1}(\Omega) \\
      -y_2^\mathrm{#1}(\Omega) & y_1^\mathrm{#1}(\Omega)
    \end{pmatrix}
  }
}

\newcommand{\colRYF}[1][]{                %
  \ifthenelse{\isempty{#1}}           %
    { \binom {R_{Y_1F}} {R_{Y_2F}}  }                   
    { \binom {R_{Y_1F}^\mathrm{#1}} {R_{Y_2F}^\mathrm{#1}} }       
}
\newcommand{\colRYFOm}[1][]{                %
  \ifthenelse{\isempty{#1}}           %
    { \binom {R_{Y_1F}} {R_{Y_2F}}  }                   
    { \binom {R_{Y_1F}^\mathrm{#1}(\Omega)} {R_{Y_2F}^\mathrm{#1}(\Omega)} }       
}

\newcommand{\col}[2]{
  \begin{bmatrix}
    #1 \\
    #2
  \end{bmatrix}
}

\newcommand{\matr}[4]{
  \begin{bmatrix}
    #1 & #2 \\
    #3 & #4
  \end{bmatrix}
}

\usepackage{cancel, soul}
\usepackage[usenames,dvipsnames,svgnames]{xcolor}

\newcommand{\SD}[1]{{\color{black}#1}}

\begin{document}

\title{Broadband detuned Sagnac interferometer for future generation gravitational wave astronomy.}

\author{N.V.~Voronchev}
\affiliation{Moscow State University, Faculty of Physics, Moscow 119991, Russia}

\author{S.P.~Tarabrin}
\affiliation{Institut f\"{u}r Gravitationsphysik, Leibniz Universit\"{a}t Hannover and \\ Max-Planck-Institut f\"{u}r Gravitationsphysik (Albert-Einstein-Institut), Callinstra\ss{}e 38, 30167 Hannover, Germany}
\affiliation{Institut f\"ur Theoretische Physik, Leibniz Universit\"at Hannover, Appelstra\ss{}e 2, 30167 Hannover, Germany}

\author{S.L.~Danilishin}
\email{stefan.danilishin@ligo.org}
\affiliation{University of Glasgow, School of Physics and Astronomy, Glasgow G12 8QQ, Scotland, UK}%
\affiliation{School of Physics, University of Western Australia, 35 Stirling Hwy, Crawley 6009, WA, Australia}

\begin{abstract}

Broadband suppression of quantum noise below the Standard Quantum Limit (SQL) becomes a top-priority problem for the future generation of large-scale terrestrial detectors of gravitational waves, as the interferometers of the Advanced LIGO project, predesigned to be quantum-noise-limited in the almost entire detection band, are phased in. To this end, among various proposed methods of quantum noise suppression or signal amplification, the most elaborate approach implies a so-called \textit{xylophone} configuration of two Michelson interferometers, each optimised for its own frequency band, with a combined broadband sensitivity well below the SQL. Albeit ingenious, it is a rather costly solution. We demonstrate that changing the optical scheme to a Sagnac interferometer with weak detuned signal recycling and frequency dependent input squeezing can do almost as good a job, as the xylophone for significantly lower spend. We also show that the Sagnac interferometer is more robust to optical loss in filter cavity,
 used for 
frequency dependent squeezed vacuum injection, than an analogous Michelson interferometer, thereby reducing building cost even more.
\end{abstract}
\maketitle

\section{Introduction}

The past decade was marked by great achievements in gravitational-wave (GW) instrumental science. The international network of gravitational wave (GW) detectors, comprising three LIGO (Laser Interferometer Gravitational-wave Observatory) detectors \cite{Abramovici1992, LIGOsite, Waldman2006} and the EGO (European Gravitational Observatory) detector Virgo \cite{VIRGOsite, Acernese2006} has reached the project sensitivity and has accomplished 6 runs of scientific data collection. Apart from setting limits on a population of various sources of gravitational radiation in our Galaxy and beyond
\cite{Phys.Rev.D.85.122001, ISI-000284075400040, ISI-000272313300005, ISI-000293135700010, ISI-000291312700002, ISI-000286983900001}, all detectors in the network have reached hitherto unseen displacement sensitivity of $\sim10^{-18}$~m/$\sqrt{\mathrm{Hz}}$, making them, perhaps, the most sensitive displacement sensors in the world. The figure $10^{-18}$~m/$\sqrt{\mathrm{Hz}}$ is remarkable by itself as it is only 1 order of magnitude above the magnitude of quantum zero-point fluctuations for a mechanical object with a mass of 10 kilogramms that is the mass of the LIGO interferometer core optics mirrors.

The second generation detectors, like Advanced LIGO \cite{Thorne2000,Fritschel2002}, Advanced Virgo \cite{Acernese2006-2}, KAGRA \cite{KAGRA_paper_Somiya} and GEO-HF \cite{Willke2006} will be already operating at the level of this quantum limit, meaning that their sensitivities will be governed by quantum fluctuations of light in the frequency band around 100~Hz wherein most of the target GW sources are expected to emit. Detection band below 100~Hz will be dominated by quantum fluctuations of light amplitude, known as \textit{quantum radiation pressure noise} (RPN), while at higher frequencies detector sensitivity will be limited by quantum phase fluctuations, usually referred to as \textit{quantum shot noise} (SN).
The best sensitivity point, where these two noise sources become equal, is known as the Standard Quantum Limit (SQL) \cite{92BookBrKh}, which, in a broader context, characterises the regime in which the quantum measurement noise (SN) becomes equal to the back action noise (RPN) --- the latter one being a direct consequence of the Heisenberg uncertainty principle.

\begin{figure*}[ht!]
  \includegraphics[width = 0.5\textwidth]{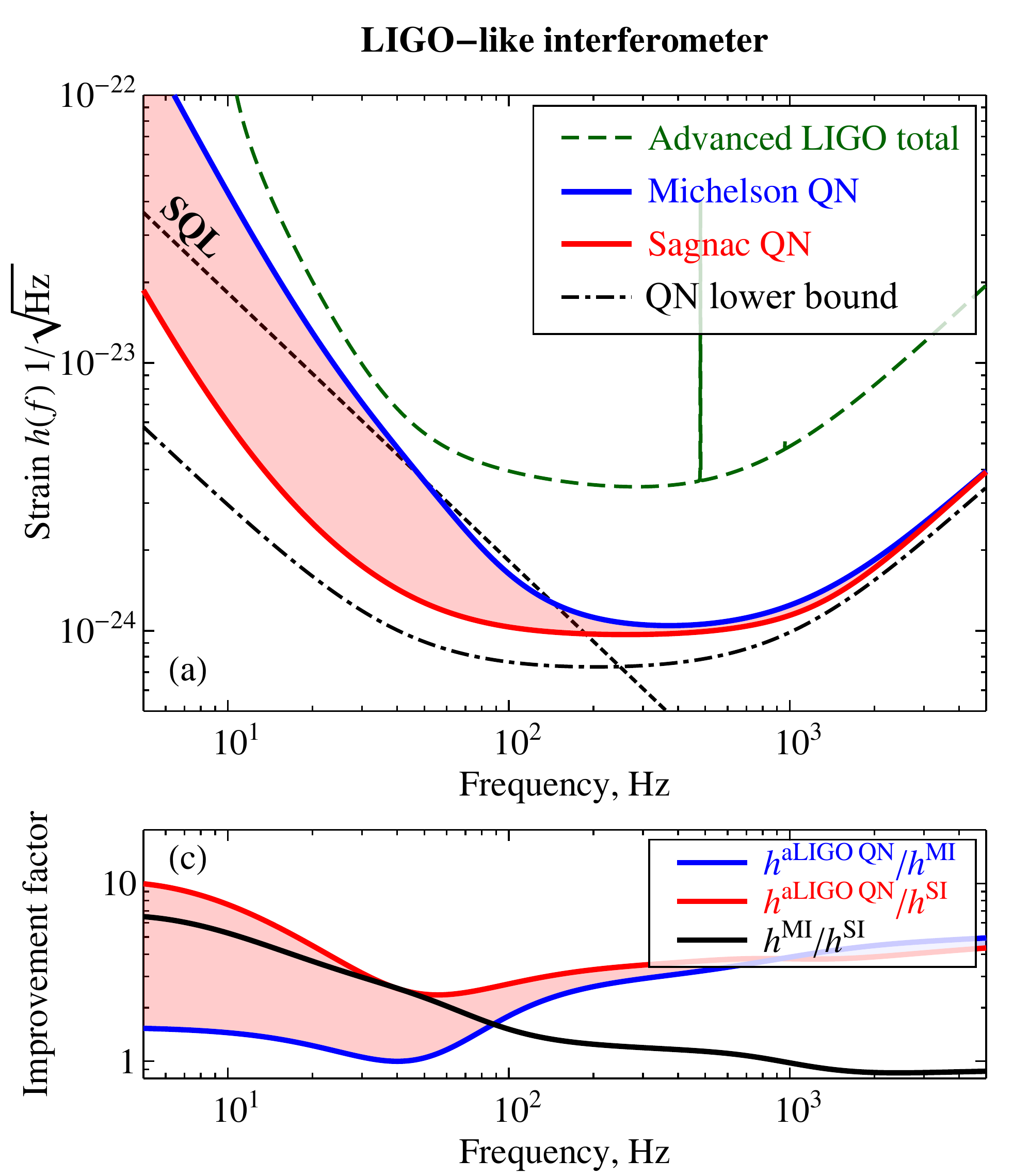}\hfill\includegraphics[width = 0.5\textwidth]{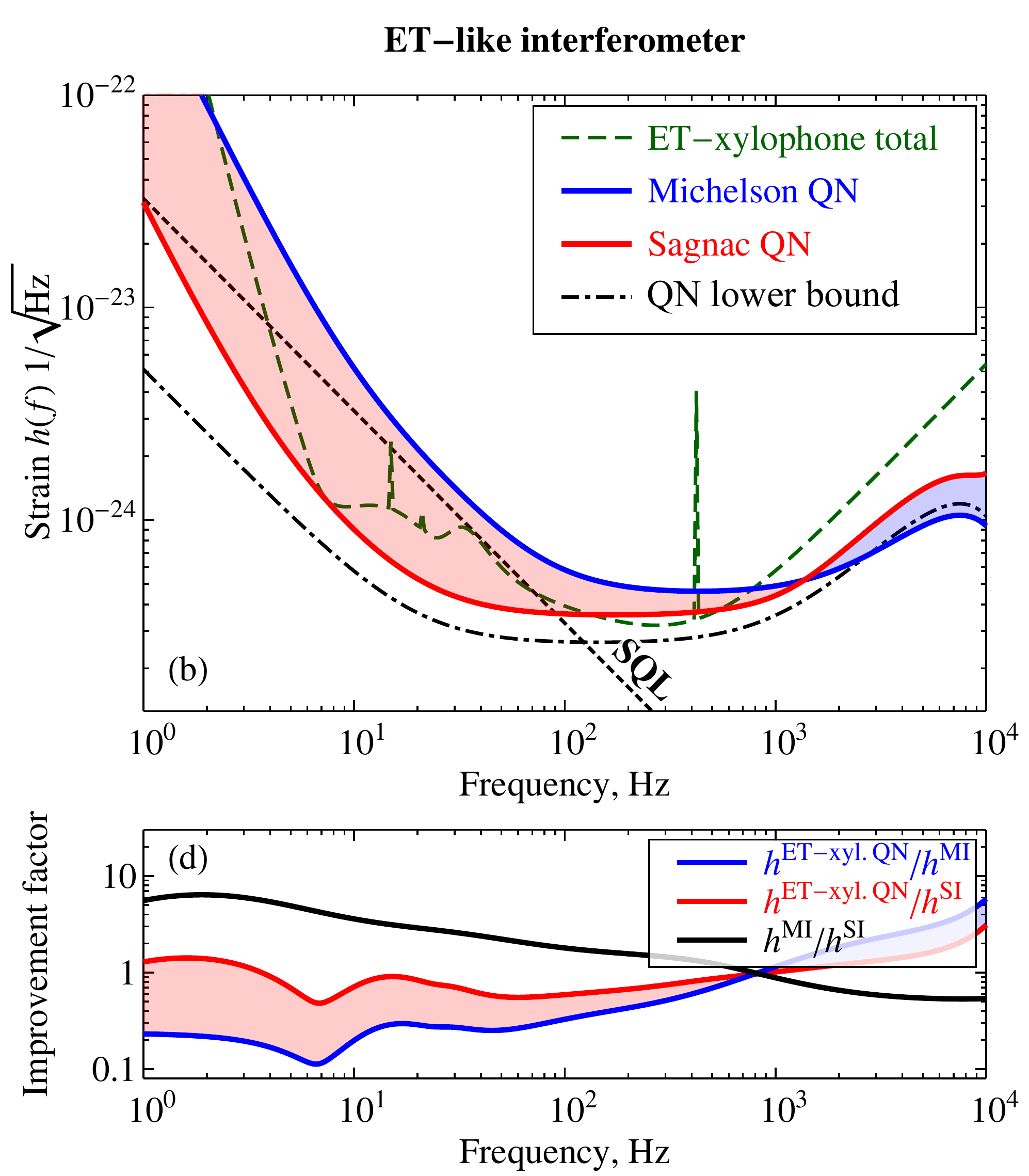}

  \caption{
  Optimal quantum noise curves of Sagnac (red line) and Michelson (blue dashed line) interferometers are shown for two variants, \textit{i.e.} LIGO-like configuration (\textit{Left panel}) features arm length $L =4$ km, mirror mass, $M = 40$ kg, and maximal circulating power in the arm, $P^{\rm max}_c = 1$~MW, while ET-like configuration (\textit{Right panel}) features arm length $L =10$ km, mirror mass, $M = 200$ kg, and maximal circulating power in the arm, $P^{\rm max}_c = 3$~MW.
  In both cases, we consider a set of QND techniques that are currently considered as the most plausible ones for implementation in the future detectors, \textit{i.e.} frequency-depended squeezed vacuum injection (filter cavity loss $A_f/L_f = 1$~ppm/m), balanced homodyne phase readout and detuned signal recycling. Optimisation of interferometer parameters is done so, as to match closely a broad-band lower bound quantum noise curve (solid black line) that is the best possible quantum noise sensitivity one can get in principle for given level of optical loss, given circulating power in the arms and fixed interferometer bandwidth (see details and discussion in Sec.~\ref{sec:RefQN}.
  For scale, we show baseline design sensitivity curves of wide-band aLIGO (\textit{Left panel, dark green dashed line}) \cite{2010_aLIGO_official_anticipated_sensitivity_curves} and of a xylophone configuration of the Einstein Telescope (\textit{Right panel, dark green dashed line}) \cite{CQG.27.1.015003_2010_Hild}. 
  Parameters of respective optical setups are given in Table~\ref{tab:SI_vs_MI_best}. 
  }\label{fig:SI_vs_MI_best}
\end{figure*}

The design and configuration of the third generation detectors remain an open question. However, there is a consensus in the community that they should have an order of magnitude better quantum noise than their predecessors \cite{CQG.27.1.015003_2010_Hild, 2011_CQG.28.9.094013_3rd_gen_design_study} in as broad frequency band as possible. At the same time, this sensitivity gain should be achieved in as economic way as possible, which implies that a single-interferometer solution should be favoured over any multi-interferometer ones. In this article, we argue that the most efficient way towards broadband quantum noise suppression goes via the change of interferometer configuration.
We show that Sagnac interferometer has superior potential for broadband sensitivity gain compared to Michelson interferometer for any given set of advanced interferometric techniques, \textit{i.e.} signal recycling \cite{Mizuno1993273,PhysRevD.38.433_1988_Vinet,PhysRevD.38.2317_1988_Meers}, squeezed vacuum injection \cite{PhysRevD.19.2888, 81a1Ca}, frequency-dependent phase rotation of squeezed vacuum phase \cite{02a1KiLeMaThVy} \textit{etc}.
We also demonstrate that better broadband performance is exhibited by Sagnac interferometer with significantly relaxed requirements to auxiliary optics and thereby at much lower cost.

\begin{table}[h]
\squeezetable
\begin{ruledtabular}
\begin{tabular}{ l c cc cc }
  Parameter                     & Notation                  & \multicolumn{2}{c}{Michelson} & \multicolumn{2}{c}{Sagnac} \\
  \hline
  Mirror mass, kg               & $m$                       &          40  &         200  &          40  &         200  \\
  Arm length, km                & $L$                       &           4  &          10  &           4  &          10  \\\
  Circ. power, MW               & $P_c$                     &         1.0  &         3.0  &         1.0  &         2.7  \\
  ITM transmittance, \%         & $T_{\rm ITM}$             &          15  &           5  &           7  &          11  \\
  Squeezing factor, dB          & $r_{\rm dB}$              &        13.2  &        13.3  &        20.0  &        19.4  \\
  \qquad (w/5\% inj. loss)      &                           &       (10.2) &       (10.3) &       (12.3) &       (12.1) \\
  Squeezing angle, deg.         & $\phi_{\rm sqz}$          & $  -6^\circ$ & $  -4^\circ$ & $ -27^\circ$ & $ -24^\circ$ \\
  Homodyne angle , deg.         & $\zeta$                   & $  84^\circ$ & $  86^\circ$ & $  63^\circ$ & $  67^\circ$ \\
  \hline \multicolumn{6}{c}{Signal recycling cavity parameters} \\ \hline 
  SRM transmittance, \%         & $T_{\rm SRM}$             &          82  &          12  &          77  &          67  \\
  SRC detuning, deg.            & $\phi_{\rm SRC}$          & $  90^\circ$ & $  90^\circ$ & $ 102^\circ$ & $  99^\circ$ \\
  \hline \multicolumn{6}{c}{Filter cavity parameters} \\ \hline 
  FC detuning, Hz               & $\delta_f$                &          34  &          13  &         525  &         456  \\
\makebox[0pt]{\hspace{-0.6em}\raisebox{0pt}[0pt][0pt]{\raisebox{-1.5ex}{$\Big[$}}}FC bandwidth, Hz              & $\gamma_{f1}$             &          34  &          17  &         922  &         767  \\
FC mirr. trans., ppm/m & $T_f/L_f$                 &         2.8  &         1.5  &        77.3  &        64.3  \\
\makebox[0pt]{\hspace{-0.6em}\raisebox{0pt}[0pt][0pt]{\raisebox{-1.5ex}{$\Big[$}}}FC loss bandwidth, Hz         & $\gamma_{f2}$             &          12  &          12  &          12  &          12  \\
FC r. trip loss, ppm/m     & $A_f/L_f$                 &        1.00  &        1.00  &        1.00  &        1.00  \\
\end{tabular}
\end{ruledtabular}
  \caption{Optimal parameters for configurations of Sagnac and Michelson interferometers which quantum noise sensitivity curves are presented in Fig.~\ref{fig:SI_vs_MI_best}. }
  \label{tab:SI_vs_MI_best}
\end{table}

In Fig.~\ref{fig:SI_vs_MI_best}, we show quantum noise sensitivities for two variants, namely for a LIGO-sized interferometer with 4-kilometre arms and 40 kilogram mirrors, and for the interferometer with parameters planned for Einstein Telescope, \textit{i.e.} for an interferometer with 10-kilometre arms and mirrors of 200 kilogram. Full list of optical parameters for these configurations that was obtained by means of optimisation procedure outlined in Sec.~\ref{sec:NumOpt} is given in Table~\ref{tab:SI_vs_MI_best}.
These results answer the question: what is the best single interferometer configuration that has quantum noise as broadband and as low as the total noise of the aLIGO broadband baseline configuration \cite{2010_aLIGO_official_anticipated_sensitivity_curves} and of the ET-D xylophone configuration \cite{CQG.27.1.015003_2010_Hild}, respectively, for a given set of advanced interferometric techniques \textit{a.k.a.} QND techniques, \textit{i.e.} frequency dependent 20 dB squeezing, detuned signal recycling and balanced homodyne detection?
Table~\ref{tab:SI_vs_MI_best} clearly demonstrates Sagnac scheme advantage over Michelson in an entire frequency band. Noteworthy is the fact that this excellent result can be achieved with a single filter cavity with rather realistic optical loss, 1 ppm/metre, that can be already achieved in the laboratory \cite{2013_PhysRevD.88.022002_RealisticFC}.

\section{On comparison of different interferometer configurations and choosing the best one.}\label{sec:RefQN}
 
Before we move on to details of quantum noise calculations of considered interferometer schemes, it is important to have an agreement on how to rate those configurations against their ability to detect GWs. Quantum noise of the 3rd generation interferometers is a complicated interplay of several advanced quantum techniques, mentioned above. It results in a quite complex dependence of the interferometer sensitivity on a multitude of parameters. This raises a question of optimisation of quantum noise curve and finding the best combination of those parameters. But what ``the best'' means in this context. How do we define the criterion that resolves what configuration is better than the others? 

Many different answers were given to this question in the literature. Some optimised the interferometer signal-to-noise ratio for specific GW sources \cite{08a1KoSiKhDa,Phys.Rev.D.86.122003}. For instance, the most popular figure of merit in GW community, the detection range, is nothing more than a renormalised SNR for the detection of GWs emitted by an etalon compact binary system (comprised of neutron stars or black holes) in the course of inspiralling phase of their evolution. More general criterion was proposed in \cite{2014_CQG_Miao_et_al}, where the optimisation seeks to provide the broadest possible total noise curve (a sum of a quantum noise and other classical noise sources), thereby seeking to include as many various astrophysical sources in the detection band of the antenna, as possible. In both cases, however, the optimisation depended heavily on the model of non-quantum noise for a specific project, \textit{i.e.} aLIGO, or Einstein Telescope. 

We argue that this approach is very restrictive and does not allow to reveal the potential of different interferometer configurations in full, for it is squeezed into a narrow frame of existing noise models. Nevertheless, we know examples when a new technology comes into play and predictions of classical noise models have to be revised dramatically \cite{2013_nat.Photon.7.8.644_Crystallyne_coatings}. 

In this paper, we suggest that quantum noise of different configurations shall be compared against the limitations of quantum origin. Namely, we suggest that a lower bound of quantum noise, which all considered configurations shall be compared against, is to be derived as a sum of limitations imposed by two main parameters for any quantum noise-limited interferometer, \textit{i.e.} the level of optical loss in the main optics and the finite optical power circulating therein.

Incoherent loss-associated vacuum fields that enter the interferometer in accordance with the fluctuation-dissipation theorem \cite{PhysRev.83.34} create random radiation pressure force that cannot be compensated by any quantum technique. If, due to various reasons, the total fraction $\epsilon_{\rm loss}$ of all photons that enter the interferometer is lost therein, and if the effective squeezing of vacuum injected in the dark port is given by a factor $e^{-r_{\rm eff}}$ [see eq. \eqref{eq: r_eff def}], the ultimate limit for residual radiation pressure noise (in GW strain spectral density units) reads (cf. Eq.~(413) in \cite{Liv.Rv.Rel.15.2012}):
\begin{equation}\label{eq:LossQlimit}
  S^h_{\rm RP\ loss}(\Omega) \simeq h^2_{\rm SQL}(\Omega) \epsilon_{\rm loss}^{1/2}e^{-r_{\rm eff}}\,,
\end{equation}
where 
\begin{equation}\label{eq:hSQL}
h_{\rm SQL}(\Omega) = \sqrt{\dfrac{8\hbar}{ML^2\Omega^2}}\,,
\end{equation}
is the Standard Quantum Limit of the interferometer in terms of GW strain\footnote{\label{footnote: dARM}Here we used a particular formula for the SQL of an interferometer with 4 test masses of equal value $M$ and the mechanical \textit{dARM}-mode defined as $x_{\rm dARM} = (x^N_{\rm ETM}-x^N_{\rm ITM}) - (x^E_{\rm ETM}-x^E_{\rm ITM})$.}
We will use this limit as a first component of our QN lower bound curve.
 
High frequency region of all interferometers is dominated by shot noise, or quantum phase fluctuations. Its rise on upper frequencies is determined by a finite bandwidth, $\gamma$, of the detector. For a simple resonance-tuned Fabry-Perot interferometer with squeezed vacuum and lossy optics, a shot noise contribution can be written as:
\begin{equation}\label{eq:ShotNoiseQlimit}
  S^h_{\rm SN}(\Omega) = \frac{h^2_{\rm SQL}(\Omega)}{2} \frac{e^{-2 r_{\rm eff}}+\epsilon_{\rm loss}}{\mathcal{K}_{\rm eff}(\Omega)}\,,
\end{equation}
where $\mathcal{K}_{\rm eff}$ is the frequency-dependent optomechanical coupling strength introduced by Kimble \textit{et al.} \cite{02a1KiLeMaThVy} and equal to:
\begin{equation}\label{eq:KFPeff}
  \mathcal{K}_{\rm eff}(\Omega) = \frac{\Theta\tau}{\Omega^2}\frac{2\,T_{\rm eff}}{1+2\sqrt{R_{\rm eff}}\cos2\Omega\tau+R_{\rm eff}}\,,
\end{equation}
with $\Theta = 4\omega_0 P_c/(McL)$ and $P_c$ is the total optical power circulating in the interferometer, $\tau = L/c$ is light travel time between the mirrors of the cavity and  $R_{\rm eff} = 1-T_{\rm eff}$ is the effective reflectivity of the cavity. According to \textit{scaling law} derived by Buonanno and Chen in \cite{2003_PhysRevD.67.062002_ScalLaw}, the Fabry-P\'erot interferometer is fully equivalent to a signal-recycled Fabry-P\'erot--Michelson interferometer in terms of quantum noise with (in the  the only substitution of $P_c = P_{\rm eff}$ where $P_{\rm eff}$ stands for light power circulating in the first one.

For fixed bandwidth $\gamma=cT_{\rm eff}/4L = 2\pi \cdot 1000$~s$^{-1}$, circulating power $P_c$ and achievable effective squeezing at the dark port of the interferometer, $e^{-r_{\rm eff}}$, Eq.~\eqref{eq:ShotNoiseQlimit} sets the ultimate high frequency limit on quantum noise for interferometers, not using active optomechanical amplification techniques \cite{2015_arXiv:1501.01349_Ma}. Therefore we will use it as the second component of our QN lower bound curve, which can now be expressed in terms of lower bound quantum noise spectral density as:
\begin{equation}\label{eq:RefSQN}
  S^h_{\rm ref}(\Omega) = S^h_{\rm RP\ loss}(\Omega)  + S^h_{\rm SN}(\Omega)\,.
\end{equation}

The resulting lower bound quantum noise curves (for LIGO-like and for ET-like interferometers) are plotted in Fig.~\ref{fig:SI_vs_MI_best} and will be used throughout the rest of the paper as yardsticks for different configurations of interferometers. 

\begin{figure*}[ht]
 \includegraphics[width=.49\textwidth]{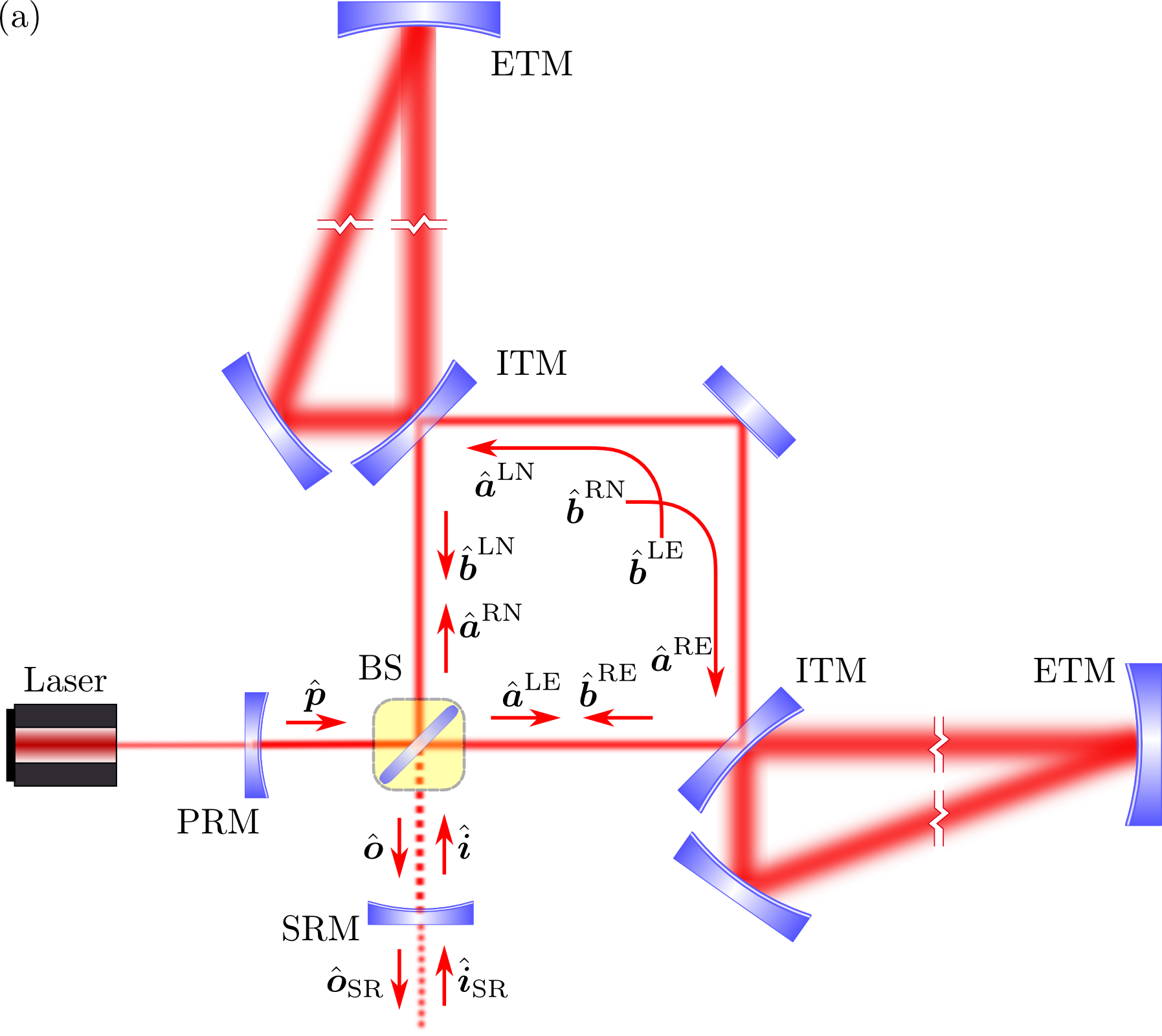} \hfill \includegraphics[width=.49\textwidth]{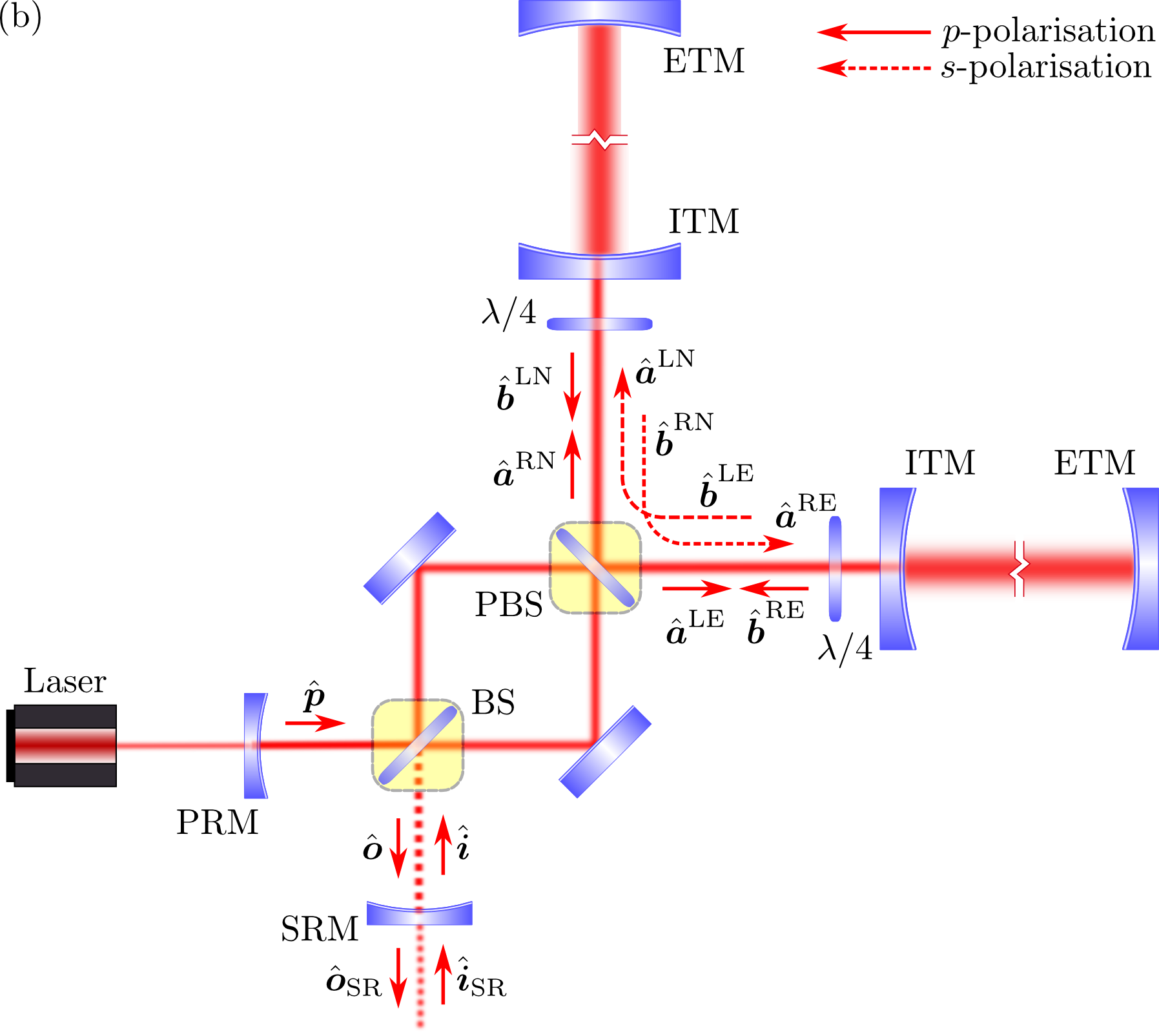} \\
 \vspace{0.5em}
 \includegraphics[width=.49\textwidth]{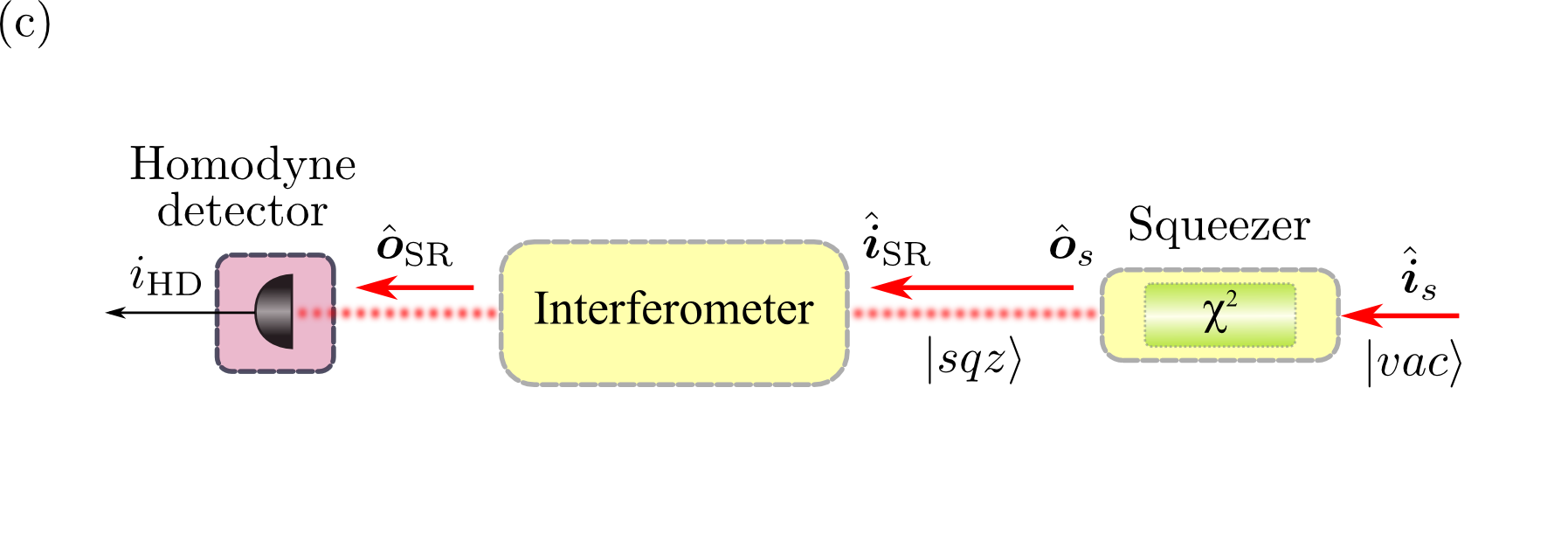} \hfill \includegraphics[width=.49\textwidth]{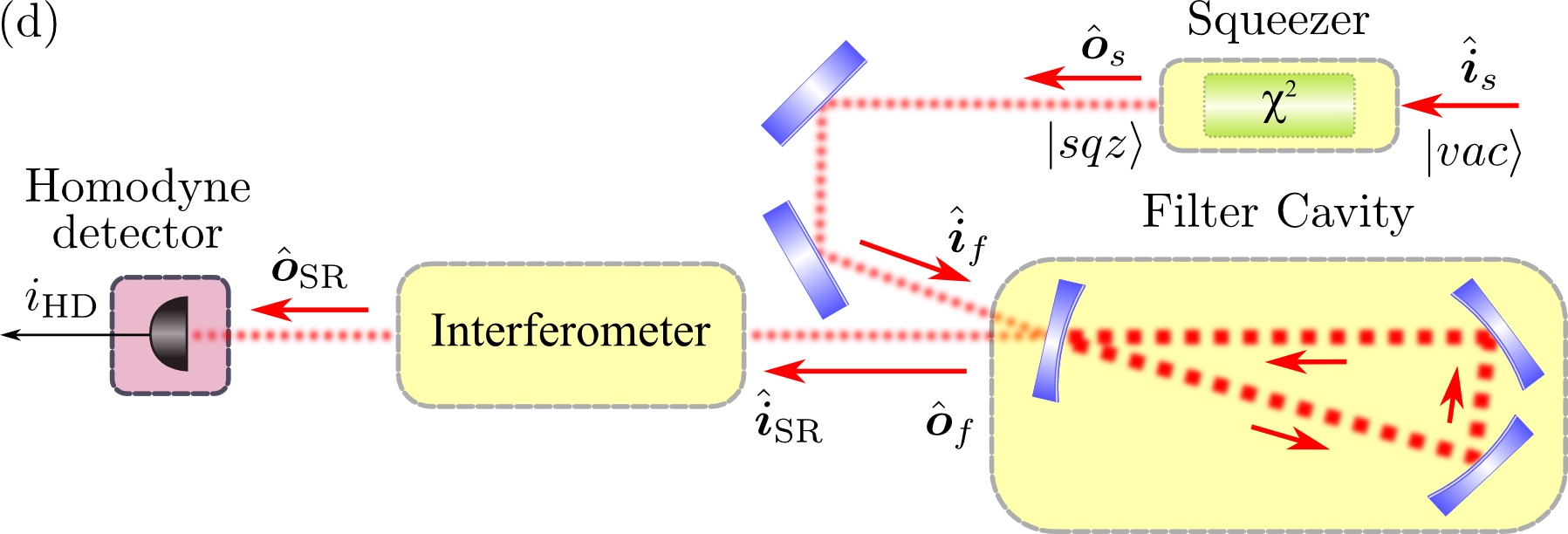}
 \caption{
  Two possible optical realisations of zero-area Sagnac speed meter.
  \textit{Upper left panel:} Sagnac interferometer with triangular ring cavities in the arms as proposed in the original work by Chen \cite{Chen2002}.
  \textit{Upper right panel:} Sagnac interferometer with linear Fabry-P\'erot cavities in the arms and polarised optics, consisting of additional polarisation beam splitter and  $\lambda/4$-plates, used to direct light from one arm to another as per proposal by Danilishin \cite{04a1Da} and later analysed by Wang \textit{et al.} in \cite{PhysRevD.87.096008}. Lower panels show simplified schematics of frequency-independent (\textit{lower left panel}) and frequency-dependent squeezed vacuum injection (\textit{lower right panel}) and serve to introduce notations for input and output fields in both situations.
}\label{fig:Sagnacfig}
\end{figure*} 

\begin{figure}[ht]
 \includegraphics[width=.49\textwidth]{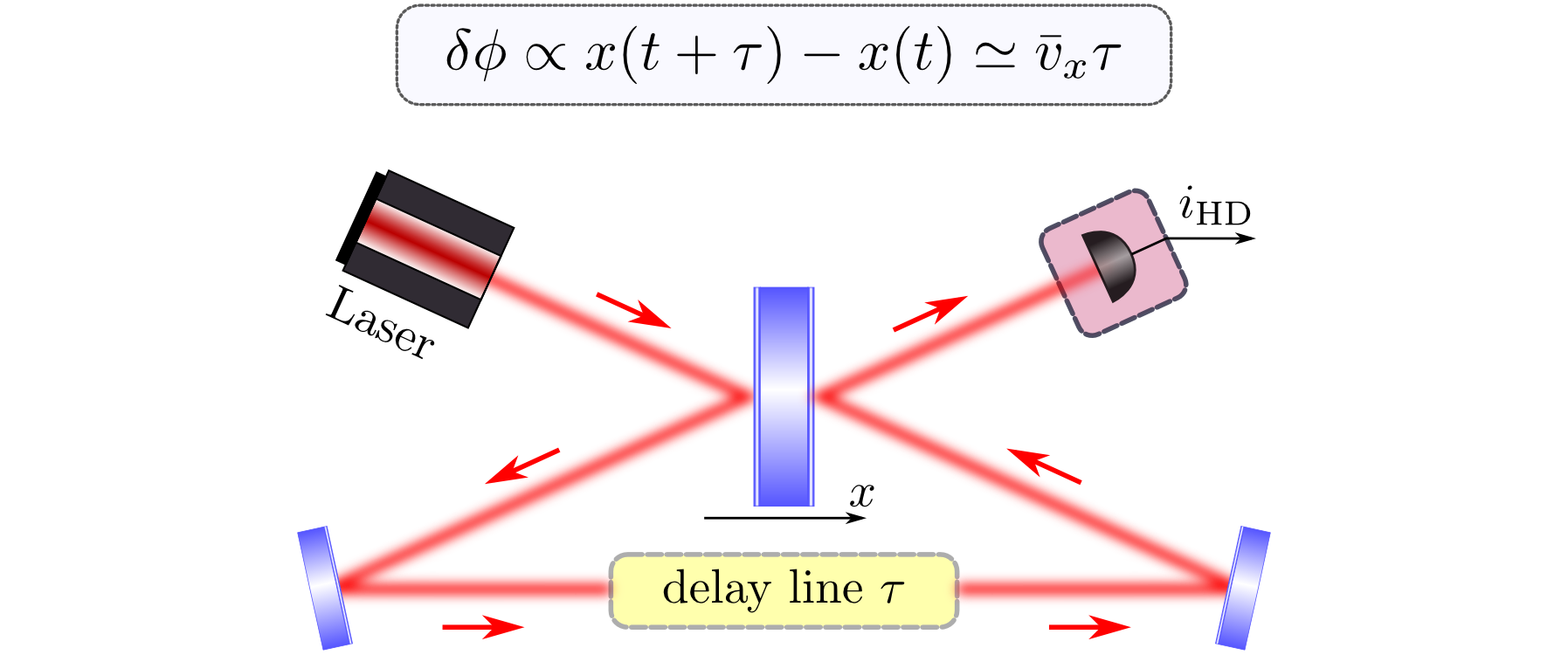}
 \caption{Thought experiment on test object speed measurement: The free mirror is sensed twice by the same laser light that is reflected from both the front and the rear surfaces thereof with a time delay $\tau$ between reflections. The phase of outgoing light is measured by, say homodyne detector, and is proportional to the the difference of the successive mirror coordinates: $\phi_{\rm out}\propto (x(t+\tau)-x(t))\simeq \bar v\tau$, where $\bar v$ stands for the mean velocity of the mirror over the interval $\tau$. }\label{fig:speed_measurement}
\end{figure} 

\section{Quantum noise of the signal-recycled Sagnac interferometer}
In this section, we concentrate solely on quantum fluctuations of light and the influence they have on the sensitivity of the GW detector based on Sagnac interferometer topology.

\subsection{Suppression of radiation pressure noise in Sagnac interferometer.}\label{ssec:RPNsuppress}
The power of Sagnac interferometer (SI) to reduce back-action noise is nested in its ability to sense the relative rate, or in other words speed of an arm cavities length variation, whereas Michelson interferometer senses arms length variation itself. The simple way to understand how a speed measurement can reduce back-action is to consider a simple thought experiment depicted in Fig.~\ref{fig:speed_measurement}. Here the free mirror is sensed twice by the same laser light that is reflected from both the front and the rear surfaces thereof with a time delay $\tau$ between reflections. The phase of outgoing light is measured by, say homodyne detector, and is proportional to the the difference of the succesive mirror coordinates: $\phi_{\rm out}\propto (x(t+\tau)-x(t))\simeq \bar v\tau$, where $\bar v$ stands for the mean velocity of the mirror over the interval $\tau$. If the signal force one seeks to measure, watching the change of the mirror velocity, has characteristic frequency $\Omega$ much smaller than 
$\tau^{-1}$, the two kicks light gives to the mirror on the consecutive reflection partly compensate each other and the resulting back-action force turns out to be depressed by a factor $\propto\Omega\tau\ll1$:
\begin{equation}\label{eq:F_RP_speedmeter}
  \hat{F}_{\rm b.a.}(\Omega) \simeq -i\Omega \tau\frac{2 \bar P_{\rm pulse}}{c}\,,
\end{equation} 
as compared to the back-action of single light pulse with an average power $\bar P_{\rm pulse}$ which one expects in a single reflection experiment sensitive to the test mass displacement.

As was realised by Chen and Khalili \cite{Chen2002,02a2Kh}, that the zero area SI \cite{1999_JOSAB.16.9.1354_Beyersdorf,1999_Opt.Lett.24.16.1112_Beyersdorf} actually implements the initial double-measurement variant of the quantum speed meter, shown in Fig.\,\ref{fig:speed_measurement}. Indeed, visiting consequently both arms (see Fig.~\ref{fig:Sagnacfig}), counter propagating light beams acquire phase shifts proportional to a sum of arms length variations $x_{N,E}(t) \equiv (x_{\rm ETM}^{N,E}(t) - x_{\rm ITM}^{N,E}(t))$ (hereinafter I(E)TM stands for Input (End) Test Mass) for  of both cavities taken with time delay equal to average single cavity storage time $\tau_{\rm arm}$:
\begin{eqnarray}
    \delta\phi_R &\propto & x_N(t) + x_E(t+\tau_{\rm arm})\,,\\
    \delta\phi_L &\propto & x_E(t) + x_N(t+\tau_{\rm arm})\,.
\end{eqnarray}
After recombining at the beam splitter and photo detection the output signal will be proportional to the phase difference of clockwise (R) and counter clockwise (L) propagating light beams:
\begin{multline}\label{phi_speedmeter}
 \delta\phi_R - \delta\phi_L \propto [x_N(t) - x_N(t+\tau_{\rm arm})] - \\
   -[x_E(t) - x_E(t+\tau_{\rm arm})]\propto \\
  \propto \dot{x}_N(t) - \dot{x}_E(t) + O(\tau_{\rm arm})
\end{multline}
that, for frequencies $\Omega \ll \tau_{\rm arm}^{-1}$, is proportional to relative rate of the interferometer arms length variation.

\subsection{Transfer matrix for calculation of quantum noise.}

As shown by Caves and Schumaker in \cite{85a1CaSch, 85a2CaSch}, quantum noise of light in any parametric optical device can be conveniently described within the frames of two-photon formalism. Namely, noise can be considered as tiny stochastic variations of quadratures of carrier field travelling through the device. Any variations of interferoemeter parameters due to signal force, {\it e.g.} differential arms length change in GW interferometer, also lead to variation of outgoing light quadratures and thus can be described using the same formalism.

In two-photon formalism, one starts with writing the input and outgoing light fields of the interferometer at some fixed location via its {\it sine} and {\it cosine} quadratures:
\begin{eqnarray}
  \hat{E}^{in}(t) &=& \mathcal{E}_0\left[(A^{in}+\hat{a}^{in}_c)\cos\omega_p t+\hat{a}^{in}_s\sin\omega_pt\right]\,,\\
  \hat{E}^{out}(t) &=& \mathcal{E}_0\left[(B^{out}+\hat{b}^{out}_c)\cos\omega_p t+\hat{b}^{out}_s\sin\omega_pt\right]\,,
\end{eqnarray}
where $A^{in}$ ($B^{out}$) is classical mean amplitude of the input (output) light at pump laser frequency $\omega_p$ and $\hat{a}^{in}_{c,s}$ ($\hat{b}^{in}_{c,s}$) are small quantum variations and variations due to signal force that can be defined through creation and annihilation operators as
\begin{equation}
  \hat{a}_c = \frac{\hat{a}+\hat{a}^\dag}{\sqrt{2}}\,,\quad \mbox{and}\quad\hat{a}_s = \frac{\hat{a}-\hat{a}^\dag}{i\sqrt{2}}\,,
\end{equation}
and similarly for outgoing fields. Note that we do not specify time as an argument here, as the same definition holds in spectral domain that we make use of in the rest of this article. These two domains are connected through Fourier transform as follows:
\begin{equation*}
  \hat{a}_{c,s}(t) = \intinfty \frac{d\Omega}{2\pi} \hat{a}_{c,s}(\Omega)e^{-i\Omega t}\,.
\end{equation*} 

So, in order to fully describe signal and noise in a (lossless) GW interferometer it is sufficient to track the transformation it does to the quadrature operators of input light that can be represented in a matrix form as transformations of 2-dimensional vectors $\vq{a} = \{\hat{a}_c,\,\hat{a}_s\}^{\rm T}$ and $\vq{b} = \{\hat{b}_c,\,\hat{b}_s\}^{\rm T}$ and GW signal $h(\Omega)$. For GW detector it can be written in general form as:
\begin{equation}\label{eq:IOlossless}
  \vq{b} = \tq{T}\cdot\vq{a} + \vs{t}\frac{\mathcal{X}}{\mathcal{X}_{\rm SQL}}\,,
\end{equation}
where
\begin{equation}\label{eq:T_def}
  \tq{T} \equiv
  \begin{bmatrix}
    T_{cc}(\Omega) & T_{cs}(\Omega)\\
    T_{sc}(\Omega) & T_{ss}(\Omega)
  \end{bmatrix}
\end{equation}
is the optical transfer matrix of the interferometer,
\begin{equation}\label{eq:t_h_def}
  \vs{t} \equiv
  \begin{bmatrix}
    t_c(\Omega)\\
    t_s(\Omega)
  \end{bmatrix}
\end{equation}
is an optical response of the interferometer on an external influence that is denoted as $\mathcal{X}$, and $\mathcal{X}_{\rm SQL}$ is the corresponding SQL for the mechanical degree of freedom of the interferometer expressed in units of $\mathcal{X}$. In GW interferometry $\mathcal{X}$ is either the displacement of the mirrors, $x$, or the external force, $F$, that causes this displacement, or, more habitually, the GW strain, $h$. In each case, the corresponding SQL applies. The relation between these three quantities is discussed 
in Sec.~4.3 of \cite{Liv.Rv.Rel.15.2012}.

Note, that the I/O-relation form \eqref{eq:IOlossless} means for optical detuning-less system the following expression:
\begin{equation*}
  \vq{b} = e^{2i\beta(\Omega)} \matr{1}{0}{-\mathcal{K}(\Omega)}{1} \vq{a} +
    e^{i\beta(\Omega)}\col{0}{\sqrt{2\mathcal{K}(\Omega)}} \frac{\mathcal{X}}{\mathcal{X}_{\rm SQL}} \,,
\end{equation*}
where $\mathcal{K}$ is a Kimble optomechanical coupling factor and $\beta$ is some frequency-dependent phase shift optical fields acquire when passing through the interferometer and that does not affect the expression for spectral density of quantum noise.

If one assumes the interferometer readout quantity $\hat{o}_\zeta$ (proportional to homodyne detector photo-current $\hat{i}^{out}_\zeta$)  as a result of a homodyne detection with a homodyne angle $\zeta$, {\it i.e}:
\begin{equation}\label{eq:o_zeta_lossless}
  \hat{o}_\zeta \equiv \hat{b}_c\cos\zeta + \hat{b}_s\sin\zeta \equiv \vs{H}_\zeta^{\rm T}\cdot\vq{b}\,,\
  \vs{H}_\zeta\equiv
  \begin{bmatrix}
    \cos\zeta\\
    \sin\zeta
  \end{bmatrix}\,,
\end{equation}
the spectral density of quantum noise at the output port of the interferometer can be obtained using the following simple rule:
\begin{equation}\label{eq:SpDens_h}
  S^h(\Omega) = h^2_{\rm SQL}\frac{\vs{H}^\tr_\zeta\cdot\tq{T}\cdot\tq{S}_{a}^{in}\cdot\tq{T}^\dag\cdot\vs{H}_\zeta}{|\vs{H}^\tr_\zeta\cdot\vs{t}_h|^2}
\end{equation}
where $\tq{S}_a^{in}$ stands for spectral density matrix of injected light and components thereof can be defined as:
\begin{multline}\label{eq:SpDens_a}
  2\pi\delta(\Omega-\Omega') \, \tq{S}_{a, ij}^{in}(\Omega) \equiv \\
    \frac12\bra{in}\hat{a}_i(\Omega)(\hat{a}_j(\Omega'))^\dag+(\hat{a}_j(\Omega'))^\dag\hat{a}_i(\Omega)\ket{in}\,,
\end{multline}
where $\ket{in}$ is the quantum state of vacuum injected in the dark port of the interferometer and $(i,j) = \{c,s\}$  (see Sec. 3.3 in \cite{Liv.Rv.Rel.15.2012} for more details).
In present article we deal with \emph{single-sided} spectral densities $S$ and hence in case of input vacuum state:
\begin{equation*}
  \ket{in} = \ket{vac} \qquad\Rightarrow\qquad \tq{S}_a^{in} = \matr{1}{0}{0}{1} \,.
\end{equation*}

Now with all necessary tools in hand we can proceed with the analysis of Sagnac interferometer and derive spectral density of quantum noise for it.

\subsection{Sagnac interferometer I/O relations.}

First we consider a bare lossless zero-area Sagnac interferometer and derive its input-output (I/O) relations.\SD{For definiteness, in this section, we stick to a configuration of Sagnac interferometer that utilises ring arm cavities (as per the left panel of Fig.~\ref{fig:Sagnacfig}), although the results we obtain are applicable to both realisations unless loss is taken into account.} 

Unlike Michelson interferometer, in Sagnac interferometer light beam visits two arm cavities before recombination with a counter-rotating beam at the beam splitter (see Fig.~\ref{fig:Sagnacfig}). At the same time, two light beams hit the cavity, one coming directly from the beam splitter and the one, that has just left another arm. In notations of Chen's paper \cite{Chen2002} quadrature operators of light enter ing and leaving the arm can be identified with two indices $IJ$, \textit{e.g.} $a_c^{IJ}$, where $I$ stands for the either of two beams, L or R, and $J$ stands for the either of two arms ($J=E,N$). Here $R$ marks the light beam that first enters North arm and then travels the interferometer in the right direction (clockwise), and $L$ marks the beam travelling the interferometer in the opposite (counterclockwise) direction after entering the interferometer through the East arm. Thus, single lossless arm 
I/O relations read, assuming high-finesse arm cavities ($\TITM\ll1$, for general case see Appendix~\ref{app:transfer matrices}):
\begin{eqnarray}
  b_c^{IJ} &=&   e^{2i\beta_{\rm arm}(\Omega)} a_c^{IJ}\,, \\ 
  b_s^{IJ} &=&   e^{2i\beta_{\rm arm}(\Omega)} [a_s^{IJ} - \SD{\mathcal{K}_{\rm arm}} (a_c^{IJ}+a_c^{\bar IJ})] \nonumber\\
           & & + e^{i\beta_{\rm arm}(\Omega)} \SD{\sqrt{2\mathcal{K}_{\rm arm}} \frac{\sqrt{2}x_J}{h_{\rm SQL}L}} 
\end{eqnarray}
with $\bar I$ indicating the other beam than $I$, \textit{i.e.} $\bar R = L$ and $\bar L = R$, $h_J = x_J^{\rm ETM}-x_J^{\rm ITM}$ is the arm elongation induced by signal force (\textit{e.g.} gravitational wave tidal force), and
\begin{eqnarray}
  & \mathcal{K}_{\rm arm} &= \SD{\dfrac{\Theta_{\rm arm}\tau}{\Omega^2} \dfrac{2\,\TITM}{1-2\sqrt{\RITM} \cos 2\Omega\tau + \RITM} \simeq}\nonumber\\
  & &\SD{\simeq \frac{2 \Theta_{\rm arm} \gamma_{\rm arm}}{\Omega^2(\gamma_{\rm arm}^2+\Omega^2)}}\,,\label{eq:Karm}\\
  & \beta_{\rm arm} &= \atan{ \dfrac{1+\sqrt{\RITM}}{1-\sqrt{\RITM}} \tan\Omega\tau }\SD{ \simeq }\nonumber\\
  & &\SD{\simeq  \atan{\Omega/\gamma_{\rm arm}}}\label{eq:Betaarm}\,,
\end{eqnarray}
with $\Theta_{\rm arm} = 4\omega_0 P_{\rm arm}/(McL)$ and $P_{\rm arm} = P_c/4$, where $P_c$ is the total optical power circulating in both arms
and $\gamma_{\rm arm} = \TITM/(4\tau)$ is the half-bandwidth of an arm cavity. The final, approximate expressions above are obtained assuming that cavity linewidth and signal frequency are much smaller than cavity free spectral range $\nu_{\rm FSR}=(2\tau)^{-1}$. This approximation nearly breaks down for detectors with arm length $\gtrsim 10$~km, like Einstein Telescope, at frequencies of the order of 10~kHz, therefore we present exact formulae as well.

Then, it is straightforward to derive full Sagnac I/O-relations, using junction equations for the fields at the output beam splitter (ring-cavity topology):
\begin{eqnarray}\label{i-o relations at BS in SI, p.1}
  \vq{a}^{\rm RN} = \dfrac{\vq{p}+\vq{i}}{\sqrt{2}} \,, &
  \vq{a}^{\rm LE} = \dfrac{\vq{p}-\vq{i}}{\sqrt{2}} \,, &
  \vq{o} = \dfrac{\vq{b}^{\rm LN}-\vq{b}^{\rm RE}}{\sqrt{2}} \,,
\end{eqnarray}
as well as continuity relations between the beams that leave one arm and enter the other:
\begin{eqnarray}\label{i-o relations at BS in SI, p.2}
  \vq{a}^{\rm RE} = \vq{b}^{\rm RN} \,, &
  \vq{a}^{\rm LN} = \vq{b}^{\rm LE} \,.
\end{eqnarray}
The resulting I/O-relations for lossless zero-area Sagnac interferometer then read:
\begin{equation}\label{eq:IOSaglossless}
  \begin{bmatrix} \hat{o}_c\\\hat{o}_s\end{bmatrix} =
  e^{2i\beta_{\rm SI}}\begin{bmatrix} 1 & 0\\-\mathcal{K}_{\rm SI} & 1\end{bmatrix}
  \begin{bmatrix} \hat{i}_c\\\hat{i}_s\end{bmatrix}+
  \begin{bmatrix}0\\\sqrt{2\mathcal{K}_{\rm SI}}\end{bmatrix}e^{i\beta_{\rm SI}}
\frac{h}{h_{\rm SQL}}\,,
\end{equation}
with coupling constant $\mathcal{K}_{\rm SI}$ defined as:
\begin{equation}
  \mathcal{K}_{\rm SI} = 4\,\mathcal{K}_{\rm arm} \sin^2\beta_{\rm arm}
  \SD{\simeq \frac{4 \Theta_{\rm SI} \gamma_{\rm arm}}{(\Omega^2+\gamma_{\rm arm}^2)^2}},
\end{equation}
where $\Theta_{\rm SI} \equiv 4\Theta_{\rm arm}$ and additional phase shift:
\begin{equation}\label{eq:SIphase}
  \beta_{\rm SI} = 2\beta_{\rm arm} + \frac{\pi}{2}
\end{equation}

One can now calculate spectral density of quantum noise of the zero-area Sagnac, using Eq.~\eqref{eq:SpDens_h}, where transfer matrix $\tq{T}$ and response vector $\vs{t}$ read:
\begin{equation}\label{eq:SI_T_and_t}
  \tq{T} = e^{2i\beta_{\rm SI}}\begin{bmatrix} 1 & 0\\-\mathcal{K}_{\rm SI} & 1\end{bmatrix}\,,\quad\vs{t} = e^{i\beta_{\rm SI}} \begin{bmatrix}0\\\sqrt{2\mathcal{K}_{\rm SI}}\end{bmatrix}\,.
\end{equation}
Therefore one gets this simple expression for spectral density (it is the same for all tuned interferometers with balanced homodyne readout of quadrature $b_\zeta$ and vacuum state at the dark port, save to the expression for $\mathcal{K}$):
\begin{equation}\label{eq:Sh_plain}
   S^h = \frac{h^2_{\rm SQL}}{2}\left\{\frac{\left[\mathcal{K}_{\rm SI}-\cot\zeta\right]^2+1}{\mathcal{K}_{\rm SI}}\right\}\,.
\end{equation}

\subsection{Signal recycling in Sagnac interferometer}\label{sec:SR_Sag}

\begin{figure}[ht!]
  \includegraphics[width = 0.5\textwidth]{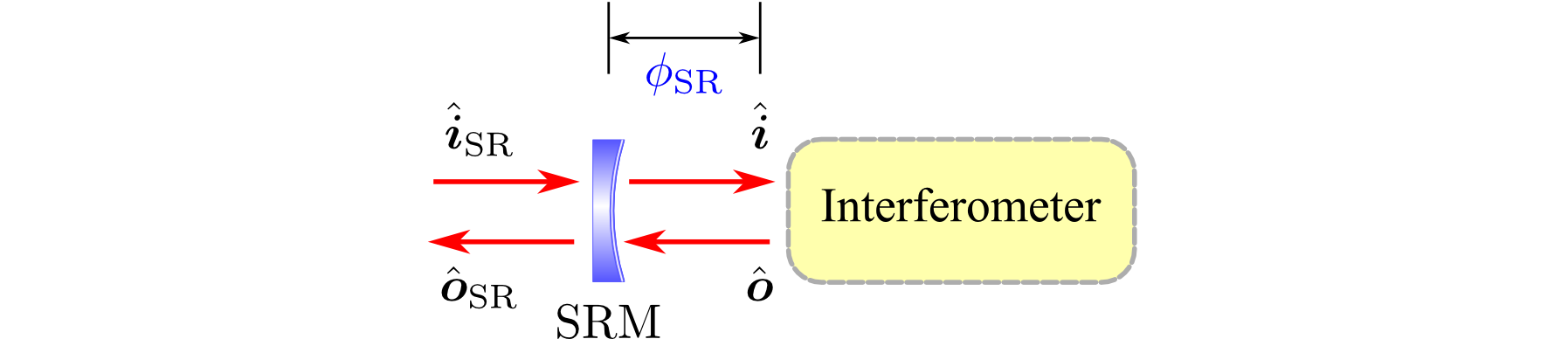} \\
  \caption{Schematic of I/O-relations for a signal-recycled interferometer.}\label{fig:SRCzoom}
\end{figure}

Consider now how an addition of signal recycling mirror in the dark port of the Sagnac interferometer changes its quantum noise. A scheme of signal recycling cavity is shown in Fig.~\ref{fig:SRCzoom}. Essentially, it acts as a feedback loop for the light leaving the main interferometer, and its sign is defined by the phase shift signal light acquires in a round trip across the signal recycling cavity (SRC). The feedback features of the SRC become evident if one writes down the I/O-relations for the lossless signal recycled interferometer in matrix form:
\begin{equation}
  \vq{o}_{\rm SR} = \tq{T}_{\rm SR}\vq{i}_{\rm SR} + \vs{t}_{\rm SR} (h/h_{\rm SQL})\,,
\end{equation}
where
\begin{subequations}\label{eq:SR_T_and_t}
\begin{align}
\tq{T}_{\rm SR} &= -\rho_{\rm SR}\tq{I}+ \tau_{\rm SR}^2\tq{R}_{\phi_{\rm SR}}(\tq{I}-\rho_{\rm SR}\tq{T}\tq{R}_{2\phi_{\rm SR}})^{-1}\tq{T}\tq{R}_{\phi_{\rm SR}}\\
\vs{t}_{\rm SR} &= \tau_{\rm SR} \tq{R}_{\phi_{\rm SR}}(\tq{I}-\rho_{\rm SR}\tq{T}\tq{R}_{2\phi_{\rm SR}})^{-1}\vs{t}
\end{align}
\end{subequations}
with $\rho_{\rm SR}$ and $\tau_{\rm SR}$ are SR mirror amplitude reflectivity and transmissivity (sign notations are given in Fig.~\ref{fig:SRCzoom}), $\tq{I}$ is identity matrix and 
\begin{equation}\label{eq: Rot Matr RR}
  \tq{R}_{\varphi} = \begin{bmatrix}\cos\varphi & -\sin\varphi \\ \sin\varphi & \cos\varphi \end{bmatrix} \,,
\end{equation}
is rotation matrix responsible for propagation of light by a distance $l_\varphi$ resulting in a phase shift $\varphi = \omega_0 l_\varphi/c$. 

These expressions are universal and allow to obtain transfer matrices for any (lossless) signal-recycled interferometer characterised by a bare transfer matrix $\tq{T}$ and response vector $\vs{t}$. A general case of arbitrary SRC detuning and lossy interferometer is considered in Appendices~\ref{app:SRIOrels}~and~\ref{app:PBS leak I/O-relations}. Here we only give expressions for two special cases known as Resonant Sideband Extraction (RSE)  \cite{Mizuno1993273,1995_PhD_Mizuno,1999_PhD_Heinzel} and Tuned Signal Recycling (TSR) \cite{PhysRevD.38.433_1988_Vinet,PhysRevD.38.2317_1988_Meers} that stand for the cases of SRC tuned in anti-resonance and in resonance, respectively. 

In terms of control theory, the first one corresponds to a negative feedback, as the light reflected off the SR mirror enters the interferometer with the opposite sign ($2\phi_{\rm SR} = \pi$ phase shift in a full round trip in the SRC). The second case gives a positive feedback (light returns to the interferometer in the same phase). Intermediate case is more complicated as it involves modification of the test masses dynamics by means of creating an optical spring \cite{2002_PhysRevD.65.042001_OptSpr1,2004_PhysRevD.69.102004_OptSpr2}.

Substituting Eq.~\eqref{eq:SI_T_and_t} into Eq.~\eqref{eq:SR_T_and_t} with $\phi_{\rm SR} = 0$ (TSR) and $\pi/2$ (RSE), one arrives at I/O-relations of a form identical to that of Eq.~\eqref{eq:SI_T_and_t}, but with modified optomechanical coupling strength and overall phase shift that read:

\begin{align}
 \mathcal{K}_{\rm SI, SR}^{\pm} &= \dfrac{\tau_{\rm SR}^2\mathcal{K}_{\rm SI}}{|1\pm\rho_{\rm SR}e^{2i\beta_{\rm SI}}|^2} = \dfrac{\tau_{\rm SR}^2\mathcal{K}_{\rm SI}}{|1\mp\rho_{\rm SR}e^{4i\beta_{\rm arm}}|^2}\,,\\
 \beta^{\rm \pm}_{\rm SI, SR} &= \atan{\dfrac{1\pm\rho_{\rm SR}}{1\mp\rho_{\rm SR}}\tan\beta_{\rm SI}} =\nonumber\\
  &= - \acot{\dfrac{1\mp\rho_{\rm SR}}{1\pm\rho_{\rm SR}}\tan2\beta_{\rm arm}}\,,
\end{align}
where we denote TSR case by "+" and RSE case by "-" in accordance with the sign of feedback it creates. 

\begin{figure*}[ht]
  \centering
  \includegraphics[width=\textwidth]{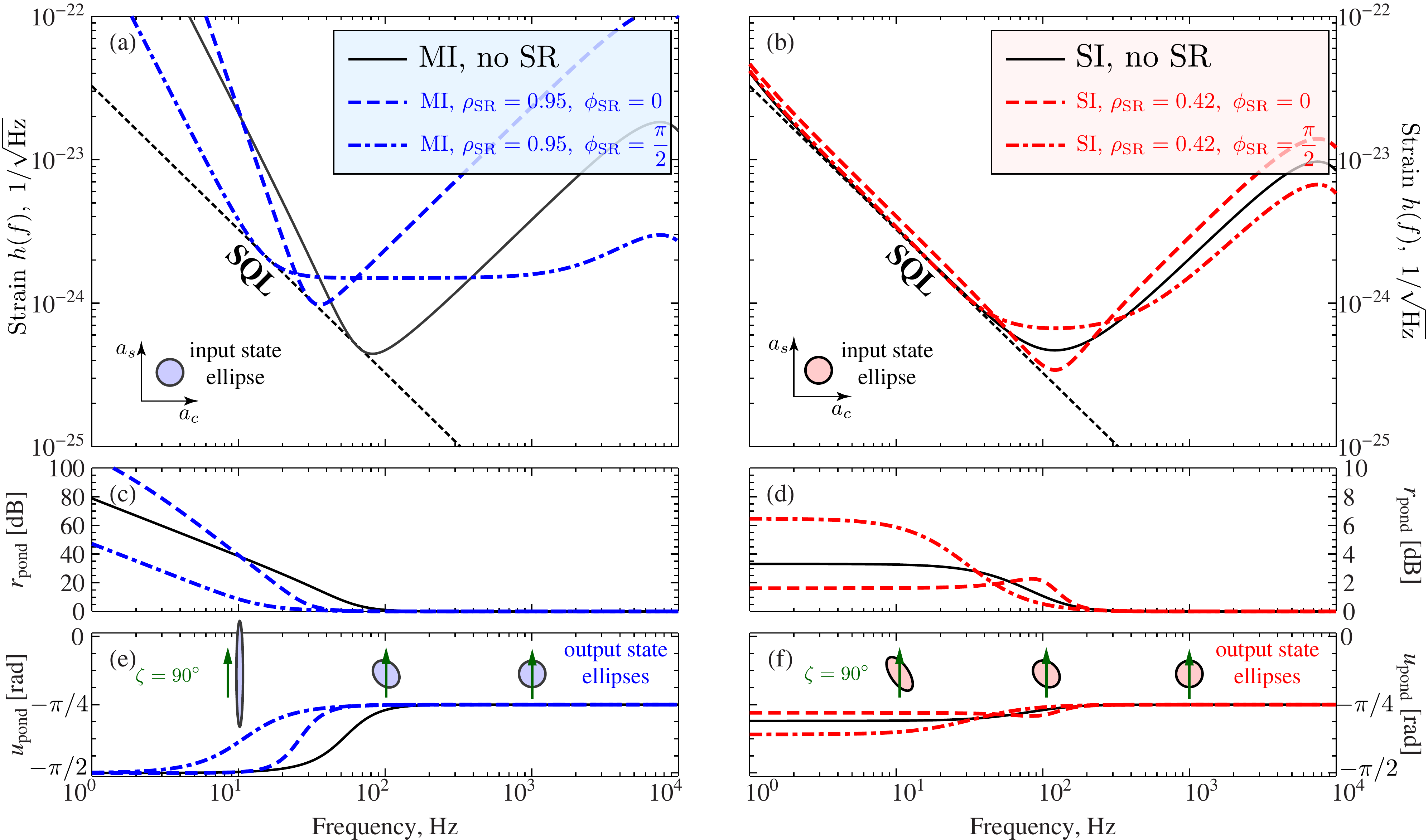}
  \caption{Effect of signal recycling on Michelson (left) and Sagnac (right) interferometers quantum noise sensitivity (first row, panels (a) and (b)), ponderomotive squeezing strength (middle row, panels (c) and (d)) and quantum state rotation angle, $u_{\rm pond}$ (bottom row, panels (e) and (f)).  Plots are drawn for ET typical parameters: arms length $L=10$~km, mirrors mass, $m=200$~kg, arm circulating power, $P_{\rm arm} = 3$~MW, phase quadrature readout, $\zeta=\pi/2$. ITM power transmissivity for Michelson is equal to $T_{\rm ITM} = 5.2\%$, and for Sagnac is $T_{\rm ITM} = 9.1\%$. Light quantum state transformation by an interferometer (w/o signal recycling) is illustrated by noise ellipses. Circles on panels (a) and (b) stand for injected vacuum noise ellipse. Ellipses on panels (e) and (f) demonstrate quantum state of the outgoing light transformed by ponderomotive squeezing at different frequencies (10, 100, and 1000 Hz). Green arrows point at the direction of readout quadrature, $\zeta = \pi/
2$.}
  \label{fig:MI_SI_SReffect}
\end{figure*}

In approximation of $\Omega\ll\nu_{\rm FSR}$, the expressions for $\mathcal{K}_{\rm SI}^{\pm}$ transform as:
\begin{equation}
  \mathcal{K}_{\rm SI, SR}^{\pm} = \frac{4 \Theta_{\rm SI} \gamma_{\rm arm}\digamma_{\pm}}{\Omega^4+2\gamma_{\rm arm}^2\Omega^2(2\digamma_\pm^2-1)+\gamma_{\rm arm}^4},
\end{equation}
where $\digamma_\pm = (1\pm\rho_{\rm SR})/(1\mp\rho_{\rm SR})$ . This approximate form clearly displays the effect of signal recycling on quantum noise: it makes the overall bandwidth of the interferometer wider at the expense of lower peak sensitivity, when feedback is negative, \textit{i.e.} for $\phi_{\rm SR} = \pi/2$ (RSE), and narrower, but with improved peak sensitivity, for positive feedback when $\phi_{\rm SR} = 0$ (TSR). Fig.~\ref{fig:MI_SI_SReffect} show the influence of signal recycling mirror on the quantum noise sensitivity of both, Michelson (panel (a)) and Sagnac (panel (b)) interferometers. 

\subsection{Ponderomotive squeezing and radiation pressure noise in recycled Sagnac interferometer.}

It is instructive to investigate how signal recycling effects the frequency dependence of $\mathcal{K}$ for Michelson and Sagnac interferometers and thereby radiation pressure noise in both interferometers. The appearance of $\mathcal{K}$ in the off-diagonal component of the light field transfer matrix $\tq{T}$ is responsible for the transformation of the quantum state of light, passing through the interferometer, known as ponderomotive squeezing.
Kimble {\it et al.} \cite{02a1KiLeMaThVy} showed that the transform \eqref{eq:IOSaglossless} is equivalent to a sequence of phase plane rotation followed by a squeeze and finally concluded by another rotation (see Appendix~\ref{app:pond_sqz} for details).
Vector of quadratures of the input field, $\{\hat{a}_c,\,\hat{a}_s\}$, undergoes a clockwise rotation by a frequency-dependent angle $v_{\rm pond} = \phi_{\rm pond}$, then gets anti-squeezed (squeezed) along the $a_c$ ($a_s$) quadrature axis by a factor $e^{r_{\rm pond}}$ ($e^{-r_{\rm pond}}$), and finally is rotated counter-clockwise by an angle $u_{\rm pond} = \phi_{\rm pond} + \theta_{\rm pond}$, where
\begin{equation}\label{eq:Theta}
  \theta_{\rm pond} = \arctan\frac{\mathcal{K}}{2}\,,
\end{equation}
and
\begin{equation}\label{eq:r_and_phi_pond}
  e^{r_{\rm pond}} = \sqrt{1+\Bigl(\frac{\mathcal{K}}{2}\Bigr)^2}+\frac{\mathcal{K}}{2}\,,\quad
  \phi_{\rm pond}  = \frac{1}{2}\mathrm{arccot}\frac{\mathcal{K}}{2}\,.
\end{equation}
Using a trigonometric identity, $\arctan{x}+\arccot{x} = \pi/2$ for $x>0$, it is straightforward to show  that $v_{\rm pond}  - u_{\rm pond} = \pi/2$.  Frequency dependences of ponderomotive squeezing factor, $r_{\rm pond}$, and angle, $u_{\rm pond}$, are given in two lower rows of Fig.~\ref{fig:MI_SI_SReffect} and illustrated by the transformation of the noise ellipses of the vacuum field at the input of the interferometer to the ponderomotively squeezed vacuum at the output thereof (low frequency noise ellipse for Michelson case is drawn not to scale to fit it into a figure). 

One can readily see the feedback effect of the signal recycling cavity on the level of radiation pressure noise. Negative feedback (RSE) expectedly makes ponderomotive squeezing smaller in both, Michelson and Sagnac cases. Positive one (TSR) has an opposite effect. The superior radiation pressure noise performance of Sagnac interferometer is vivid from these plots too.  

Finally, we can see that squeezing angle, $u_{\rm pond}$, varies less in Sagnac than in Michelson. This, as we will see later, allows to relax requirements for filter cavities for frequency dependent squeezing angle rotation in Sagnac case. 

\subsection{Injection of squeezed vacuum and frequency dependent squeezing.}

Injection of squeezed vacuum into a signal port of the interferometer in order to reduce its quantum noise has been first proposed by Unruh \cite{PhysRevD.19.2888} and then further developed by Caves \cite{81a1Ca}.
\begin{figure*}[ht]
  \centering
  \includegraphics[width=\textwidth]{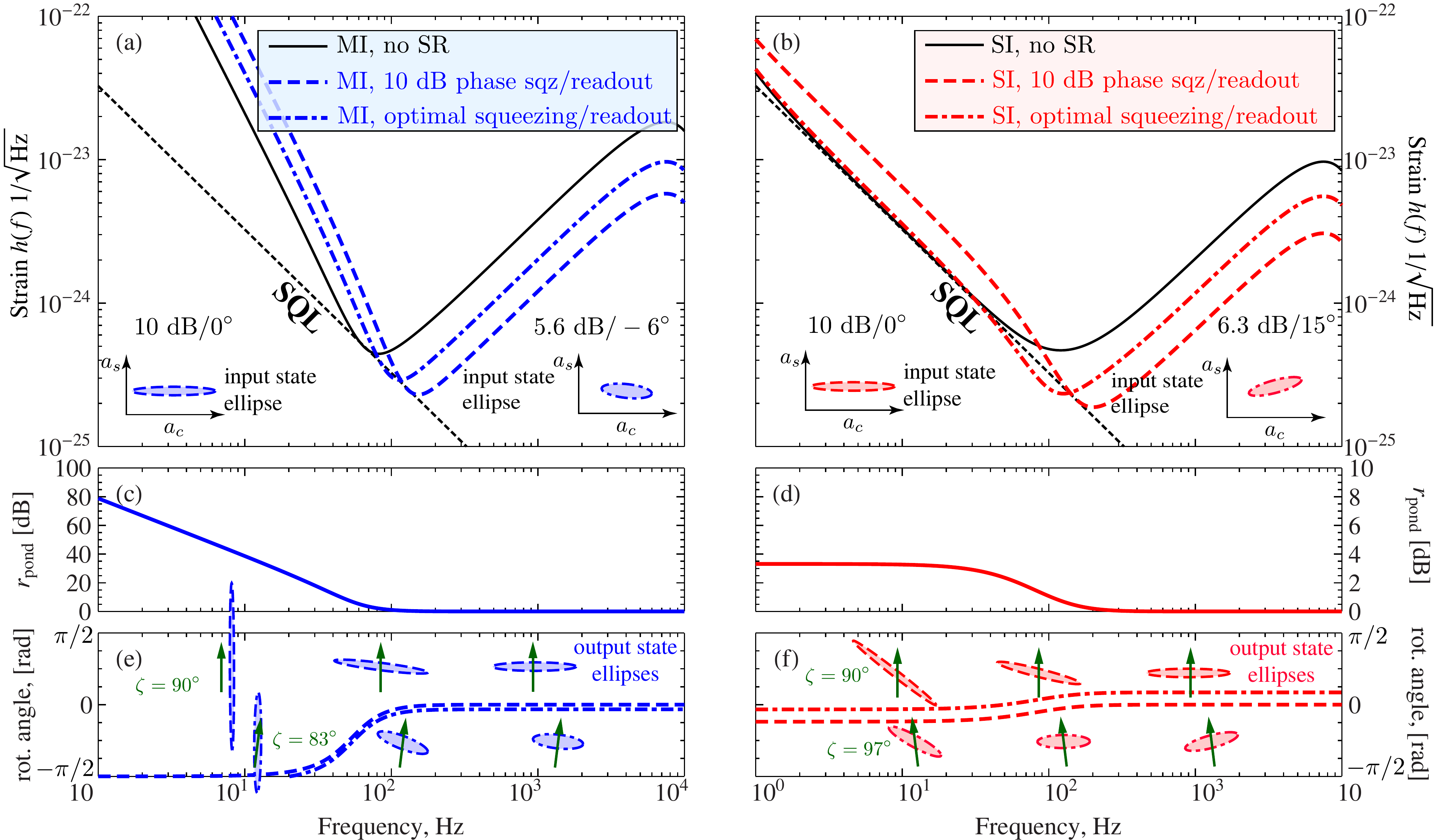}
  \caption{Quantum noise sensitivity of Michelson (left, panel (a)) and Sagnac (right, panel (b)) interferometers quantum in case of constant phase squeezed vacuum injection. \SD{Panels (c) and (d) show ponderomotive squeezing factor, $r_{\rm pond}(\Omega)$, while panels (e) and (f) show quantum noise ellipse overall rotation angle, \textit{i.e.} $\mathrm{rot.\ angle} = \lambda + u_{\rm pond} + v_{\rm pond}$ of Michelson and Sagnac interferometers, respectively.}  Solid black lines of quantum noise sensitivity of bare interferometers (w/o recycling and squeezing) are given for reference. Dashed lines demonstrate the effect of constant phase squeezing ($r=10$~dB, $\lambda=0$) with phase quadrature readout ($\zeta=\pi/2$), dash-dotted lines show quantum noise for optimised squeezing angle and readout quadrature. Parameters used for these plots are given in Table~\ref{tab:fixed_sqz_injection}. Light's quantum state transformation by an interferometer with squeezed vacuum injection is illustrated by noise 
ellipses. Ellipses on panels (a) and (b) stand for injected squeezed vacuum noise ellipses (left one  for 10 dB phase squeezed input, right one for optimised squeezed input). Ellipses on panels (e) and (f) demonstrate quantum state of the outgoing light transformed by ponderomotive squeezing at 
different frequencies (10, 100, and 1000 Hz). Green arrows point at the direction of readout quadrature, $\zeta$. An upper row of ellipses on both lower panels refers to the 10 dB phase squeezed vacuum injection case. A lower row stands for the optimised fixed angle squeezed vacuum injection case.}
  \label{fig:fixed_sqz_injection}
\end{figure*}

\paragraph{Constant squeezing.}

Squeezed vacuum used in this technique is a product of nonlinear interaction of light with a Kerr medium of a nonlinear crystal. The crystal is pumped at double carrier frequency $2\omega_p$, and vacuum photons at carrier frequency are produced from the pumping photons in the process known as parametric down conversion (PDC). Since there is a strong correlation between the produced photons, quantum fluctuations of the vacuum field turned out to be "squeezed" in one quadrature and is amplified by the same factor (ideally) in an orthogonal one. From the mathematical point of view this transformation is similar to the ponderomotive squeezing and thus can be described by expressions given in Appendix~\ref{app:pond_sqz}. For now, we are interested in the spectral density matrix, $\tq{S}^{\rm sqz}_{a}$, of a squeezed vacuum with arbitrary squeezing factor $r$ and angle $\lambda$. It is readily obtainable from Eq.~\ref{app.eq:sqz_state}  and taking into account that in a vacuum state $\ket{in}=\ket{vac}$ the 
spectral density matrix \eqref{eq:SpDens_a} equals to identity matrix, one can calculate the spectral density matrix for squeezed light from \eqref{eq:SqzTr} as:
\begin{equation}\label{eq:Sin_sqz}
  \tq{S}_a^{\rm sqz} = \tq{R}_\lambda\cdot\tq{S}[2r]\cdot\tq{R}^\dag_{\lambda}\,.
\end{equation}  
Substituting it into \eqref{eq:SpDens_h} and using Eqs.~\eqref{eq:SI_T_and_t}, one obtains a universal expression for quantum noise spectral density of an interferometer with injected squeezed vacuum:
  \begin{multline}\label{eq:Sh_tuned_w_arb_sq}
  S^h_{\rm sqz} = \frac{h^2_{\rm SQL}}{2\mathcal{K}}\left\{e^{2r}(\sin\lambda-
\cos\lambda\left[\mathcal{K}-\cot\zeta\right])^2+\right.\\ \left.+e^{-2r}(\cos\lambda+
\sin\lambda\left[\mathcal{K}-\cot\zeta\right])^2\right\}\,.
\end{multline}
This expression is valid for any interferometer with a transfer matrix and a response vector represented in the form of Eqs.~\eqref{eq:SI_T_and_t}, meaning for us, it is true for both, Michelson and Sagnac configurations, with or without resonance signal recycling.

Injection of squeezed vacuum pursues a clear goal of reducing quantum noise in the readout quadrature. Were it not for radiation pressure noise and ponderomotive squeezing, the optimal squeezing angle would be $\lambda = \zeta+\pi/2$, resulting in $e^{-2r}$ lower shot noise. For example, for phase quadrature readout, $\zeta=\pi/2$, one requires a phase-squeezed vacuum, that is $\lambda=0$. 

Radiation pressure noise, however, makes constant squeezing not that beneficial. Imagine, we feed the Michelson interferometer with phase squeezed vacuum. Ponderomotive squeezing will transform it according with plots shown in panels (c) and (e) of Fig.~\ref{fig:fixed_sqz_injection}. At higher frequencies (where $\mathcal{K}$ is close to 0) there are no changes to the quantum state as there is no ponderomotive squeezing. At lower frequencies, $r_{\rm pond}\gg r$ is huge and the injected vacuum is squeezed even more along the $a_s$-axis and anti-squeezed along the $a_c$-axis. Then it is followed by a rotation by an angle $u_{\rm pond} = -\pi/2$ effectively making it a strongly amplitude squeezed vacuum, increasing quantum noise in phase quadrature by a factor $e^{2(r_{\rm pond}+r)}\gg1$.  

For Sagnac, the story is a bit more complex, as one has to account for non-zero rotation represented by $v_{\rm pond} = u_{\rm pond}+\pi/2$ and a much smaller ponderomotive squeezing factor, $r_{\rm pond}$. Consider the case of no signal recycling with phase squeezed vacuum injection shown by dashed red lines in panels (b) and (f) of Fig.~\ref{fig:fixed_sqz_injection}. First, it gets rotated counterclockwise by 
$v_{\rm pond} \simeq \pi/5 \simeq 35^\circ$ at LF and $v_{\rm pond} \simeq \pi/4 \simeq 45^\circ$ at HF
which is followed by a weak, $r_{\rm pond}\simeq 3.25$~dB ponderomotive squeezing (we look at low frequencies $\sim10$~Hz). The latter stretches the rotated ellipse along the $a_c$-axis and squeezes it along the $a_s$-axis by the same factor of $~2$. The resulting stretched ellipse inclines more towards the $a_s$-axis (new angle $~8^\circ$). Finally, it is rotated clockwise by 
$u_{\rm pond} \simeq -\pi/3 \simeq -55^\circ$ at LF and $u_{\rm pond} \simeq -\pi/4 \simeq -45^\circ$ at HF 
ending up with an ellipse at almost $45^\circ$ to both axes, thereby having equal amount of noise in both, phase and amplitude quadrature. At higher frequencies quantum noise remains squeezed at phase quadrature.

Two conclusions can be derived from the above qualitative discussion: (i) because of very strong ponderomotive squeezing, Michelson interferometer can benefit from squeezing injection only if squeezing angle varies with frequency (frequency dependent squeezing), or the optimal readout quadrature is chosen at each frequency (variational readout), and (ii) in Sagnac interferometer, however, one can find the right values of squeezing factor, squeezing angle and readout quadrature to partially compensate for ponderomotive squeezing. 
The resulting sensitivity curves for both configurations are shown in Fig.~\ref{fig:fixed_sqz_injection} and optimal parameters are given in Table~\ref{tab:fixed_sqz_injection}. 

\begin{table}
	\begin{ruledtabular}
		\begin{tabular}{ l c cc }
		  Parameter                     & Notation                  & Michelson    & Sagnac       \\
		  \hline
		  Mirror mass, kg               & $m$                       &         200  &         200  \\
		  Arm length, km                & $L$                       &          10  &          10  \\\
		  Circ. power, MW               & $P_c$                     &         3.0  &         3.0  \\
		  ITM transmittance, \%         & $T_{\rm ITM}$             &           5  &           9  \\
		  Squeezing factor, dB          & $r_{\rm dB}$              &         5.6  &         6.3  \\
		  Squeezing angle, deg.         & $\phi_{\rm sqz}$          & $  -6^\circ$ & $  15^\circ$ \\
		  Homodyne angle , deg.         & $\zeta$                   & $  83^\circ$ & $  97^\circ$ \\
		\end{tabular}
	\end{ruledtabular}
  \caption{Optimal parameters for configurations plotted in Fig.~\ref{fig:fixed_sqz_injection}.}\label{tab:fixed_sqz_injection}
\end{table}

For constant squeezing angle, $\lambda$, and readout quadrature, $\zeta$, the optimal broadband configuration requires squeezing factor to be limited. If $\lambda$ and $\zeta$ were optimal at each frequency, \textit{i.e.} satisfied the following relation (cf. Eq.~(16) of \cite{PhysRevD.68.042001_2003_Harms}) that delivers minimum to Eq.~\eqref{eq:Sh_tuned_w_arb_sq}:
\begin{equation}\label{eq:optFDsqz}
  \cot\zeta + \tan\lambda = \mathcal{K}(\Omega)\,,
\end{equation}
there would be no such limitation and the resulting quantum noise would decrease by the factor $e^{-2r}$:
\begin{equation}\label{eq:Sh_tuned_w_FD_sq}
  S^h = \frac{h^2_{\rm SQL}e^{-2r}}{2}\left\{\frac{1+
\left[\mathcal{K-\cot\zeta}\right]^2}{\mathcal{K}}\right\}\,.
\end{equation}
However, this can be achieved only by having either $\lambda$, or $\zeta$, or both frequency dependent to satisfy Eq.~\eqref{eq:optFDsqz} at each frequency $\Omega$.

\begin{figure*}[ht]
  \centering
  \includegraphics[width=\textwidth]{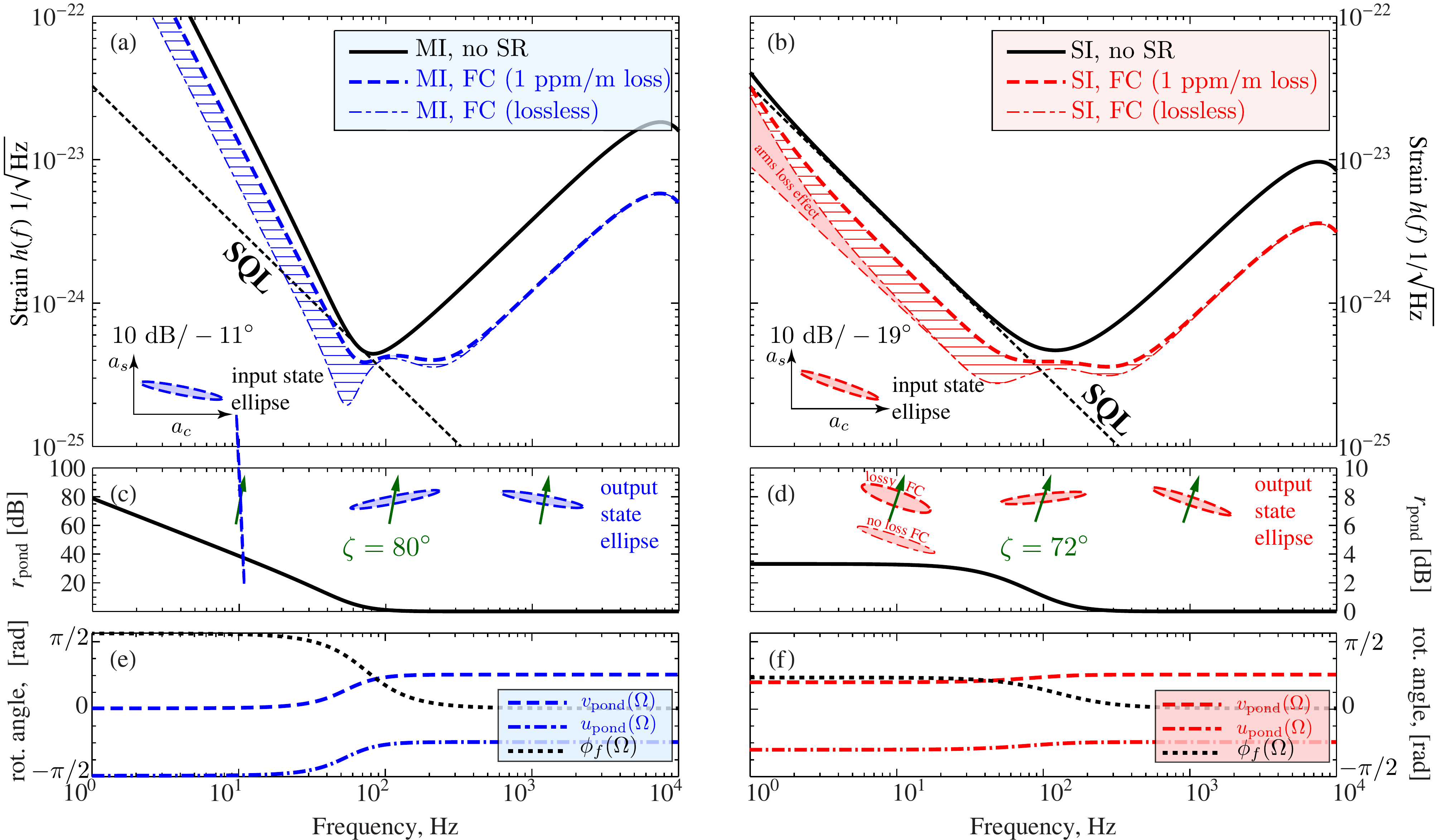}
  \caption{Quantum noise of Michelson (left, panel (a)) and Sagnac (right, panel (b)) interferometers with frequency dependent squeezing injection: solid black lines show quantum noise without squeezing, thin dash-dotted lines show quantum noise with lossless single filter cavity with parameters given in Table~\ref{tab:FD_sqz_injection}, and thick dashed lines show quantum noise with account for optical loss in the FC.
  Hatched regions demonstrate sensitivity deterioration due to FC loss. Panels (c) and (d) show ponderomotive squeezing factor vs. frequency, and panels (e) and (f) illustrate frequency dependent phase space rotation of the light noise ellipse in the interferometer at all stages, \textit{i.e.} before the ponderomotive squeezing is applied (dashed lines), after it (dash-dotted lines), and the overall rotation angle (black dotted lines). Corresponding output noise ellipses are shown for both interferometers in panels (c) and (d) at 3 different frequencies (10, 100 and 1000 Hz).
  Readout quadrature direction, given by angle $\zeta$, is illustrated by green arrows. Quantum noise at each frequency can be conveniently represented by a projection of the corresponding noise ellipse on the readout quadrature direction. These plots show that a single filter cavity is able to provide only a quasi-optimal phase rotation. The effect of this deviation from optimal dependence is most pronounced near the medium frequencies (around $~100$ Hz).
  Effect of optical loss in the arms is not negligible for Sagnac interferometer and shown by pink colour-filled area at low frequencies ($T_{\rm ETM} = 40$~ppm).}
  \label{fig:FD_sqz_injection}
\end{figure*}
\paragraph{Frequency dependent squeezing.}
Optimal frequency dependence can be followed quite closely using a phase dispersion of an ordinary optical cavity.
This idea was raised by Kimble {\it et al.} \cite{02a1KiLeMaThVy} who suggested to introduce deliberately constructed optical cavities between the dark port of the interferometer and a squeezed vacuum generator, thus making frequency dependent $\lambda(\Omega)$, or before the readout stage to make frequency dependent $\zeta(\Omega)$. 
 
The latter idea known as \textit{variational readout} is more beneficial, in theory, as setting readout quadrature to $\zeta = \arccot[\mathcal{K}(\Omega)]$ makes radiation pressure contribution to the final spectral density effectively zero (see, \textit{e.g.}, Eq.~\eqref{eq:Sh_plain}). Closer look, accounting for all losses, has shown that this advantage vanishes and the modification of the alternative of frequency dependent squeezing injection has proven to pay off more (see Sec. 6.1 of review \cite{Liv.Rv.Rel.15.2012} and references therein).

Omitting loss, filter cavity I/O-relations can be written as:
\begin{equation}\label{eq:FC_IO}
  \vq{o}_{f}(\Omega) = \tq{T}_f(\Omega)\vq{i} \equiv e^{\beta_f(\Omega)}\tq{R}_{\phi_{f}}(\Omega)\vq{i} ,
\end{equation}
where transfer matrix, $\tq{T}_f(\Omega)  \equiv e^{\beta_f(\Omega)}\tq{R}_{\phi_{f}}(\Omega)$, represents a simple phase shift, $\beta_f(\Omega)$ of no interest to us, plus phase plane rotation with frequency dependent rotation angle
\begin{equation*}
  \phi_f(\Omega) = \arctan\frac{2\gamma_{f1}\delta_f}{\gamma_{f1}^2-\delta_f^2+\Omega^2}\,,
\end{equation*}
where $\gamma_{f1} = cT_f/(4L_f)$ and $\delta_f$ are the filter cavity half-bandwidth and detuning, respectively, and $T_f$ stands for a filter input mirror power transmissivity while $L_f$ is its length).

Using \eqref{eq:Sin_sqz}, quantum state of the light after the filter cavity can be written in the form of spectral density matrix as: 
\begin{equation}
  \tq{S}_{o_f}^{in} = \tq{R}_{\lambda+\phi_{f}}(\Omega)\cdot \tq{S}[2r]\cdot\tq{R}^\dag_{\lambda+\phi_{f}}(\Omega) \,,
\end{equation}  
which can then be substituted into the general equation \eqref{eq:SpDens_h} to yield spectral density of the interferometer with frequency dependent squeezing injection. The result, in the absence of loss, will be a formula similar to \eqref{eq:Sh_tuned_w_arb_sq} but with substitution $\lambda\to\lambda+\phi_f$.

\begin{table}
 \begin{ruledtabular}
\begin{tabular}{ l c cc }
  Parameter                     & Notation                  & Michelson    & Sagnac       \\
  \hline
  Mirror mass, kg               & $m$                       &         200  &         200  \\
  Arm length, km                & $L$                       &          10  &          10  \\\
  Circ. power, MW               & $P_c$                     &         3.0  &         3.0  \\
  ITM transmittance, \%         & $T_{\rm ITM}$             &           5  &           9  \\
  Squeezing factor, dB          & $r_{\rm dB}$              &        10.0  &        10.0  \\
  Squeezing angle, deg.         & $\phi_{\rm sqz}$          & $ -11^\circ$ & $ -19^\circ$ \\
  Homodyne angle , deg.         & $\zeta$                   & $  80^\circ$ & $  72^\circ$ \\
  \hline \multicolumn{4}{c}{Filter cavity parameters} \\ \hline 
  FC detuning, Hz               & $\delta_f$                &          57  &          44  \\
\makebox[0pt]{\hspace{-0.6em}\raisebox{0pt}[0pt][0pt]{\raisebox{-1.5ex}{$\Big[$}}}FC bandwidth, Hz              & $\gamma_{f1}$             &          48  &         117  \\
FC mirr. transmittance, ppm/m & $T_f/L_f$                 &           4  &          10  \\
\makebox[0pt]{\hspace{-0.6em}\raisebox{0pt}[0pt][0pt]{\raisebox{-1.5ex}{$\Big[$}}}FC loss bandwidth, Hz         & $\gamma_{f2}$             &          12  &          12  \\
FC round trip loss, ppm/m     & $A_f/L_f$                 &        1.00  &        1.00  \\
\end{tabular}
\end{ruledtabular}
  \caption{Optimal parameters for configurations plotted in Fig.~\ref{fig:FD_sqz_injection}.}\label{tab:FD_sqz_injection}  
\end{table}

If filter cavity parameters, $\gamma_{f1}$ and $\delta_f$ are chosen properly, as well as the initial squeezing angle, $\lambda$, one gets a frequency dependent rotation of the injected noise ellipse such as to almost completely compensate the ponderomotive squeezing. Fig.~\ref{fig:FD_sqz_injection} demonstrates how it happens. Quantum noise spectral densities for Michelson and Sagnac interferometers with single input FC are shown (panels (a-b)). Transformation of input phase-squeezed vacuum state in the interferometers is illustrated by noise ellipses for output light at different frequencies (10, 100 and 1000~Hz) and by frequency dependence of ponderomotive squeezing factors and state rotation angles. 

However, an overall quantum noise suppression shown in Fig.~\ref{fig:FD_sqz_injection} is not homogeneous over frequencies. This is due to optical loss in the filter cavity, taken to be 1 ppm/metre for these plots which is achievable by current technology \cite{2013_PhysRevD.88.022002_RealisticFC}. Loss in the cavity causes an incoherent mix in of vacuum photons pertaining thereto, with squeezed vacuum fields entering the cavity, thereby degrading its squeezing factor. This effect is most pronounced within the filter cavity bandwidth which explains worse than expected improvement of quantum noise sensitivity at low frequencies ($<100$~Hz) \cite{10a1Kh,Liv.Rv.Rel.15.2012,2014_CQG_Miao_et_al}. The detailed account for optical loss in various elements of the interferometer and its influence on the overall sensitivity is briefly outlined in the section below and considered in detail in Appendix~\ref{app:transfer matrices}

\subsection{Optical loss in gravitational wave interferometers}

As we argued in the Introduction,  optical loss makes the major constraint on the achievable sub-SQL sensitivity for any scheme of GW detector relying on quantum correlations for beating the SQL \cite{GRG.43.2.671_2011_Chen}, which includes use of squeezed light and optimal quadrature readout. Its influence on detector sensitivity is quite diverse depending on where it originates in the optical layout. Nevertheless, the way quantum correlations are degraded by optical loss is the same.  Incoherent vacuum fields that due to Fluctuation-Dissipation Theorem of Callen and Welton \cite{PhysRev.83.34} accompany loss sources in any system admix with squeezed light travelling through the interferometer. 

\paragraph{Losses in the readout train.} 
It is well understood that optical loss in the elements of readout optical train, including the non-unity quantum efficiency of photodetectors, can be reduced to a single, frequency independent coefficient of equivalent quantum efficiency, $\eta_d = 1 - \epsilon_d < 1$, \SD{where $\epsilon_d < 1$ can be thought of as a fractional photon loss at the photodetector} \cite{02a1KiLeMaThVy,2014_CQG_Miao_et_al,Liv.Rv.Rel.15.2012}. Frequency dependence can be safely omitted here, for any resonant optical element in the readout train, including output mode cleaners (OMC), has bandwidth much larger than the detection band of the interferometer. 

The expression \eqref{eq:o_zeta_lossless} for an output observable of the GW interferometer is modified in the presence of readout losses as follows:
\begin{multline}\label{eq:o_zeta_lossy}
  \hat{o}_{\zeta}^{\rm loss} \equiv
    \sqrt{1-\epsilon_d} \left(\hat{b}_c\cos\zeta + \hat{b}{}_s\sin\zeta\right) \\
    + \sqrt{\epsilon_d} \left(\hat{n}_{d;\;c}\cos\zeta + \hat{n}_{d;\;s}\sin\zeta\right) \\
  \equiv
    \sqrt{1-\epsilon_d}\,\vs{H}_\zeta^{\rm T}\cdot\vq{b}+\sqrt{\epsilon_d}\,\vs{H}_\zeta^{\rm T}\cdot\vq{n}{}_d \, ,
\end{multline}
where $\vq{n}{}_d = \{\hat{n}_{d;\;c},\,\hat{n}_{d;\;s}\}^{\rm T}$ stands quadrature vector of loss-associated vacuum fields with unity spectral density matrix.

Spectral density formula \eqref{eq:SpDens_h} in lossy readout case will be simply:
\begin{equation}\label{eq:SpDens_h_loss_PD}
  S^h_{\rm PD\,loss}(\Omega) = h^2_{\rm SQL}\frac{\vs{H}^\tr_\zeta \cdot 
    \left[ \tq{T}\cdot\tq{S}^{in}_{a}\cdot\tq{T}^\dag+\xi^2_d \right]\cdot\vs{H}_\zeta}{|\vs{H}^\tr_\zeta\cdot\vs{t}_h|^2} \, ,
\end{equation}
where $\xi_d = \sqrt{\epsilon_d/(1-\epsilon_d)}$.

\paragraph{Optical loss in a squeezing injection optics.}
Recent successes in generating low frequency strong squeezing and using it to suppress quantum noise of the GW interferometers GEO 600 and LIGO below the vacuum level \cite{Nat.Phys.7.962_2011,Aasi2013NatPhot} have indicated optical loss in the injection train as the main barrier for getting highly squeezed vacuum entering the GW detector dark port. A complicated, multi-element optical setup is required to make the squeezed light leaving a squeezer match perfectly with the spatial profile of the carrier mode of the interferometer. This results in reasonably high level of integral optical loss in this installation, which can be characterised by an integral, frequency-independent injection power loss coefficient, $\epsilon_{\rm sqz}$. Following the same chain of argument as for the readout train loss, the I/O-relation for the injection train can be written as:
\begin{equation}
\vq{i}_{\rm dark\ port} = \sqrt{1-\epsilon_{\rm sqz}}\,\vq{i}_{\rm sqz}+\sqrt{\epsilon_{\rm sqz}}\,\vq[sqz]{n}\,,
\end{equation}
where $\vq{i}_{\rm dark\ port}$ stands for the light field, entering the dark port of the detector (or the filter cavity in case of frequency dependent squeezing injection), and $\vq[sqz]{i}$ and $\vq[sqz]{n}$ are the field generated by a squeezer and a vacuum field due to injection loss, respectively. If the squeezer is capable of generating squeezed vacuum with $e^{-r}(e^r)$ (anti-)squeezing factor, the effective (anti-)squeezing factor at the dark port reads:
\begin{eqnarray}\label{eq: r_eff def}
e^{-2r_{\rm eff}} &=& (1-\epsilon_{\rm sqz})e^{-2r}+\epsilon_{\rm sqz}\,,\nonumber\\
\Bigl(e^{2 r_{\rm a.s. eff}} &=& (1-\epsilon_{\rm sqz})e^{2r}+\epsilon_{\rm sqz}\Bigr)\,.
\end{eqnarray}

\paragraph{Optical loss in the arms and in filter cavities.}

Optical loss in Fabry-P\'erot cavities, such as arm cavities and filter cavities, is known to have frequency dependence with the major impact at low sideband frequencies within the cavity optical bandwidth. A  very illuminating discussion on this subject is given in \cite{2014_CQG_Miao_et_al} where optical loss in filter cavities is studied in detail. The main source of such loss in large suspended cavities is the scattering of light off the mirror surface imperfections of microscopic (micro-roughness) and relatively macroscopic ("figure error") size \cite{2013_OE.21.30114_Loss_in_FC_Isogai, 2014_CQG_Miao_et_al}. 

In general, optical loss in the cavity depends on the cavity length in an involved way (see Appendix C in \cite{2014_CQG_Miao_et_al}). However, for a fixed filter cavity length ($L_f = 40$ metres in our case) and a fixed value of total photon loss per metre ($\epsilon_f = 1$~ppm/m), a conventional description of Fabry-P\'erot cavity with one lossy mirror (usually, an ETM one) and another lossless one (an ITM, respectively), works perfectly fine. A detailed derivation of lossy cavity I/O-relation is given in Appendix~\ref{app:lossy FP I/O-rels}. Here we only present its general form which reads:
\begin{equation}\label{eq:lossyFP}
  \vq[arm]{b} = \tq[arm]{T}\vq[arm]{a} + \tq[arm]{N}\vq[arm]{n} + \vs[arm]{t}\frac{h}{h_{\rm SQL}}.
\end{equation}
with $\tq[arm]{T} = \tq[arm]{T}^{\rm s.n.}+\tq[arm]{T}^{\rm b.a.}$ a transfer matrix for input fields, $\vq[arm]{a}$,  $\tq[arm]{N} = \tq[arm]{N}^{\rm s.n.}+\tq[arm]{N}^{\rm b.a.}$  is a transfer matrix for loss-associated vacuum fields, $\vq[arm]{n}$, and $\vs[arm]{t}$ is an optomechanical response function of the cavity defined by Eq.~\eqref{eq:t_h_def}. We wrote transfer matrices $\tq[arm]{T} $ and $\tq[arm]{N} $ as sums of shot-noise component, $\tq[arm]{T}^{\rm s.n.}/\tq[arm]{N}^{\rm s.n.}$ and back-action component, $\tq[arm]{T}^{\rm b.a.}/\tq[arm]{N}^{\rm b.a.}$, to discern Fabry-P\'erot cavities with movable mirrors, as in the arms, from the ones with fixed mirrors, as in filter cavity. In the latter case, the back-action components equal to zero, as well as the optomechanical response function $\vs[arm]{t}=0$. The exact expressions for these matrices are derived in Appendix~\ref{app:lossy FP I/O-rels}.
In a special case of filter cavity, the I/O-relations can be simplified by omitting back-action and signal parts in the above formula:
\begin{equation}\label{eq:IOlossyFC}
  \vq{o}_f(\Omega) =  \tq{T}_f\vq{i}(\Omega) + \tq{N}_f\,\vq{n}\,,
\end{equation}
where $\vq{i}$ and $\vq{o}$ stand for input and output fields of the FC, respectively, and $\vq{n}$ represents vacuum fields due to loss. Transfer matrices for filter cavity are defined as:
\begin{equation}
\tq{T}_f = \tq[arm]{T}^{\rm s.n.}(\Omega)\,,\quad\tq{N}_f = \tq[arm]{N}^{\rm s.n.}(\Omega)\,,
\end{equation}
with expressions for $\tq[arm]{T}^{\rm s.n.}$ and $\tq[arm]{N}^{\rm s.n.}$ given by Eqs.~\eqref{eq_app:FP_Tsn} and \eqref{eq_app:FP_Nsn}, respectively.  

Using this simplified formula, quantum noise spectral density for interferometer with lossy input filter cavities can be obtained by substituting into \eqref{eq:SpDens_h_loss_PD} the following expression for input field spectral density matrix:
\begin{equation}\label{eq:Sin_FDsqz_loss}
 \tq{S}_{o_f,\,loss}^{in} = \tq{T}_f\cdot\tq{R}_{\lambda}\cdot \tq{S}[2r]\cdot\tq{R}^\dag_{\lambda}\cdot\tq{T}_f^\dag + \tq{N}_f\cdot\tq{N}_f^\dag\,,
\end{equation}
instead of \eqref{eq:Sin_sqz}. The last term here peaks near the resonant frequency of the cavity which thereby decreases squeezing of the vacuum fields entering the cavity. But the off-resonant squeezed vacuum fields reflect off the FC without deterioration. This explains why optical loss in the cavities have major impact at low frequencies within the FC linewidth. The effect of lossy FC on interferometer sensitivity is illustrated in panels (a) and (b) of Fig.~\ref{fig:FD_sqz_injection} by hatched region between lossy and lossless FC sensitivity curves.

\paragraph{Optical loss in Michelson and in Sagnac interferometers}

Depending on the topology of an interferometer the influence of loss differs. In Michelson, the loss in each arm cavity sum up incoherently to produce additional, loss-induced quantum noise at the output port of the interferometer. The quantum noise I/O-relations for Michelson interferometer are the same to that of a single Fabry-P\'erot cavity,,  a fact known as "scaling law" \cite{2003_PhysRevD.67.062002_ScalLaw,Liv.Rv.Rel.15.2012}. Using I/O-relations at the beam splitter for the fields entering and leaving the interferometer at the dark port and for the fields at north (N) and east (E) arm cavities:
\begin{eqnarray}
  \vq{a}^{\rm N} = \dfrac{\vq{p}+\vq{i}}{\sqrt{2}}\,, &
  \vq{a}^{\rm E} = \dfrac{\vq{p}-\vq{i}}{\sqrt{2}}\,, &
  \vq{o} = \dfrac{\vq{b}^{\rm N}-\vq{b}^{\rm E}}{\sqrt{2}}\,,
\end{eqnarray}
one can derive I/O-relations for lossy interferometer that read
\begin{equation}\label{eq: MI_IO_rels}
  \vq[MI]{o} = \tq[MI]{T}\vq{i} + \tq[MI]{N}\vq{n} + \vs[MI]{t}\frac{h}{h_{\rm SQL}}\,,
\end{equation}
where
\begin{equation}
  \tq[MI]{T} = \tq[arm]{T}, \;\;
  \tq[MI]{N} = \tq[arm]{N}, \;\;
  \vs[MI]{t} = \sqrt{2}\vs[arm]{t},
\end{equation}
and $\vq{n} = \left(\vq{n}^{\rm N} - \vq{n}^{\rm E}\right)/\sqrt{2}$ is a loss-associated quantum noise. The only difference with a single Fabry-P\'erot is in a $\sqrt{2}$ higher response to GW signal.

Quantum noise spectral density matrix for a lossy Michelson interferometer can be written as 
\begin{equation}
\tq{S}_{i,\,loss}^{\rm MI} = \tq[MI]{T}\cdot \tq{S}^{in}_i\cdot\tq[MI]{T}^\dag + \tq[MI]{N}\cdot\tq[MI]{N}^\dag\,,
\end{equation}
and the resulting spectral density of the interferometer quantum noise in terms of GW metric variation reads:
\begin{multline}\label{eq:SpDens_h_loss_MI}
  S^{h\,,loss}_{\rm MI}(\Omega) = \dfrac{h^2_{\rm SQL}}{|\vs{H}^\tr_\zeta\cdot\vs[MI]{t}|^2} \times \\ 
    \vs{H}^\tr_\zeta \cdot \left\{ \tq[MI]{T} \cdot \tq{S}^{in}_a \cdot \tq[MI]{T}^\dag + \tq[MI]{N} \cdot \tq[MI]{N}^\dag + \xi^2d \right\} \cdot \vs{H}_\zeta \, ,
\end{multline}
where we included losses in the readout train as well. The input state spectral density matrix, $\tq{S}^{in}_a$, is substituted either from Eq.~\eqref{eq:Sin_sqz} for fixed squeezing input, or from Eq.~\eqref{eq:Sin_FDsqz_loss} for frequency dependent squeezing input with realistic filter cavity, or can be just an identity matrix for vacuum injection.

In the case of small arm cavity round-trip loss, which can be reduced to ETM transmissivity $T_{\rm ETM} \ll T_{\rm ITM}$, one can get the following closed expression for Michelson quantum noise spectral density: 
\begin{equation}\label{eq:Sh_MI_lossy_small_loss}
S^{h,\ loss}_{\rm MI} = S^h_{\rm sqz} + \frac{h^2_{\rm SQL}}{2}\Biggl\{\frac{\xi_d^2}{\mathcal{K}_{\rm MI}\sin^2\zeta}+\xi^2_{\rm arm}
\mathcal{K}_{\rm MI}\Biggr\}\,,
\end{equation}   
where $S^h_{\rm sqz}$ is the lossless quantum noise spectral density defined in Eq.~\eqref{eq:Sh_tuned_w_arb_sq} with substitution $\mathcal{K}\to \mathcal{K}_{\rm MI}$, and $\xi_{\rm arm} = \sqrt{T_{\rm ETM}/R_{\rm ETM}}$. Note that the last term in the expression above is responsible for an additional radiation pressure created by the arm loss-associated fields $\vq{n}$. This contribution is usually neglected, for it is much smaller than the main radiation pressure term in the Michelson, as $\xi^2_{\rm arm} \simeq T_{\rm ETM} \sim 10^{-5} \ll 1$. In Sagnac, however, as we will see below, this approximation breaks down.

In case of the Sagnac interferometer, two counter propagating light beams visit two arm cavities in sequence on their way to the output port of the interferometer. On each pass, a new loss-associated vacuum noise is generated and contributes to both, shot noise and radiation pressure components of the total quantum noise. The I/O relations \eqref{eq:IOSaglossless} shall be modified to: 
\begin{equation}\label{eq:SI_IO_rels}
  \vq{o} = \tq[SI]{T}\vq{i} + \tq[SI]{N}^{\rm I}\vq[I]{n} + \tq[SI]{N}^{\rm II}\vq[II]{n} + \vs[SI]{t}\frac{h}{h_{\rm SQL}},
\end{equation}
where, for a Sagnac scheme based on ring arm cavities
\begin{equation}\label{eq: n_I, n_II in SI}
  \vq[I]{n}  = \frac{\vq{n}^{\rm LN} - \vq{n}^{\rm RE}}{\sqrt{2}}, \quad
  \vq[II]{n} = \frac{\vq{n}^{\rm RN} - \vq{n}^{\rm LE}}{\sqrt{2}}, 
\end{equation}
are the combined noise fields at the dark port originating from the first (I) and the second (II) pass. In case of polarisation-base Sagnac scheme, the above definitions of  $\vq[I]{n}$ and $\vq[II]{n}$ has to be multiplied by $-1$. Transfer matrices, $\tq[SI]{T}$, $\tq[SI]{N}^{\rm I}$, $\tq[SI]{N}^{\rm II}$ and signal response function, $ \vs[SI]{t}$ are defined in Eqs.~\eqref{app eq: lossy SI matrices}, respectively. 

There is a significant difference between contributions to the sum quantum noise from the first pass, $\tq[SI]{N}^{\rm I}$, and from the second one, $\tq[SI]{N}^{\rm II}$. This difference comes mainly from the different behaviour of back-action components of these two. For back action force created by the first-pass fields, $\vq[I]{n}$, speedmeter radiation pressure suppression, as in \eqref{eq:F_RP_speedmeter} in Sec.~\ref{ssec:RPNsuppress}, works perfectly fine, as these fields share the same optical path as the dark port input field. For the second-pass fields, $\vq[II]{n}$, this mechanism does not work as they only travel inside one arm cavity before leaving the interferometer through the dark port. This results in uncompensated Michelson-like radiation pressure force that shows up as a steeper ($\propto\Omega^{-2}$) elevation of noise at low frequencies.

Quantum noise spectral density matrix for a lossy Sagnac interferometer can be written as 
\begin{equation}
\tq{S}_{i,\,loss}^{\rm SI} = \tq[SI]{T}\cdot \tq{S}^{in}_i\cdot\tq[SI]{T}^\dag + \tq[SI]{N}^{\rm I}\cdot(\tq[SI]{N}^{\rm I})^\dag+ \tq[SI]{N}^{\rm II}\cdot(\tq[SI]{N}^{\rm II})^\dag\,,
\end{equation}
and the resulting spectral density of the interferometer quantum noise in terms of GW metric variation reads:
\begin{multline}\label{eq:SpDens_h_loss_SI}
S^{h\,,loss}_{\rm SI}(\Omega) = \dfrac{h^2_{\rm SQL}}{|\vs{H}^\tr_\zeta\cdot\vs[SI]{t}|^2}\times\vs{H}^\tr_\zeta\cdot\left\{\tq[SI]{T}\cdot \tq{S}^{in}_a\cdot\tq[SI]{T}^\dag +\right.\\\left.+ \tq[SI]{N}^{\rm I}\cdot(\tq[SI]{N}^{\rm I})^\dag+ \tq[SI]{N}^{\rm II}\cdot(\tq[SI]{N}^{\rm II})^\dag\right\}\cdot\vs{H}_\zeta\,,
\end{multline}
which can be written in a closed, simplified form for the case of small loss ($A_{\rm loss}\ll T_{\rm ITM}$):
\begin{equation}\label{eq:Sh_SI_lossy_small_loss}
S^{h,\ loss}_{\rm SI} = S^h_{\rm sqz} + \frac{h^2_{\rm SQL}}{2}\Biggl\{\frac{\xi_d^2}{\mathcal{K}_{\rm SI}\sin^2\zeta}+\xi^2_{\rm arm}
\mathcal{K}_{\rm MI}\Biggr\}\,,
\end{equation}   
where $S^h_{\rm sqz}$ is the lossless quantum noise spectral density defined in Eq.~\eqref{eq:Sh_tuned_w_arb_sq} with substitution $\mathcal{K}\to \mathcal{K}_{\rm SI}$. Note the same last term here, as in Eq.~\eqref{eq:Sh_MI_lossy_small_loss}. Now this term cannot be neglected as $\mathcal{K}_{\rm MI}(\Omega\to 0)\gg\mathcal{K}_{\rm SI}(\Omega\to 0)$, which physically means that radiation pressure in Michelson is way too high compared to Sagnac interferometer, and even for fairly small arm loss, $T_{\rm ETM} \sim 10^{-5}$, this term causes noticeable effect at low frequencies as shown by pink colour-filled area on the panel (b) of Fig.~\ref{fig:FD_sqz_injection}.

\subsection{Combined signal recycling and frequency dependent squeezing injection.}

Finally, we consider combined signal recycling and frequency dependent squeezing injection. Above we saw that Sagnac interferometer benefits a way more from additional filter cavity than Michelson, because of much weaker ponderomotive squeezing due to back-action. However, a non-perfect match of spectra of ponderomotive squeezing angles, $v_{\rm pond}(\Omega)$ $u_{\rm pond}(\Omega)$, and that of the filter cavity phase transfer function, $\phi_f(\Omega)$, results in worsened sensitivity at medium frequencies, $\Omega/2\pi\sim 100$~Hz. Panel (d) of Fig.~\ref{fig:FD_sqz_injection} displays the result of this mismatch in a clear way as a non-optimal angle between the readout quadrature direction, $\zeta$, and the orientation of the noise ellipse at around this frequency.
  
\begin{figure*}[ht]
  \centering
  \includegraphics[width=\textwidth]{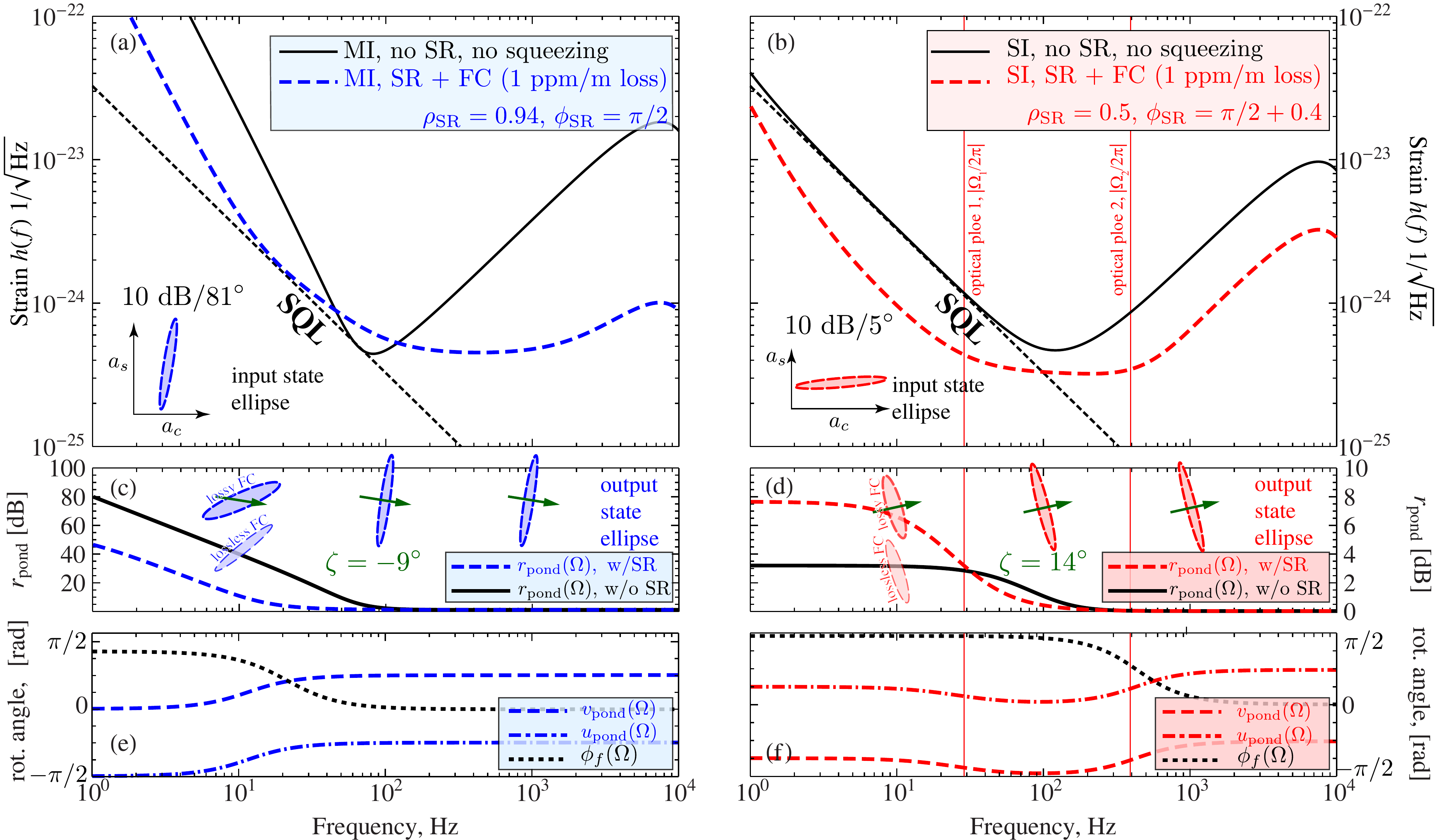}
  \caption{Quantum noise sensitivity of Michelson (left, panel (a)) and Sagnac (right, panel (b)) interferometers in case of combined frequency dependent squeezed vacuum injection 
and optimal signal recycling. Red vertical lines denote the location of two optical poles of the SI defined in Eq.~\eqref{eq:SI_poles}. \SD{Solid black lines in panels (c) and (d) show 
ponderomotive squeezing in interferometers w/o signal recycling mirror, while red dashed lines demonstrate how it changes if an optimally detuned signal-recycling is employed. 
Panels (e) and (f) show frequency dependence of ponderomotive angles (red lines) and the optimal FC phase shift (black dashed curve) for a detuned signal-recycled 
interferometer. Light quantum state transformation by an interferometer with squeezed vacuum injection is illustrated by noise ellipses. Ellipses on panels (a) and (b) stand for 
injected squeezed vacuum noise ellipses for a 10 dB optimally squeezed vacuum. Ellipses on panels (c) and (d) demonstrate quantum state of the outgoing light transformed by 
ponderomotive squeezing at different frequencies (10, 100, and 1000 Hz). Green arrows point at the direction of readout quadrature, $\zeta$. Lower ellipse at 10 Hz represents a 
would-be output state, were there no optical loss in the filter cavity. All the relevant parameters for these plots are given in Table~\ref{tab:SR+FDsqueezing}.}
}
  \label{fig:SR+FDsqueezing}
\end{figure*}

\begin{table}
	\begin{ruledtabular}
		\begin{tabular}{ l c cc }
		  Parameter                     & Notation                  & Michelson    & Sagnac       \\
		  \hline
		  Mirror mass, kg               & $m$                       &         200  &         200  \\
		  Arm length, km                & $L$                       &          10  &          10  \\\
		  Circ. power, MW               & $P_c$                     &         3.0  &         3.0  \\
		  ITM transmittance, \%         & $T_{\rm ITM}$             &           5  &           9  \\
		  Squeezing factor, dB          & $r_{\rm dB}$              &        10.0  &        10.0  \\
		  Squeezing angle, deg.         & $\phi_{\rm sqz}$          & $  -9^\circ$ & $ -36^\circ$ \\
		  Homodyne angle , deg.         & $\zeta$                   & $  81^\circ$ & $  55^\circ$ \\
		  \hline \multicolumn{4}{c}{Signal recycling cavity parameters} \\ \hline 
		  SRM transmittance, \%         & $T_{\rm SRM}$             &          12  &          76  \\
		  SRC detuning, deg.            & $\phi_{\rm SRC}$          & $  90^\circ$ & $ 112^\circ$ \\
		  \hline \multicolumn{4}{c}{Filter cavity parameters} \\ \hline 
		  FC detuning, Hz               & $\delta_f$                &          13  &         315  \\
		\makebox[0pt]{\hspace{-0.6em}\raisebox{0pt}[0pt][0pt]{\raisebox{-1.5ex}{$\Big[$}}}FC bandwidth, Hz              & $\gamma_{f1}$             &          17  &         307  \\
		FC mirr. transmittance, ppm/m & $T_f/L_f$                 &         1.4  &          26  \\
		\makebox[0pt]{\hspace{-0.6em}\raisebox{0pt}[0pt][0pt]{\raisebox{-1.5ex}{$\Big[$}}}FC loss bandwidth, Hz         & $\gamma_{f2}$             &          12  &          12  \\
		FC round trip loss, ppm/m     & $A_f/L_f$                 &        1.00  &        1.00  \\
		\end{tabular}
	\end{ruledtabular}
  \caption{Optimal parameters for configurations plotted in Fig.~\ref{fig:SR+FDsqueezing}.}\label{tab:SR+FDsqueezing}  
\end{table}

It turns out that for a Sagnac this mismatch can be reduced by means of detuned signal recycling. In this case, the non-monotonous dependence of ponderomotive angle on frequency allows for more effective use of the filter cavity (see panel ((d) in Fig.~\ref{fig:SR+FDsqueezing}). Such a dependence comes from a structure of the detuned signal-recycled Sagnac interferometer response function. As a system of two coupled Fabri-P\'erot cavities, it, quite expectedly, has two optical resonances manifesting themselves
as two poles of both, the optical and the optomechanical transfer functions \cite{MuellerEbhardt2009}, that is $\{\tq{T}_{\rm SI,\,SR},\,\vs{t}_{\rm SI,\,SR}\}\propto\mathcal{D}^{-1}(\Omega)$
where
\begin{align}
\mathcal{D}(\Omega) &= (\Omega-\Omega_1)(\Omega+\Omega_1^*)(\Omega-\Omega_2)(\Omega+\Omega_2^*)\,,\\
\Omega_{1,2} &= \delta_{1,2}+i\gamma_{1,2}\,,\label{eq:SI_poles}\\
\delta_{1,2} &= \delta_{\rm arm}+\pm \gamma_{\rm arm}\dfrac{2\sqrt{\rho_{\rm SR}}\cos\phi_{\rm SR}}{1+\rho_{\rm SR}\pm2\sqrt{\rho_{\rm SR}}\sin\phi_{\rm SR}}\,,\\ 
\gamma_{1,2} &= \gamma_{\rm arm}\dfrac{1-\rho_{\rm SR}}{1+\rho_{\rm SR}\pm2\sqrt{\rho_{\rm SR}}\sin\phi_{\rm SR}} + \gamma_{\rm loss}\,, 
\end{align}%
and $\delta_{\rm arm}$ and $\gamma_{\rm arm}$ are the arm cavities detuning and bandwidth and $\gamma_{\rm loss} = c A_{\rm loss}/(4 L)$ is the additional bandwidth of the arm cavity due to a fractional round-trip photon loss $A_{\rm arm}$ therein. 

\begin{figure}[ht]
  \centering
  \includegraphics[width=.5\textwidth]{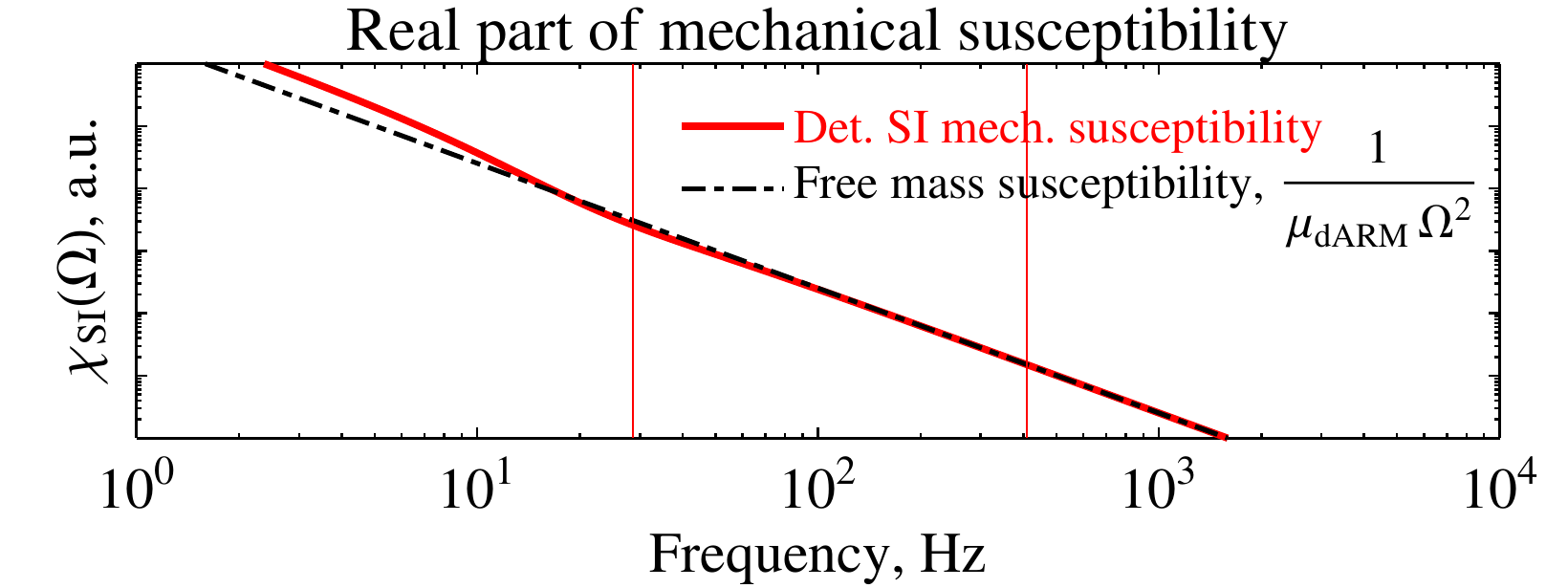}
  \caption{Real part of the mechanical susceptibility of the Sagnac interferometer dARM mode with detuned signal recycling whose spectral density is plotted in the right panel of Fig.~\ref{fig:SR+FDsqueezing}. Black dashed line represents the susceptibility of a free mass and, by coincidence, is the  mechanical susceptibility of the Michelson interferometer dARM. 
 Red vertical lines denote the location of two optical poles of the SI. 
}
  \label{fig:chi_SR_Sag}
\end{figure}

Similar to Michelson interferometer, the detuned signal recycling causes test mass dynamics variation in Sagnac interferometers. However, unlike MI, the frequency dependent optical rigidity that arises in detuned Sagnac interferometer has no DC component and thus creates no mechanical pole in the mechanical response function. As shown in \cite{MuellerEbhardt2009}, it has the first non-vanishing component in the Taylor expansion that is $\propto\Omega^2$ thereby adding an inertial term to the test mass dynamics, \textit{i.e.}
\begin{multline*}
  K_{\rm SI}(\Omega) = -\dfrac{m\Theta_{\rm arm} \Omega^2(\delta_1+\delta_2)}{2\mathcal{D}(\Omega)} \simeq \\
    -\Omega^2\dfrac{m\Theta_{\rm arm} (\delta_1+\delta_2)}{2|\Omega_1|^2|\Omega_2|^2}+\mathcal{O}(\Omega^4)\,,
\end{multline*}
where the coefficient in front of $\Omega^2$ acts akin to additional mass, \textit{i.e.} $K_{\rm SI}(\Omega) = -m_{\rm opt}(\Omega)\Omega^2$. This is the reason, why the detuned SI has no opto-mechanical pole (see Sec. 6.4.3 of \cite{Liv.Rv.Rel.15.2012} and references therein for more details). 

The prospects of using optical inertia in Sagnac remains an open question, for its sign changes between the optical resonances, $|\Omega_1|\leqslant\Omega\leqslant|\Omega_2|$, which makes it hard to provide a stable broadband sensitivity improvement. The broadband optimisation favours diminishing optical inertia. In Fig.~\ref{fig:chi_SR_Sag}, we plot the (real part of) mechanical susceptibility of a detuned signal-recycled Sagnac interferometer, $\chi_{\rm SI}(\Omega) = -(\mu_{\rm dARM}\Omega^2-K_{\rm SI}(\Omega))^{-1}$, along with the free mass susceptibility, $\chi_{\rm f.m.} = -(\mu_{\rm dARM}\Omega^2)^{-1}$, pertaining to the RSE configuration of the Michelson interferometer. Here, $\mu_{\rm dARM} = m/4$ is the reduced mass of the dARM mode of the interferometer. The two curves diverge slightly only at frequencies below 10~Hz. This means the influence of the opto-mechanical interaction on the optical poles is negligible and the above expressions \eqref{eq:SI_poles} can be used.

Comparing panels (b) of Figs.~\ref{fig:FD_sqz_injection} and \ref{fig:SR+FDsqueezing}, one can notice that the optimal location of the two optical poles encompasses the region of middle frequencies, where frequency dependent squeezing alone fails to compensate the ponderomotive squeezing rotation completely, yielding a hump-like behaviour of the quantum noise spectral density. Separation between optical poles provided by detuning allows to shape the frequency dependence of ponderomotive rotation angles, $u_{\rm pond}(\Omega)$ and $v_{\rm pond}(\Omega)$, in such a way to match the filter cavity phase rotation characteristic almost perfectly. The result of this combined action is illustrated by the plots of $u_{\rm pond}(\Omega)$, $v_{\rm pond}(\Omega)$ and $\phi_f(\Omega)$, as well as noise ellipse diagrams in panel (d) of Fig.~\ref{fig:SR+FDsqueezing}. One can see that the readout quadrature angle, $\zeta$, now fits the orientation of the 100~Hz noise ellipse perfectly.

Note also that for Michelson interferometer, the broadband optimisation favours RSE configuration without additional rigidity and, therefore, without additional optomechanical pole. This can be explained by the unwelcome effect of the optical spring, turning the dARM mode into an oscillator, on the rotation of the noise ellipse around the new mechanical resonance frequency it introduces. The ensuing frequency dependence $u_{\rm pond}(\Omega)$ and $v_{\rm pond}(\Omega)$ could hardly be matched by a single filter cavity.

\section{Numerical optimisation}\label{sec:NumOpt}

\begin{table}
  \begin{ruledtabular}
  \begin{tabular}{lccc}
    Parameter                           & Notation        & LIGO                   & ET             \\
    \hline
    Low frequency detection limit, Hz   & $f_{\rm min}$   & 5                      & 1              \\
    High frequency detection limit, kHz & $f_{\rm max}$   & 5                      & 10             \\
    Optical pump wavelength, nm         & $\lambda_p$     & \multicolumn{2}{c}{$1064$}              \\
    Mirrors mass, kg                    & $m$             & $40$                   & $200$          \\
    Interferometer arms length, km      & $L$             & $4$                    & $10$           \\
    Circulating optical power, MW       & $P_{\rm arm}$   & $\leqslant 1$          & $\leqslant 3$  \\
    Optical poser passes through BS, kW & $P_{\rm BS}$    & \multicolumn{2}{c}{$\leqslant 80$}      \\
    Squeezing, dB                       & $r_{\rm dB}$    & \multicolumn{2}{c}{$\leqslant 20$ }     \\
    Arm cavity round trip losses, ppm   & $T_{\rm ETM}$   & \multicolumn{2}{c}{$40$}                \\
    Photodetector power losses, \%      & $\epsilon_d$    & \multicolumn{2}{c}{1}                   \\
    Squeezer power loss, \%             & $\epsilon_{\rm sqz}$ & \multicolumn{2}{c}{5}              \\
    FC round trip losses, ppm/m         & $A_f/L_f$       & \multicolumn{2}{c}{$1$, $0.1$, $0.01$}  \\
  \end{tabular}
  \end{ruledtabular}
  \caption{The main parameters and their numerical values.}\label{tab:params}
\end{table}

\subsection{Optimisation protocol and figure of merit}

In this section we discuss the optimisation protocol we applied to obtain optimal broadband configurations described above. We start with the definition of the parameter space for each configuration. For a simple signal-recycled interferometer w/o squeezing injection the parameter space consists of four-parameter vectors  $\mathbf{p}_{\rm SR} = \left\{P_{\rm arm}, \RITM, \rho_{\rm SR}, \phiSR \right\}$, including arm circulating power, $P_{\rm arm}$, power reflectivity of the ITM, $\RITM$, and two parameters of the SR cavity, the SRM amplitude reflectivity, $\rho_{\rm SR}$ and the SRC phase shift, $\phi_{\rm SR}$. Introduction of fixed input squeezing extends the parameter space by two more dimensions, namely squeezing factor, $r$, and angle, $\lambda$, thus yielding $\mathbf{p}_{\rm sqz} = \left\{P_{\rm arm}, \RITM, \rho_{\rm SR}, \phiSR, r, \lambda \right\}$. Finally, we include filter cavity to account for frequency-dependent squeezing injection, which adds two more parameters to the parameter vector, \
textit{i.e.} the filter cavity detuning, $\delta_f$, and bandwidth, $\gamma_f$, giving the 8-dimensional parameter vector $\mathbf{p}_f = \left\{P_{\rm arm}, \RITM, \rho_{\rm SR}, \phiSR, r, \lambda, \gamma_{f1}, \delta_f \right\}$. 

Other parameters of the interferometer, such as arm length, mirror mass, loss coefficients \textit{etc.} were fixed and their values used for optimisation are given in Table~\ref{tab:params}. This table also outlines the constraints we put on the parameters of the interferometer. Another limitation we set is an upper bound of 80 kW on the power at the beam splitter, $P_{\rm BS}$, to mitigate detrimental effect of thermal lensing. As $P_{\rm BS} = 2 P_{\rm arm} \frac{1-\sqrt{\RITM}}{1+\sqrt{\RITM}}$, this bound sets a limit on $\RITM$ for given $P_{\rm arm}$ and the other way round.

The figure of merit we use for optimisation is the logarithmic cost function for broadband detection proposed in \cite{2014_CQG_Miao_et_al}, namely 
\begin{equation}
  C(\mathbf{p}) = \int_{f_{\rm min}}^{f_{\rm max}} \log_{10}\left[ S^h(2\pi f;\mathbf{p}) + S^h_{\rm ref}(2\pi f)\right] d\log_{10}f\,.
\end{equation}
 The set of parameters $\mathbf{p}$ that provides a global minimum for $C(\mathbf{p})$ is looked for numerically, using Nelder-Mead simplex method \cite{GSLsite}.
 Here $S^h(2\pi f;\mathbf{p})$ is quantum noise spectral density for the considered configuration of the interferometer and $S^h_{\rm ref}(2\pi f)$ is a reference curve discussed above in Sec.~\ref{sec:RefQN} and defined in \eqref{eq:RefSQN}. By construction, this protocol seeks to minimise the area under the sum of these two curves in logarithmic scale.

\subsection{Results of optimisation.}

The main optimisation results are summarised in Fig.~\ref{fig:SI_vs_MI_best} and in Table~\ref{tab:SI_vs_MI_best}. We evaluated here the potential of using Sagnac interferometer topology in conjunction with other prospective quantum-noise-enhancement techniques, such as injection of frequency-dependent squeezed vacuum, detuned signal recycling and  balanced homodyne detection. Results of our study clearly show that for all things being equal, Sagnac interferometer quantum noise has much higher potential for improvement compared to the Michelson topology.

In this section, we focus on two particular questions: (i) how optimal sensitivity depends on the level of loss in the filter cavity and (ii) is there a significant benefit in changing mirror mass, thereby decreasing susceptibility of the interferometer to radiation pressure force. 

\begin{figure*}
  \includegraphics[width = 0.999\textwidth]{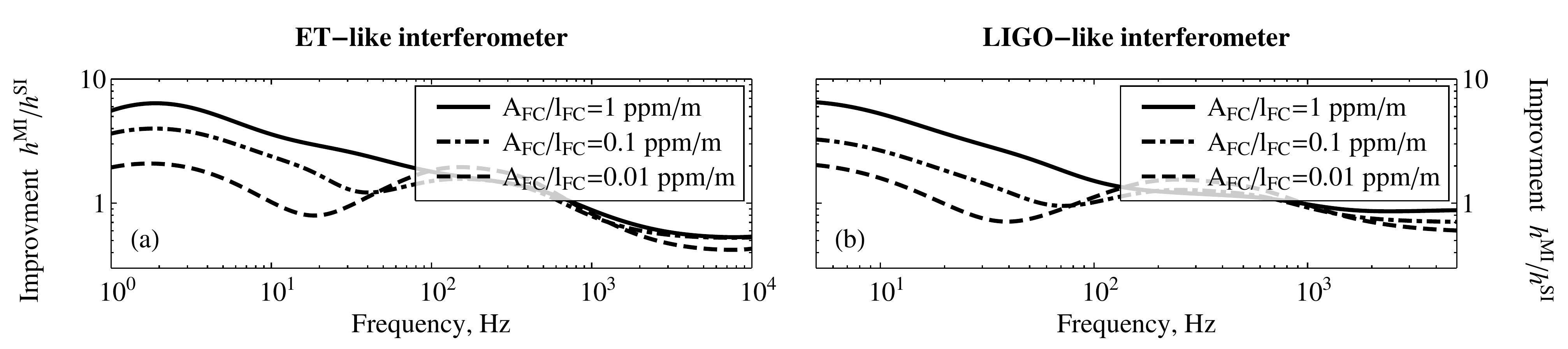}
  \caption{Improvement factor of the Sagnac interferometer with respect to the Michelson interferometer for different filter cavity loss levels. \textit{Left panel:} ET-like detectors (cf. lower panels of Fig.~\ref{fig: LIGO, misc lFC}); \textit{Right panel:} LIGO-like detector (cf. upper panels of Fig.~\ref{fig: LIGO, misc lFC}).}
  \label{fig: impr factor}
\end{figure*}

We vary loss of the filter cavity only for its impact on quantum noise is known to be dominating \cite{10a1Kh,Liv.Rv.Rel.15.2012,2013_PhysRevD.88.022002_RealisticFC}. The results are presented in Figs.~\ref{fig: LIGO, misc lFC}. Sagnac interferometer is proven by this result to be robust to FC loss influence to much greater extent than its rival. The reason behind this result is quite simple. The adequate measure of optical loss in FC is the ratio of input mirror transmissivity to the round-trip absorption therein. Since much weaker back action in Sagnac interferometer allows for much broader band of FC than that that is required for Michelson interferometer, this ratio is much greater for the former one. Hence much weaker dependence on optical loss. Most clear illustration of this advantage is shown in Fig.~\ref{fig: impr factor} where the ratios of Michelson quantum noise amplitude spectral density and that of the Sagnac interferometer are plotted vs. GW 
frequency for different levels of optical loss. The larger is the loss the stronger is low-frequency improvement of the Sagnac interferometer over the Michelson one. Optimal parameters for considered configurations are listed in Table~ \ref{table: ET, misc lFC}.
 
Variation of the mirror mass, to the contrary, does not favour any of the two considered configurations, as one can see in the Fig.~\ref{fig: LIGO, misc mass}. These plots clearly demonstrate that since the mechanical susceptibility of the interferometr to an external force (including radiation pressure noise) changes identically for both, the Michelson and the Sagnac interferometers, the impact of mass variation on the overall quantum noise sensitivity must be the same for both of them. Optimal parameter values of considered configurations are given in Table~\ref{table: ET, misc mass}.

\begin{figure*}
  \includegraphics[width = 0.999\textwidth]{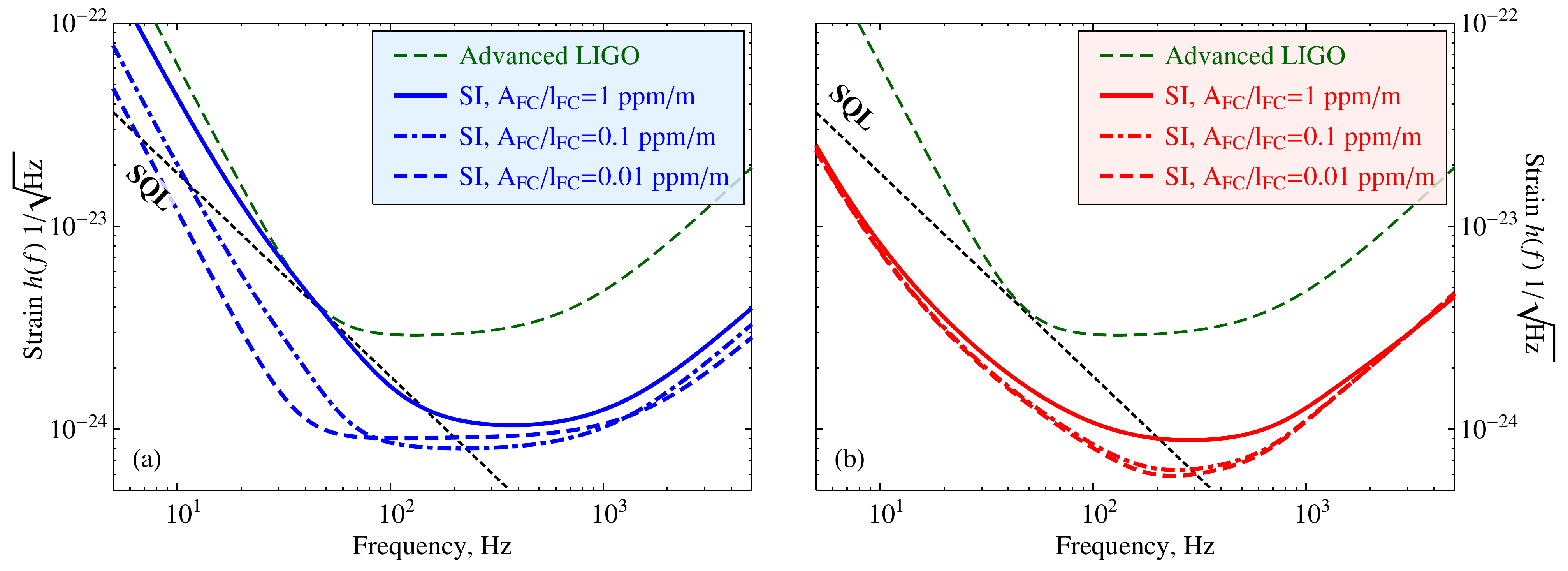}\\
   \includegraphics[width = 0.999\textwidth]{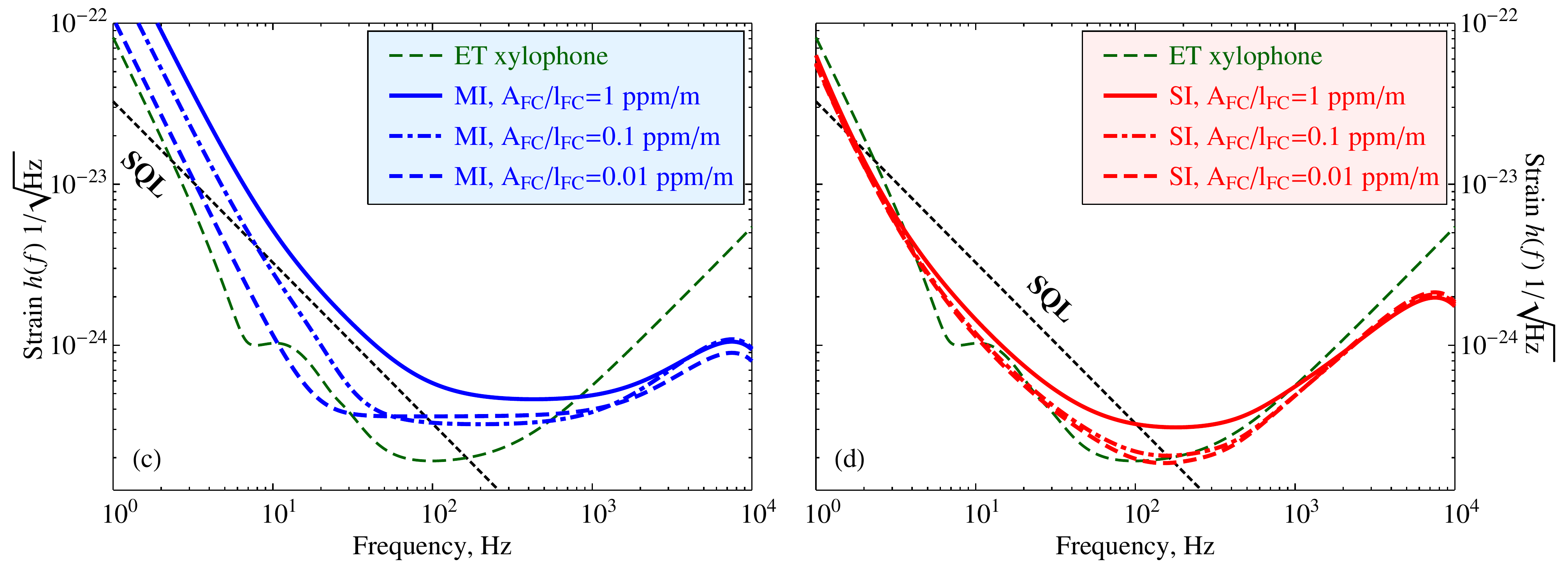}
  \caption{Quantum noise spectral densities for LIGO-like (upper panels) and ET-like (lower panels) interferometers with different levels of optical loss in the filter cavity. Parameters for all shown configurations are listed in Table~\ref{table: ET, misc lFC}.}
  \label{fig: LIGO, misc lFC}
\end{figure*}

\begin{figure*}
  \includegraphics[width = 0.999\textwidth]{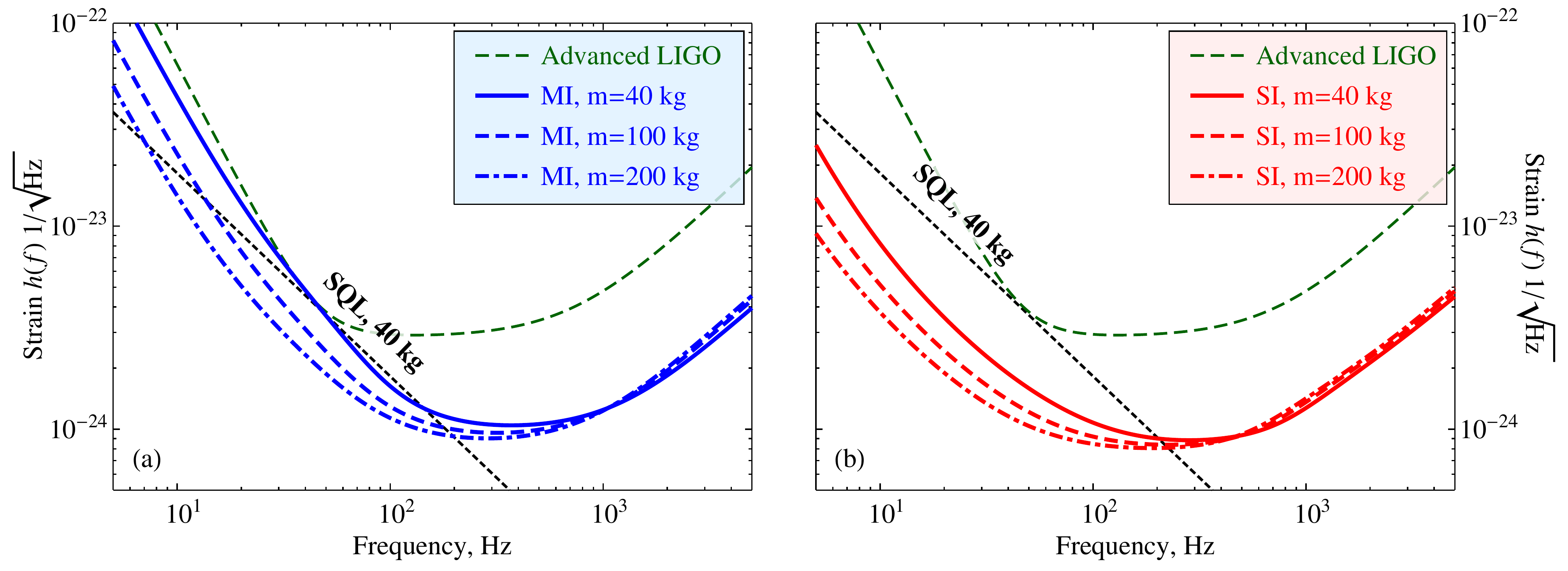}\\
    \includegraphics[width = 0.999\textwidth]{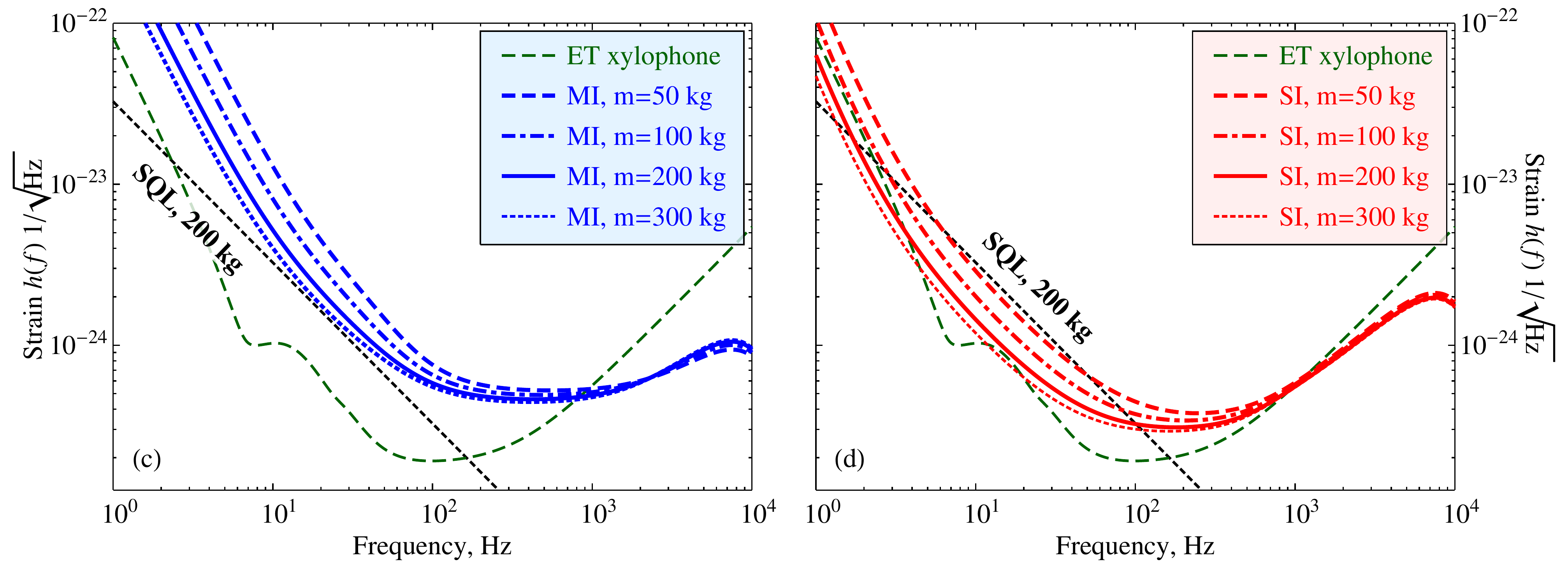}
  \caption{Quantum noise spectral densities for  LIGO-like (upper panels) and ET-like (lower panels) interferometers with different mirror masses. Parameters for all shown configurations are listed in Table~\ref{table: ET, misc mass}.}
  \label{fig: LIGO, misc mass}
\end{figure*}

\begin{table*}
	  \begin{ruledtabular}
		\begin{tabular}{ l c | ccc|ccc | ccc|ccc }
		  \multirow{2}{*}{Parameter}    & \multirow{2}{*}{Notation} & \multicolumn{6}{c|}{LIGO} & \multicolumn{6}{c}{ET} \\
		                                &                           & \multicolumn{3}{c}{Michelson} & \multicolumn{3}{c|}{Sagnac}& \multicolumn{3}{c}{Michelson} & \multicolumn{3}{c}{Sagnac} \\
		  \hline
		  Mirror mass, kg               & $m$                       &          40  &          40  &          40  &          40  &          40  &          40  &          50  &         100  &         300  &          50  &         100  &         300  \\
		  Arm length, km                & $L$                       &           4  &           4  &           4  &           4  &           4  &           4  &          10  &          10  &          10  &          10  &          10  &          10  \\
		  Circ. power, MW               & $P_c$                     &         1.0  &         1.0  &         1.0  &         1.0  &         1.0  &         1.0  &         3.0  &         3.0  &         3.0  &         1.8  &         2.2  &         2.8  \\
		  ITM transmittance, \%         & $T_{\rm ITM}$             &          15  &          15  &          15  &           8  &           7  &           7  &           5  &           5  &           5  &          16  &          14  &          11  \\
		  Squeezing factor, dB          & $r_{\rm dB}$              &        13.2  &        19.4  &        20.0  &        17.9  &        18.6  &        18.7  &        13.3  &        13.2  &        13.5  &        18.5  &        18.2  &        18.2  \\
		  Squeezing angle, deg.         & $\phi_{\rm sqz}$          & $  -6^\circ$ & $  -9^\circ$ & $ -10^\circ$ & $ -17^\circ$ & $ -12^\circ$ & $ -11^\circ$ & $  -7^\circ$ & $  -5^\circ$ & $  -4^\circ$ & $ -13^\circ$ & $ -13^\circ$ & $ -11^\circ$ \\
		  Homodyne angle , deg.         & $\zeta$                   & $  84^\circ$ & $  81^\circ$ & $  80^\circ$ & $  73^\circ$ & $  78^\circ$ & $  79^\circ$ & $  83^\circ$ & $  85^\circ$ & $  86^\circ$ & $  78^\circ$ & $  77^\circ$ & $  79^\circ$ \\
		  \hline \multicolumn{14}{c}{Signal recycling cavity parameters} \\ \hline 
		  SRM transmittance, \%         & $T_{\rm SRM}$             &          82  &          82  &          70  &          89  &          95  &          96  &           9  &          11  &          13  &          93  &          90  &          87  \\
		  SRC detuning, deg.            & $\phi_{\rm SRC}$          & $  90^\circ$ & $  90^\circ$ & $  90^\circ$ & $ 101^\circ$ & $ 104^\circ$ & $ 105^\circ$ & $  90^\circ$ & $  90^\circ$ & $  90^\circ$ & $ 100^\circ$ & $  99^\circ$ & $  98^\circ$ \\
		  \hline \multicolumn{14}{c}{Filter cavity parameters} \\ \hline 
		  FC detuning, Hz               & $\delta_f$                &          34  &          33  &          28  &         269  &         172  &         154  &          24  &          18  &          10  &         155  &         157  &         147  \\
		\makebox[0pt]{\hspace{-0.6em}\raisebox{0pt}[0pt][0pt]{\raisebox{-1.5ex}{$\Big[$}}}FC bandwidth, Hz              & $\gamma_{f1}$             &          34  &          28  &          24  &         806  &         587  &         544  &          25  &          21  &          16  &         644  &         600  &         539  \\
		FC mirr. transmittance, ppm/m & $T_f/L_f$                 &         2.8  &         2.3  &         2.0  &          68  &          49  &          46  &         2.1  &         1.7  &         1.3  &          54  &          50  &          45  \\
		\makebox[0pt]{\hspace{-0.6em}\raisebox{0pt}[0pt][0pt]{\raisebox{-1.5ex}{$\Big[$}}}FC loss bandwidth, Hz         & $\gamma_{f2}$             &          12  &         1.2  &         0.1  &          12  &         1.2  &         0.1  &          12  &          12  &          12  &          12  &          12  &          12  \\
		FC round trip loss, ppm/m     & $A_f/L_f$                 &        1.00  &        0.10  &        0.01  &        1.00  &        0.10  &        0.01  &        1.00  &        1.00  &        1.00  &        1.00  &        1.00  &        1.00  \\
		\end{tabular}
	\end{ruledtabular}
  \caption{Optimal parameters for LIGO-like and ET-like detectors for different values of optical loss in filter cavity.}
  \label{table: ET, misc lFC}
\end{table*}

\begin{table*}
	\begin{ruledtabular}
		\begin{tabular}{ l c | cc|cc | ccc|ccc }
		  \multirow{2}{*}{Parameter}    & \multirow{2}{*}{Notation} & \multicolumn{4}{c|}{LIGO} & \multicolumn{6}{c}{ET} \\
		                                &                           & \multicolumn{2}{c}{Michelson} & \multicolumn{2}{c|}{Sagnac}& \multicolumn{3}{c}{Michelson} & \multicolumn{3}{c}{Sagnac} \\
		  \hline
		  Mirror mass, kg               & $m$                       &         100  &         200  &         100  &         200  &          50  &         100  &         300  &          50  &         100  &         300  \\
		  Arm length, km                & $L$                       &           4  &           4  &           4  &           4  &          10  &          10  &          10  &          10  &          10  &          10  \\
		  Circ. power, MW               & $P_c$                     &         1.0  &         1.0  &         1.0  &         1.0  &         3.0  &         3.0  &         3.0  &         1.8  &         2.2  &         2.8  \\
		  ITM transmittance, \%         & $T_{\rm ITM}$             &          15  &          15  &           6  &           5  &           5  &           5  &           5  &          16  &          14  &          11  \\
		  Squeezing factor, dB          & $r_{\rm dB}$              &        13.0  &        13.1  &        18.1  &        18.3  &        13.3  &        13.2  &        13.5  &        18.5  &        18.2  &        18.2  \\
		  Squeezing angle, deg.         & $\phi_{\rm sqz}$          & $  -4^\circ$ & $  -3^\circ$ & $ -15^\circ$ & $ -13^\circ$ & $  -7^\circ$ & $  -5^\circ$ & $  -4^\circ$ & $ -13^\circ$ & $ -13^\circ$ & $ -11^\circ$ \\
		  Homodyne angle , deg.         & $\zeta$                   & $  86^\circ$ & $  87^\circ$ & $  75^\circ$ & $  77^\circ$ & $  83^\circ$ & $  85^\circ$ & $  86^\circ$ & $  78^\circ$ & $  77^\circ$ & $  79^\circ$ \\
		  \hline \multicolumn{12}{c}{Signal recycling cavity parameters} \\ \hline 
		  SRM transmittance, \%         & $T_{\rm SRM}$             &          88  &          91  &          87  &          85  &           9  &          11  &          13  &          93  &          90  &          87  \\
		  SRC detuning, deg.            & $\phi_{\rm SRC}$          & $  90^\circ$ & $  90^\circ$ & $  99^\circ$ & $  98^\circ$ & $  90^\circ$ & $  90^\circ$ & $  90^\circ$ & $ 100^\circ$ & $  99^\circ$ & $  98^\circ$ \\
		  \hline \multicolumn{12}{c}{Filter cavity parameters} \\ \hline 
		  FC detuning, Hz               & $\delta_f$                &          23  &          17  &         217  &         179  &          24  &          18  &          10  &         155  &         157  &         147  \\
		\makebox[0pt]{\hspace{-0.6em}\raisebox{0pt}[0pt][0pt]{\raisebox{-1.5ex}{$\Big[$}}}FC bandwidth, Hz              & $\gamma_{f1}$             &          26  &          22  &         710  &         648  &          25  &          21  &          16  &         644  &         600  &         539  \\
		FC mirr. transmittance, ppm/m & $T_f/L_f$                 &         2.2  &         1.9  &          60  &          54  &         2.1  &         1.7  &         1.3  &          54  &          50  &          45  \\
		\makebox[0pt]{\hspace{-0.6em}\raisebox{0pt}[0pt][0pt]{\raisebox{-1.5ex}{$\Big[$}}}FC loss bandwidth, Hz         & $\gamma_{f2}$             &          12  &          12  &          12  &          12  &          12  &          12  &          12  &          12  &          12  &          12  \\
		FC round trip loss, ppm/m     & $A_f/L_f$                 &        1.00  &        1.00  &        1.00  &        1.00  &        1.00  &        1.00  &        1.00  &        1.00  &        1.00  &        1.00  \\
		\end{tabular}
	\end{ruledtabular}
  \caption{Optimal parameters for LIGO-like and ET-like detectors with different mirrors' masses.}
  \label{table: ET, misc mass}
\end{table*}

\section{Conclusion}
In our article, we conduct a detailed study of Sagnac interferometer as an alternative configuration for the 3-rd generation GW detectors. Our analysis included the most robust quantum non-demolition technologies planned to be used in the second generation detectors, namely, frequency-dependent squeezed vacuum injection and detuned signal recycling. We made a fair comparison of equivalent Michelson and Sagnac interferometers using basic parameters for two prospective projects of future detectors, namely, Advanced LIGO and Einstein Telescope. In both cases, we clearly demonstrated a much stronger potential for broadband quantum noise suppression by Sagnac interferometer. We derived optimal parameter sets for both projects. 

In case of aLIGO-like scheme, \textit{i.e.} for arm length of $4$ km and maximum mirror mass of $40$ kg, the quantum noise of Sagnac interferometer is consistently better by factor of 3 to 10 times than the quantum noise of the currently planned baseline aLIGO broadband configuration (see solid red line in left panels (a) and (c) of Fig.~\ref{fig:SI_vs_MI_best}). Michelson interferometer with the same base parameters shows much poorer performance at lower frequencies, as shown by blue dashed lines in left panels (a) and (c) of Fig.~\ref{fig:SI_vs_MI_best}. 

For a more ambitious European project Einstein Telescope that features xylophone configuration of two Michelson interferometers which individual sensitivities are tailored so, as to provide best low- and high-frequency performance, respectively, we demonstrate that a single Sagnac interferometer can do almost equally well, or even better in terms of quantum noise for much lower price.  This is clearly seen from the solid red lines in the right panels (b) and (d) of Fig.~\ref{fig:SI_vs_MI_best}. 

Two interesting conclusions can be derived from the comparison of optimal parameters for Michelson and Sagnac interferometers given in Table~\ref{tab:SI_vs_MI_best}. Firstly, due to inherently low radiation pressure noise at low frequencies, requirements to filter cavities for Sagnac configuration are much relaxed compared to that for Michelson one. In the former case the optimal filter cavity has more than an order of magnitude broader bandwidth (thus lower finesse) than in the latter one, which results in much weaker dependence of sensitivity on optical loss in these devices (see Fig.~\ref{fig: LIGO, misc lFC}). Secondly, strong radiation pressure at low frequencies precludes Michelson interferometer to benefit from stronger squeezing (optimal squeezing factor is lower than the upper limit allowed by the optimisation). to the contrary, Sagnac interferometer requires as strong squeezing as possible, thereby making full use of this experimentally most developed QND technology.

Finally, we have shown that slightly detuned from resonance, low-finesse signal recycling cavity allows to improve medium-frequency quantum-noise sensitivity of Sagnac interferometer by tailoring the ponderomotive rotation angles frequency dependence so as to match the filter cavity's phase response function, thereby providing almost optimal output squeezed state rotation angle.  
 
\section*{Acknowledgements}
Authors are grateful to Haixing Miao, Christian Gr\"af and Stefan Hild for fruitful and elucidating discussions, as well as to all the members of the Macroscopic Quantum Mechanics discussion group for helpful comments and suggestions on improvement of our manuscript. S. D. is grateful to the University of Western Australia for support under the UWA Postdoctoral Research Fellowship scheme.

\appendix

\section{SQL for multi-mirror mechanical modes of an interferometer}\label{app:SQL}

Consider an effective mechanical mode $\hat{x}_\mu$ of the interferometer that can be defined in most common form as:
\begin{equation}\label{app eq: x_mu def}
 \hat{x}_\mu(\Omega) = \sum_{j = 1}^N \frac{\hat{x}_j(\Omega)}{\alpha_j} \,,
\end{equation}
with $\hat{x}_j$ standing for individual displacements of each of the $N$ interferometer's moving mirrors with individual masses $M_j$. Coefficients $\alpha_j$ are arbitrary constants different authors prefer to choose individually (in most LSC papers it is assumed 1). Note that all $\hat{x}_j$ in this formula are taken with "+" sign unlike the definition of $x_{\rm dARM}$ in the footnote on Page~\pageref{footnote: dARM}. This is simply because we choose $x$-axis for each mirror so, as to absorb the sign information in the definition of $\hat{x}_j$.

Assuming all test masses free, one writes their equations of motion in frequency domain as $\chi_{j;\,0}^{-1}(\Omega)\,\hat{x}(\Omega) = \hat{F}_j$ with $\chi_{j;\,0}(\Omega) = -1/(M_j\Omega^2)$ is the mechanical susceptibility of a free mass and $\hat{F}_j$ is the external force acting on it. The equation of motion for $\hat{x}_\mu$ then reads:
\begin{equation}\label{app eq: chi0_mu and F_mu def}
\begin{gathered}
 \chi_{\mu;\,0}^{-1}(\Omega) \, \hat{x}_\mu(\Omega) = \hat{F}_\mu(\Omega) \,, \\
 \mathrm{where}\ \chi_{\mu;\,0}(\Omega) = \sum_{j=1}^N \frac{\chi_{j;\,0}(\Omega)}{\alpha_j^2} \,, \\
\mathrm{and}\  \hat{F}_\mu(\Omega) = \frac{1}{\chi_{\mu;\,0}(\Omega)} \sum_{j=1}^N \frac{\chi_{j;\,0}(\Omega) \hat{F}_j(\Omega)}{\alpha_j} \,.
\end{gathered}
\end{equation}
Here index $0$ in $\chi_{\mu;\,0}$ indicates that this is the mechanical susceptibility of the mode $\hat{x}_\mu$ without any rigidities, either mechanical, or optical. By definition $\chi_{\mu;\,0}^{-1}(\Omega) = - \mu\Omega^2$ thus yielding the following expression for the reduced mass $\mu$
\begin{equation}\label{app eq: mu def}
 \mu = \left( \sum_{j=1}^N \frac{1}{\alpha_j^2\,M_j} \right)^{-1} .
\end{equation}

In GW interferometers, when only quantam noise is considered, external force $\hat{F}_j$ comprises of two components, the GW-induced one
\begin{equation}\label{app eq: G_j def}
  G_j(\Omega) = \chi_{j;\,0}^{-1}(\Omega) \cdot \frac{L_j h(\Omega)}{2} \,,
\end{equation}
with $h$ the GW-strain amplitude, and the radiation pressure one (or back-action one) that can be written as:
\begin{equation*}
  \hat{F}_j^{\rm pond}(\Omega) = \hat{F}_j^{\rm b.a.\;fl}(\Omega) + K_j(\Omega)\,\hat{x}_j(\Omega) \,,
\end{equation*}
where $\hat{F}_j^{\rm b.a.\;fl}$ is a fluctuational part of back action force and $K_j$ is an optical rigidity. The requirement that all $K_j$ shall be combined into a single optical rigidity $K_\mu$ of the effective mode $\hat{x}_\mu$ sets a constraint on coefficients $\alpha_j$:
\begin{equation*}
  \frac{\alpha_j}{\alpha_i} = \frac{\chi_{j;\,0}(\Omega)\,K_j(\Omega)} {\chi_{i;\,0}(\Omega)\,K_i(\Omega)} \,, \quad
  \forall i,j \,.
\end{equation*}
This is why, for instance, for a Fabry-P\'erot arm cavities with high-reflectivity mirrors all $\alpha_i$ are equal, \textit{i.e.} $\alpha_i = \alpha$.

It can be shown that expressions \eqref{app eq: x_mu def}-\eqref{app eq: mu def} conserve the form of the system Lagrangian when it is rewritten for the effective mode. The conservation of traditional form of the Robertson–Schr{\"o}dinger uncertainty relation [see eq. (148) in \cite{Liv.Rv.Rel.15.2012}] is one more indication of this fact:
\begin{equation*}
 S_{xx}(\Omega)S_{FF}(\Omega) - \abs{S_{xF}(\Omega)}^2 \geqslant \frac{\hbar^2}{2}\,,
\end{equation*}
Here $S_{xx}$, $S_{FF}$ and $S_{xF}$ are single-sided spectral densities of displacement measurement noise, back action noise and their cross-correlation spectral density for the effective mode $\hat x_\mu$, correspondingly. Therefore one can say that this single degree of freedom describes collective motion of all the test masses correctly.

The single-sided force SQL spectral density for $\hat x_\mu$ reads:
\begin{equation*}
 S_{\rm SQL}^F(\Omega) = 2\hbar \abs{\chi_\mu(\Omega)}^{-1} \,,
\end{equation*}
where susceptibility function $\chi_\mu(\Omega) = \left[ \chi_{\mu;\,0}^{-1}(\Omega) + K_\mu(\Omega) \right]^{-1}$ now includes optical rigidity $K_\mu(\Omega)$. To get the SQL spectral density normalised to GW-strain, $S_{\rm SQL}^h$, one must calculate the factor $\mathfrak{K}$, which relates tidal force $G_\mu(\Omega)$ to the GW-strain amplitude $h(\Omega)$:
\begin{equation*}
 G_\mu(\Omega) = \mathfrak{K}(\Omega) \, h(\Omega), \qquad
 S_{\rm SQL}^h(\Omega) = \frac{S_{\rm SQL}^F(\Omega)}{\abs{\mathfrak{K}(\Omega)}^2}.
\end{equation*}

In this paper, we assume $\alpha_j = \alpha$, each mirror mass $M_j = M$, $L_j|_{\rm EMT} = L$, where $L$ stands for arms length, and $G_j|_{\rm IMT} = 0$. Hence,
\begin{equation*}
 \mathfrak{K}_{2\times\rm a} = - \frac{\alpha}{4} M L \Omega^2 \,,
\end{equation*}
which follows from  eq.~\eqref{app eq: chi0_mu and F_mu def}, \eqref{app eq: G_j def}.

In case of free system, with $\chi_\mu(\Omega) = \chi_{\mu;\,0}(\Omega)$, and stands the free mass $M$ force SQL by $f_{\rm SQL}^2 \equiv S_{\rm f.m.\;SQL}^F(\Omega)$ one can get the following:
\begin{equation*}
 S_{\rm 4 \times M \; SQL}^F(\Omega) = \frac{\alpha^2}{4} f_{\rm SQL}^2 \,,\qquad
 f_{\rm SQL}^2 = 2\hbar M \Omega^2 \,.
\end{equation*}
Therefore, the GW-strain single-sided SQL spectral density reads:
\begin{equation*}
 S_{\rm 4 \times M \; SQL}^h(\Omega) = \frac{4 \, f_{\rm SQL}^2}{M^2 L^2 \Omega^4} = h_{\rm SQL}^2
\end{equation*}
where $h_{\rm SQL}$ is defined in Eq. \eqref{eq:hSQL}. Note that GW-strain SQL does not depend on exact mechanical mode selection, i.e. on $\alpha$.

\section{Ponderomotive squeezing in GW interferometers}\label{app:pond_sqz}

Ponderomotive squeezing that takes place in a tuned lossless Michelson interferometer can be written as a sequence of 3 unitary transformations -- rotation, squeezing and second rotation \cite{02a1KiLeMaThVy}:
\begin{equation}\label{app.eq:sqz_state}
  \ket{out} = e^{2i\beta}\hat{R}(u_{\rm pond})\hat{S}(r_{\rm pond})\hat{R}(v_{\rm pond})\ket{in}.
\end{equation}
where $\beta$ is a scheme-specific complex frequency-dependent phase shift which does not change the noise spectral density, the rotation operator $\hat{R}(\alpha)$ and the squeezing operator $\hat{S}(r)$ are defined in Section~3.2 of \cite{Liv.Rv.Rel.15.2012}. Action of these operators on the vector of light quadratures, $\hat{\vb{a}} = \{\hat{a}_1,\,\hat{a}_2\}^{\rm T}$, results in a new vector, $\vq{b} = \{\hat{b}_1,\,\hat{b}_2\}^{\rm T}$, that reads:
\begin{equation}\label{app.eq: pond sqz matr transformation}
  \vb{b} = \tq{T}\,\vq{a} = e^{2i\beta}\,\mathbb{R}[u_{\rm pond}]\,\mathbb{S}[r_{\rm pond}]\,\mathbb{R}[v_{\rm pond}]\,\vq{a}\,,
\end{equation}
with $\tq{R}$ the rotation matrix defined in Eq.~\eqref{eq: Rot Matr RR} and $\tq{S}$ the squeezing matrix equal to:
\begin{equation}\label{eq:SqzTr}
  \tq{S}[r] = \matr{e^r}{0}{0}{e^{-r}} \,.
\end{equation}

For a general optomechanical system without loss, the transfer matrix (TM) has a specific structure, namely, the optical TM is $\tq{T}^{\rm meas} = e^{2i\beta}\tq{R}[\psi]$ and the radiation pressure one in proportional to $\tq{T}^{\rm b.a.} \propto \vs{t}\left(\sigma_1\vs{t}\right)^\tr$, where $\sigma_1$ is the Pauli`s matrix, also known as $\sigma_x$. This structure of TM guarantees that the covariance matrix of the output field, $\tq{V}_b = \tq{T}\,\tq{V}_a\tq{T}^\dag$, will remain real like the input $\tq{V}_a$, which are corresponded to gaussian $\ket{in}$ and $\ket{out}$ states. Factoring out common complex phase $e^{2i\beta}$, one ends up with a real matrix $\tq{T}^{\rm Re} = e^{-2i\beta}[\tq{T}^{\rm meas}+\tq{T}^{\rm b.a.}]$, the singular value decomposition of which can be written as:
\begin{equation*}
  \tq{T}^{\rm Re} = \mathbb{R}[u_{\rm pond}]\,\mathbb{S}[r_{\rm pond}]\,\mathbb{R}[v_{\rm pond}] \,,
\end{equation*}
that proves that Eq. \eqref{app.eq: pond sqz matr transformation} is indeed correct.

In order to get the expressions for $r_{\rm pond}$, $u_{\rm pond}$ and $v_{\rm pond}$, one can expand $\tq{T}^{\rm Re}$ in Pauli matrices:
%
\begin{equation*}
  \tq{T}^{\rm Re} = \til{z}_0\tq{I} + \til{z}_1\sigma_1 + \til{z}_2\sigma_2 + \til{z}_3\sigma_3
\end{equation*}
where $\til{z}_{0,1,2,3}$ are complex coefficients.

Symmetries of the TM immediately allow to see that $\til{z}_3 = 0$ and the $\til{z}_0 = \tq{T}^{\rm Re}_{cc} = \tq{T}^{\rm Re}_{ss}$.
Since all elements of $\tq{T}^{\rm Re}$ are real, the following relations hold for the remaining coefficients:
\begin{equation*}
  \til{z}_1 = -\frac{ \tq{T}^{\rm Re}_{cs} + \tq{T}^{\rm Re}_{sc} }{2} = z_1, \quad
  \til{z}_2 = i \frac{ \tq{T}^{\rm Re}_{cs} - \tq{T}^{\rm Re}_{sc} }{2} = i \cdot z_2 \,,
\end{equation*}
which means $z_1,z_2$ are real. 

Then singular values can be calculated:
\begin{equation*}
  s_{1,2} = \abs{ \abs{z_1} \pm \sqrt{z_0^2 + z_2^2} } \,.
\end{equation*}
Assuming $e^{r_{\rm pond}} = \max\{s_1,s_2\}$ and $e^{-r_{\rm pond}} = \min\{s_1,s_2\}$ (i.e. $r_{\rm pond} > 0$) one can get the following expression:
\begin{equation*}
  \sinh r_{\rm pond} = 
  \left\{
  \begin{aligned}
  &\abs{z_1},               & & \text{if } \det\tq{T}^{\rm Re} = 1  \,, \\
  &\sqrt{z_0^2 + z_2^2} \,, & & \text{if } \det\tq{T}^{\rm Re} = -1 \,.
  \end{aligned}
  \right.
\end{equation*} 

The expression for angles $u_{\rm pond}$ and $v_{\rm pond}$ are:
\begin{align*}
  u_{\rm pond} &= - \frac{1}{2} \arctan\frac{z_2}{z_0} - \mathrm{sgn}\left[z_1\right]\frac{\pi}{4} \,, \\
  v_{\rm pond} &= - \frac{1}{2} \arctan\frac{z_2}{z_0} + \mathrm{sgn}\left[z_1\right]\frac{\pi}{4} \,.
\end{align*}

\section{Transfer matrices for interferometers with losses}\label{app:transfer matrices}
Here we derive the I/O-relations for considered interferometers beyond narrow-band approximation.

\subsection{Fabry-P\'erot interferometer with moving mirror}\label{app:lossy FP I/O-rels}
I/O relations for a single Fabry-P\'erot arm cavity without any additional assumptions about its bandwidth can be written as:

%
\begin{equation}\label{app. matr.: i-o realtion for arm}
  \vq[arm]{b} = \tq[arm]{T}\vq[arm]{a} + \tq[arm]{N}\vq[arm]{n} + \vs[arm]{t}\frac{h}{\sqrt{2}\,h_{\rm SQL}} \,,
\end{equation}
where transfer matrices and signal response function read:
\begin{equation}\label{app eq: TT, NN and t for arm}
\begin{aligned}
  \tq[arm]{T} &= \sqrt{\TITM}\tq[arm]{M}\tq[\tauArm]{P}\tq[ETM]{T}\tq[\tauArm]{P}\tq[ITM]{N} - \sqrt{\RITM}\tq{I} \,, \\
  \tq[arm]{N} &= \sqrt{\TITM}\tq[arm]{M}\tq[\tauArm]{P}\tq[ETM]{N} \,, \\
  \vs[arm]{t} &= \sqrt{2\,\TITM}\tq[arm]{M}\tq[\tauArm]{P}\vs[ETM]{t} \,, \\
  &\tq[arm]{M} = \left[ \tq{I} - \tq[\tauArm]{P}\tq[ETM]{T}\tq[\tauArm]{P}\tq[ITM]{T} \right]^{-1} \,,
\end{aligned}
\end{equation}
with
\begin{equation}\label{app eq: TT, NN and t for ETM}
\begin{aligned}
  \tq[ETM]{T} &= \sqrt{\RETM} \left( \tq{I} + \tq[ETM]{M} \right) \,, \\
  \tq[ETM]{N} &= \sqrt{\TETM} \left( \tq{I} + \tq[ETM]{M} \right) \,, \\
  \tq[ETM]{M} &= \matr{0}{0}{-\RETM\mathcal{K}_{\rm TM}}{0} \,,       \\
  \vs[ETM]{t} &= \sqrt{2 \RETM \mathcal{K}_{\rm TM}} \col{0}{1} \,.
\end{aligned}
\end{equation}
Here $\tq[\tauArm]{P} = e^{i\Omega\tauArm} \tq{R}[\omega_p\tauArm]$ describes the free propagation of light between the mirrors of the arm cavity. 
\begin{equation*}
  \mathcal{K}_{\rm TM} = \frac{8 \omega_p P_c}{M c^2 \Omega^2} = \dfrac{2\Theta_{\rm arm}\tau_{\rm arm}}{\Omega^2}\,,
\end{equation*}
is an optomechanical coupling factor for a single perfectly reflective free mirror and $P_c$ stands for the full power of incident light beam. Note that here $h_{\rm F.P.\;SQL} = \sqrt{2}\,h_{\rm SQL}$ stands for the SQL of a Fabry-P\'erot cavity with 2 movable mirrors of mass $M$ each, thus the factor $\sqrt{2}$ in front.

%
%
\subsubsection{Tuned cavity}
In the case of resonance tuned cavity, the above expressions simplify significantly and can be written as:
\begin{align*}
  \tq[arm]{T} &= T e^{2i\beta_{\rm arm}} \matr{1}{0}{-\sqrt{\RETM}\mathcal{K}_{\rm arm}}{1}, \\
  \tq[arm]{N} &= N e^{i\beta_{\rm arm}}  \matr{1}{0}{-\mathcal{N}}{1}, \\
  \vs[arm]{t} &= t e^{i\beta_{\rm arm}}  \sqrt{\frac{4\,\RETM\mathcal{K}_{\rm arm}}{1+\RITM}} \col{0}{1},
\end{align*}
where in small optical losses approximation ($\RETM \ll 1$):
\begin{align*}
  \mathcal{K}_{\rm arm} &= \frac{\left(1+\RITM\right)\TITM\mathcal{K}_{\rm TM}}{1-2\sqrt{\RITM} \cos 2\Omega\tau + \RITM} \\
  \beta_{\rm arm}       &= \atan{ \dfrac{1+\sqrt{\RITM}}{1-\sqrt{\RITM}} \tan\Omega\tau }, \\
  N           &= \sqrt{\frac{1-\RETM}{1+\RITM}\frac{\mathcal{K}_{\rm arm}}{\mathcal{K}_{\rm TM}}}, \quad T = t = 1, \\
  \mathcal{N} &= \sqrt{\frac{\mathcal{K}_{\rm arm}\mathcal{K}_{\rm TM}\RETM}{1-\RITM^2}}
                 \frac{1 + e^{2i\Omega\tauArm}\RITM^{3/2}}{e^{-i\beta_{\rm arm}+i\Omega\tauArm}}
\end{align*}

Following the reasoning after Eq.~\ref{eq:lossyFP} one can derive the following expressions for shot-noise and back-action components of the transfer matrices:  
\begin{gather}
  \tq{T}^{\rm meas} = e^{2i\beta_{\rm arm}} \tq{I}, \quad
  \tq{T}^{\rm b.a.} = e^{2i\beta_{\rm arm}} \matr{0}{0}{-\sqrt{\RETM}\mathcal{K}_{\rm arm}}{0},\label{eq_app:FP_Tsn}\\
  \tq{N}^{\rm meas} = e^{i\beta_{\rm arm}} N \tq{I}, \quad
  \tq{N}^{\rm b.a.} = e^{i\beta_{\rm arm}} N \matr{0}{0}{-\mathcal{N}}{0}. \label{eq_app:FP_Nsn}
\end{gather}

%
%
\subsubsection{Detuned narrow-band filter cavity.}

In case of filter cavities, mirrors can be assumed fixed and no radiation pressure effects are to be considered due to an absence of any significant classical light component therein. One can also make a so called narrow-band approximation, assuming $\Omega L_f / c \ll 1$, where $L_f$ is the filter cavity length and $T_f \ll 1$ is input mirror power transmissivity. Then one can write transfer matrices as:
\begin{align*}
  \tq[FC]{T} &= \frac{1}{\mathcal{D}} \matr{t_1}{t_2}{-t_2}{t_1}, \;
  \begin{aligned}
    t_1 &= \gamma_{f1}^2 - \gamma_{f2}^2 - \delta_f^2 + \Omega^2 + 2i\Omega\gamma_{f2}, \\
    t_2 &= - 2 \gamma_{f1} \delta_f,
  \end{aligned}
  \\
  \tq[FC]{N} &= \frac{2\sqrt{\gamma_{f1}\gamma_{f2}}}{\mathcal{D}} \matr{\gamma_f-i\Omega_1}{-\delta_f}{\delta_f}{\gamma_f-i\Omega},
\end{align*}
where $\mathcal{D} = (\gamma_f-i\Omega)^2 + \delta_f^2$, $\gamma_f = \gamma_{f1} + \gamma_{f2}$ is a full cavity half-bandwidth and $\delta_f$ if its detuning. Here $\gamma_{f1} = c T_f / (4L_f)$ is a half-bandwidth part depending on input mirror transmissivity and $\gamma_{f2} = c A_f / (4L_f)$ is the loss-associated part of bandwidth with $A_f \ll 1$ being the total round-trip fractional photon loss.

\subsection{Michelson/Fabry-P\'erot interferometer}
\subsubsection{Fabry-P\'erot--Michelson (FPM) interferometer w/o signal recycling.}

I/O-relations of a Michelson/Fabry-P\'erot interferometer can be obtained by completing the above ones for the single arm with junction relations at the beam splitter:
\begin{eqnarray*}
  \vq{a}^{\rm N} = \dfrac{\vq{p}+\vq{i}}{\sqrt{2}}\,, &
  \vq{a}^{\rm E} = \dfrac{\vq{p}-\vq{i}}{\sqrt{2}}\,, &
  \vq{o} = \dfrac{\vq{b}^{\rm N}-\vq{b}^{\rm E}}{\sqrt{2}}\,,
\end{eqnarray*}
where $\vq{a}^{\rm N,E} \equiv \vq[arm]{a}^{\rm N,E}$, $\vq{b}^{\rm N,E} \equiv \vq[arm]{b}^{\rm N,E}$ stand for the input and output fields of the  $N$ and $E$ arms, respectively. Hence, the Michelson interferometer I/O-relations read:
\begin{equation}\label{app. matr.: i-o for tuned Mich}
  \vq{o} = \tq[MI]{T}\vq{i} + \tq[MI]{N}\vq{n} + \vs[MI]{t}\frac{h}{h_{\rm SQL}}\,,
\end{equation}
where
\begin{equation*}
  \tq[MI]{T} = \tq[arm]{T} \,, \;\;
  \tq[MI]{N} = \tq[arm]{N} \,, \;\;
  \vs[MI]{t} = \vs[arm]{t} \,. 
\end{equation*}
Here $\vq{n} = \left(\vq[arm]{n}^{\rm N} - \vq[arm]{n}^{\rm E}\right)/\sqrt{2}$ represents effective vacuum fields associated with optical loss in the arm cavities.

In case of small losses the interferometer is described by opto-mechanical factor $\mathcal{K}_{\rm MI} = \mathcal{K}_{\rm arm}$ and phase $\beta_{\rm MI} = \beta_{\rm arm}$.

\subsubsection{Signal-recycled Fabry-P\'erot--Michelson (FPM) interferometer.}

Along the same lines as is done in Sec.~\ref{sec:SR_Sag}, the I/O-relations of signal recycling interferometer can be obtained from the following equations written for light fields on a signal recycling mirror (SRM):
\begin{equation}\label{app. matr.: SRM i-o in Mich}
  \left\{
  \begin{aligned}
    \vq[SR]{o} &= \tq[\alpha]{P} \left( \sqrt{1-\rho_{\rm SR}^2}\tq[SR]{P}\vq{o} + \rho_{\rm SR}\tq[\alpha]{P}\vq[SR]{i} \right) \\
    \vq{i}     &= \tq[SR]{P} \left( \rho_{\rm SR}\tq[SR]{P}\vq{o} + \sqrt{1-\rho_{\rm SR}^2}\tq[\alpha]{P}\vq[SR]{i} \right),
  \end{aligned}
  \right.
\end{equation}
where an additional phase shift $\alpha_{\rm SR}$ is introduced to satisfy the Scaling Law of \cite{2003_PhysRevD.67.062002_ScalLaw} which maps the signal-recycled FPM interferometer and a single detuned Fabry-P\'erot cavity:
\begin{gather*}
  \tq[SR]{P} = e^{i\frac{\Omega l_{\rm SR}}{c}} \tq{R}[\phiSR] \simeq \tq{R}[\phiSR] \,, \quad
  \phiSR = \frac{\omega_p l_{\rm SR}}{c} \,, \\
  \tq[\alpha]{P} \simeq \tq{R}[\alpha_{\rm SR}] \,, \quad
  \alpha_{\rm SR} = \arctan\left(\frac{\rho_{\rm SR}-1}{\rho_{\rm SR}+1}\tan\phiSR\right) \,.
\end{gather*}


The solution of \eqref{app. matr.: i-o for tuned Mich} and \eqref{app. matr.: SRM i-o in Mich} gives the following:
\begin{equation*}
  \vq[SR]{o} = \tq[MI\,SR]{T}\vq[SR]{i} + \tq[MI\,SR]{N}\vq{n} + \vs[MI\,SR]{t}\frac{h}{\hSQL} \,, \\
\end{equation*}
where
\begin{align*}
  \tq[MI\,SR]{T} &= \tq[\alpha]{P} \bigl[ (1-\rho_{\rm SR}^2)\tq[SR]{P}\tq[MI\,SR]{M}\tq[MI]{T}\tq[SR]{P} \nonumber \\
                 &  - \rho_{\rm SR}\tq{I} \bigr] \tq[\alpha]{P} \,, \\
  \tq[MI\,SR]{N} &= \sqrt{1-\rho_{\rm SR}^2} \tq[\alpha]{P} \tq[SR]{P} \tq[MI\,SR]{M} \tq[arm]{N} \,, \\
  \vs[MI\,SR]{t} &= \sqrt{1-\rho_{\rm SR}^2} \tq[\alpha]{P} \tq[SR]{P} \tq[MI\,SR]{M} \vs[MI]{t} \,, \\
  &\tq[MI\,SR]{M} = \left[ \tq{I} - \rho_{\rm SR} \tq[arm]{T} \tq[SR]{P}^2 \right]^{-1} \,.
\end{align*}

\subsection{Sagnac interferometer}\label{app:SRIOrels}

Full I/O-relation for the Sagnac interferometer are given in \eqref{eq:SI_IO_rels} and we repeat them here:
\begin{equation*}\label{app. matr.: i-o in Sagnac with ring cavities}
  \vq{o} = \tq[SI]{T}\vq{i} + \tq[SI]{N}^{\rm I}\vq[I]{n} + \tq[SI]{N}^{\rm II}\vq[II]{n} + \vs[SI]{t}\frac{h}{h_{\rm SQL}} \,.
\end{equation*}
The I/O-relations for each arm cavity of the Sagnac interferometer in general case read:
\begin{multline}\label{app. matr.: Sagnac arm i-o}
  \vq{b}^{IJ} = \tq[arm]{T}\vq{a}^{IJ} + \tq[arm]{N}\vq{n}^{IJ} +
                \left(-1\right)^J \vs[arm]{t} \frac{h}{\hSQL} + \\ +
                \tq[arm]{T}^{\rm b.a.}\vq{a}^{\bar{I}J} + \tq[arm]{N}^{\rm b.a.}\vq{n}^{\bar{I}J} \,,
\end{multline}
which accounts for an additional back action introduced by a counter-propagating beam into the response matrix of the beam under study. Here $\left(-1\right)^N \equiv 1$, $\left(-1\right)^E \equiv -1$ and $\vq{n}^{IJ} \equiv \vq[arm]{n}^{IJ}$. The effective vacuum fields that correspond to the optical losses in the arms in case of ring-cavity Sagnac interferometer are given by Eq.~\eqref{eq: n_I, n_II in SI}. For the polarisation Sagnac interferometer the definition of $\vq[I]{n}$ and $\vq[II]{n}$ differs only by the sign (minus).

\subsubsection{Ring-arm-cavities Sagnac interferometer w/o signal recycling.}

Starting with the case w/o signal recycling, one can calculate transfer matrices using standard techniques and the result reads:
\begin{equation}\label{app eq: lossy SI matrices}
\begin{aligned}
  \tq[SI]{T}          &= \tq[arm]{T}^{\rm b.a.} - \tq[arm]{T}\tq[SI]{M}\tq[arm]{T} \,, \\
  \tq[SI]{N}^{\rm I } &= \tq[arm]{N}            - \tq[arm]{T}\tq[SI]{M}\tq[arm]{N}^{\rm b.a.} \,, \\
  \tq[SI]{N}^{\rm II} &= \tq[arm]{N}^{\rm b.a.} - \tq[arm]{T}\tq[SI]{M}\tq[arm]{N} \,, \\
  \vs[SI]{t}          &= - \sqrt{2} \left(\tq{I} - \tq[arm]{T}\tq[SI]{M}\right) \vs[arm]{t} \,, \\
  \tq[SI]{M}          &= \left[\tq{I} + \tq[arm]{T}^{\rm b.a.}\right]^{-1} = \tq{I} - \tq[arm]{T}^{\rm b.a.} \,.
\end{aligned}
\end{equation}
For small optical loss, \textit{i.e.} $\RETM \ll 1$, the SI transfer matrix and response function read:
\begin{align*}
  \tq[SI]{T} &= e^{2i\beta_{\rm SI}} \matr{1}{0}{-\sqrt{\RETM}\mathcal{K}_{\rm SI}}{0} \,, \\
  \vs[SI]{t} &= e^{i\beta_{\rm SI}} \sqrt{\frac{4\,\RETM\mathcal{K}_{\rm SI}}{1+\RITM}} \col{0}{1} \,,
\end{align*} 
where $\beta_{\rm SI}$ and $\mathcal{K}_{\rm SI}$ are given by Eqs.~\eqref{eq:SIphase} and \eqref{eq:SI_T_and_t}. 
Note the difference with the Michelson case, \textit{i.e.} the double amount of power circulating in the arms of the Sagnac interferometer compred to that of an equivalent Michelson one. This results in an additional factor of 2 in the definition of the circulating power in the interferometer, $P_c = 4P_{\rm arm}$, where $P_{\rm arm}$ is understood as the single optical beam power, circulating in the arm cavity.

\subsubsection{Ring-arm-cavities Sagnac interferometer with signal recycling.}

To get now the transfer matrices and response for a signal recycled Sagnac interferometer, one have to solve equations \eqref{app. matr.: SRM i-o in Mich} and \eqref{eq:SI_IO_rels} jointly. This leads to the following expressions:
\begin{align}\label{app eq: lossy SR SI matrices}
  \tq[SI\,SR]{T}              &= \tq[\alpha]{P} \bigl[ (1-\rho_{\rm SR}^2)\tq[SR]{P}\tq[SI\,SR]{M}\tq[SI]{T}\tq[SR]{P} -
                                                       \rho_{\rm SR}\tq{I} \bigr] \tq[\alpha]{P} \,, \\
  \tq[SI\,SR]{N}^{\rm I,\,II} &= \sqrt{1-\rho_{\rm SR}^2} \tq[\alpha]{P} \tq[SR]{P} \tq[SI\,SR]{M} \tq[SI]{N}^{\rm I,\,II} \,, \\
  \vs[SI\,SR]{t}              &= \sqrt{1-\rho_{\rm SR}^2} \tq[\alpha]{P} \tq[SR]{P} \tq[SI\,SR]{M} \vs[SI]{t} \,, \\
  \mbox{where } &\tq[SI\,SR]{M}              = \left[ \tq{I} - \rho_{\rm SR} \tq[\alpha]{P} \tq[SI]{T} \tq[SR]{P}^2 \right]^{-1} \,.
\end{align}
Here again, $\tq[\alpha]{P} \simeq \tq{R}[\alpha_{\rm SR}]$, where $  \alpha_{\rm SR} = \arctan\left(\frac{\rho_{\rm SR}-1}{\rho_{\rm SR}+1}\tan\phiSR\right)$ is the auxiliary phase shift added to satisfy the Scaling Law theorem \cite{2003_PhysRevD.67.062002_ScalLaw}.

\subsection{Polarisation Sagnac interferometer with polarisation beam splitter (PBS) leakage}\label{app:PBS leak I/O-relations}

In this Appendix, we analyse in detail the input output relations of a polarisation Sagnac interferometer with lossy PBS, following the analysis of Wang \textit{et al.} \cite{PhysRevD.87.096008}, but expanding their analysis to arbitrary signal-recycled Sagnac interferometers not limited by the approximation of narrow band of arm cavities.
\begin{figure}[ht]
  \includegraphics[width=.49\textwidth]{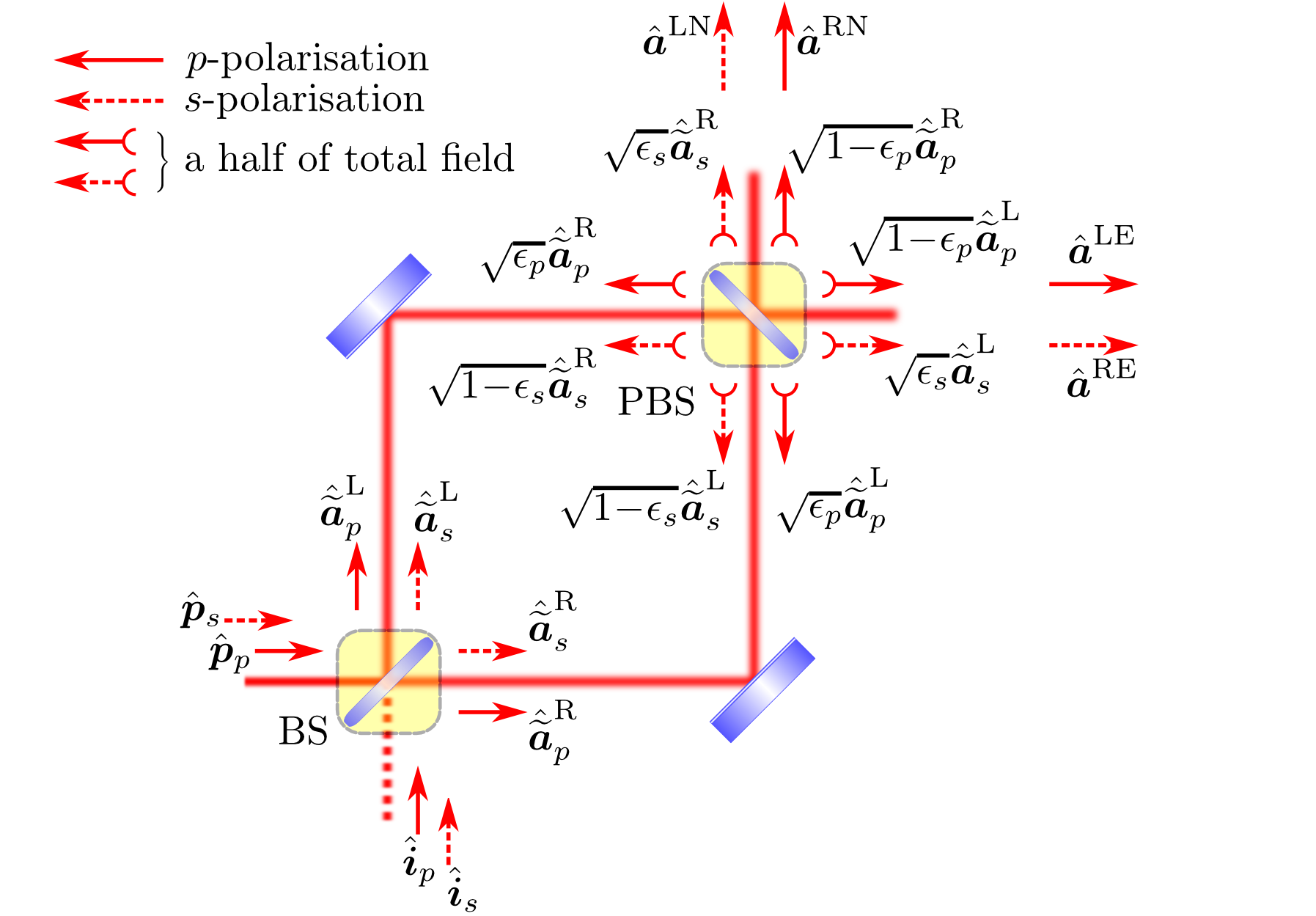}
  \includegraphics[width=.49\textwidth]{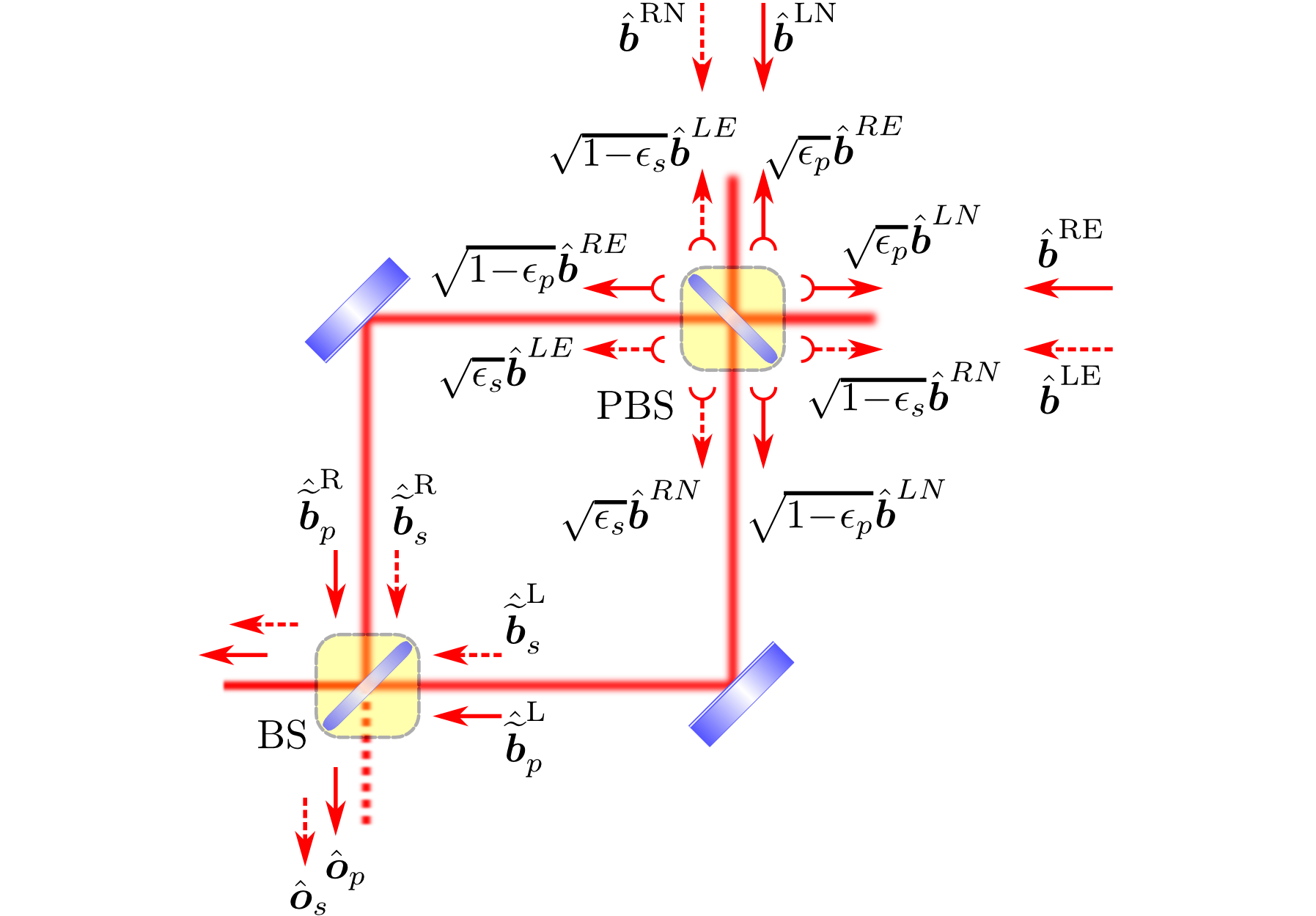}
  \caption{
Schematics of the field transformations on a lossy beam splitter (BS) and polarisation beam splitter (PBS). The straight arrows correspond to total fields of orthogonal polarisations, while the arrows with semi-circular tails represent the two components comprising the outgoing field, \textit{i.e.} the nominal polarisation component and the leaked component.
  }
\end{figure}

An ideal PBS is an optical element transmits 100\% of $p$-polarised light and reflects by 90$^\circ$ the orthogonally $s$-polarised one. 
However, in reality, some fraction of the $s$-component gets transmitted and some fraction of the $p$-component gets reflected. To characterise this leakage one usually introduces leakage coefficients, $\sqrt{\epsilon_p}$ for reflected fraction of the $p$-component, and $\sqrt{\epsilon_s}$ for transmitted fraction of the $s$-component. Physically, $\epsilon_p$ and $\epsilon_s$ show the percentage of unwanted polarised light power mixed into the desired one. Keeping to the system of notations for the clockwise and counterclockwise propagating beams adopted in the main text, one denotes fields leaving the beam splitter toward the main interferometer and PBS as $\vqtil{a}^{R,L}$, and the ones leaving it and mixing at the beam splitter to finally leave the interferometer for the readout port as $\vqtil{b}^{R,L}$. Hence we can get the following junction equations at the BS:
\begin{eqnarray*}
  \vqtil{a}^{\rm L}_{p,s} = \dfrac{\vq{p}_{p,s}+\vq{i}_{p,s}}{\sqrt{2}} \,, &
  \vqtil{a}^{\rm R}_{p,s} = \dfrac{\vq{p}_{p,s}-\vq{i}_{p,s}}{\sqrt{2}} \,, &
  \vq{o}_{p,s}            = \dfrac{\vqtil{b}^{\rm R}_{p,s}-\vqtil{b}^{\rm L}_{p,s}}{\sqrt{2}} \,,
\end{eqnarray*}
where for $IJ = \{RE,\,LN\}$ beams one has
\begin{equation*}
  \vq{a}^{\rm IJ}     = \sqrt{1-\epsilon_s}\vq{b}^{\rm I\bar{J}}  +  \sqrt{\epsilon_s}\vqtil{a}^{\rm \bar{I}}_s \,, \quad
  \vqtil{b}^{\rm I}_p = \sqrt{1-\epsilon_p}\vq{b}^{\rm IJ}        +  \sqrt{\epsilon_p}\vqtil{a}^{\rm I}_p \,,
\end{equation*}
and for $IJ = \{LE,\,RN\}$
\begin{equation*}
  \vq{a}^{\rm IJ}           = \sqrt{1-\epsilon_p}\vqtil{a}^{\rm I}_p        +  \sqrt{\epsilon_p}\vq{b}^{\rm I\bar{J}} \,, \quad
  \vqtil{b}^{\rm \bar{I}}_s = \sqrt{1-\epsilon_s}\vqtil{a}^{\rm \bar{I}}_s  +  \sqrt{\epsilon_s}\vq{b}^{\rm IJ} \,.
\end{equation*}
The above expressions together with I/O-relations for the arms \eqref{app. matr.: Sagnac arm i-o} give us the full system of equations whose solution can be written in the following form with $P = \{s,p\}$:
\begin{equation*}
  \vq{o}_P = \tq[SI]{T}^{PP}\vq{i}_P + \tq[SI]{T}^{P\bar{P}}\vq{i}_{\bar{P}} + 
             \tq[SI]{N}^{PP}\vq{n}{}_P + \tq[SI]{N}^{P\bar{P}}\vq{n}{}_{\bar P} +
             \vs[SI]{t}\frac{h}{\hSQL}.
\end{equation*}
The correspondence of the above expression to the lossless Sagnac interferometer I/O-relations of Eq.~\eqref{eq:SI_IO_rels} is straightforward. The ``signal'' polarisation is $p$-polarisation, therefore $\vq{i}_p$ and $\vq{o}_p$ serve as an input and an output fields of the interferometer, and $\vq{n}_{p,\,s}$ corresponds to $\vq[I,\,II]{n}$.
\subsubsection{Polarisation Sagnac interferometer w/o signal recycling.}

Transfer matrices and response functions for both polarisations can be shown to have the following form:
\begin{align*}
  \tq[SI]{T}^{pp} &= \left(1-\epsilon_p\right)\tq[l]{M} \left[\tq[arm]{T}^{\rm b.a.} - \sqrt{1-\epsilon_s}\tq[SI]{M}\tq[arm]{T}^2\right] - \sqrt{\epsilon_p}\tq{I} \,,\\
  \tq[SI]{T}^{ss} &= \epsilon_s\tq[l]{M} \left[\tq[arm]{T}^{\rm b.a.} - \sqrt{\epsilon_p}\tq[SI]{M}\tq[arm]{T}^2\right] - \sqrt{1-\epsilon_s}\tq{I} \,,\\
  \tq[SI]{T}^{ps} &= \tq[SI]{T}^{sp} = \sqrt{1-\epsilon_p}\sqrt{\epsilon_s} \tq[l]{M} \tq[SI]{M} \tq[arm]{T} \,, \\
  \tq[SI]{N}^{pp} &= \sqrt{1-\epsilon_p} \tq[l]{M} \left[ \tq[SI]{M}^p\tq[arm]{N} - \sqrt{1-\epsilon_s}\tq[arm]{T}\tq[arm]{N}^{\rm b.a.} \right] \,, \\
  \tq[SI]{N}^{ps} &= \sqrt{1-\epsilon_p} \tq[l]{M} \left[ \tq[arm]{N}^{\rm b.a.}  - \sqrt{1-\epsilon_s}\tq[SI]{M}\tq[arm]{T}\tq[arm]{N}  \right] \,, \\
  \tq[SI]{N}^{ss} &= \sqrt{\epsilon_s}   \tq[l]{M} \left[ \tq[SI]{M}^s\tq[arm]{N} - \sqrt{\epsilon_p}\tq[arm]{T}\tq[arm]{N}^{\rm b.a.} \right] \,, \\
  \tq[SI]{N}^{sp} &= \sqrt{\epsilon_s}   \tq[l]{M} \left[ \tq[arm]{N}^{\rm b.a.}  - \sqrt{\epsilon_p}\tq[SI]{M}\tq[arm]{T}\tq[arm]{N}  \right] \,, \\
  \vs[SI]{t}^p &= - \sqrt{2\left(1-\epsilon_p\right)} \tq[l]{M} \left[\tq{I}-\sqrt{1-\epsilon_s}\tq[arm]{T}\right] \vs[arm]{t} \,, \\
  \vs[SI]{t}^s &= - \sqrt{2\epsilon_s} \tq[l]{M} \left[\tq{I}-\sqrt{\epsilon_p}\tq[arm]{T}\right] \vs[arm]{t} \,.
\end{align*}
where
\begin{align*}
  \tq[l]{M}    &= \left[ \tq{I} - \sqrt{1-\epsilon_s}\sqrt{\epsilon_p} \tq[SI]{M} \tq[arm]{T}^2 \right]^{-1} \,, \\
  \tq[SI]{M}   &= \tq{I} - \left(\sqrt{\epsilon_p}+\sqrt{1-\epsilon_s}\right) \tq[arm]{T}^{\rm b.a.} \,, \\
  \tq[SI]{M}^p &= \tq{I} - \sqrt{\epsilon_p}\tq[arm]{T}^{\rm b.a.} \,, \quad
  \tq[SI]{M}^s  = \tq{I} - \sqrt{1-\epsilon_s}\tq[arm]{T}^{\rm b.a.} \,.
\end{align*}
The following relations that follow from the specific structure of transfer matrices can be used to simplify the final expressions:
\begin{gather*}
  \tq[arm]{T}^{\rm b.a.}\tq[arm]{T}^{\rm b.a.} = \tq[arm]{N}^{\rm b.a.}\tq[arm]{N}^{\rm b.a.} = 
    \tq[arm]{T}^{\rm b.a.}\tq[arm]{N}^{\rm b.a.} = \tq[arm]{N}^{\rm b.a.}\tq[arm]{T}^{\rm b.a.} = 0 \,, \\
  \tq[SI]{M}^p\tq[SI]{M}^s = \tq[SI]{M}^s\tq[SI]{M}^p = \tq[SI]{M} \,, \\
  \left[ \tq[arm]{T}, \tq[arm]{T}^{\rm b.a.} \right] = \left[ \tq[SI]{M}^*, \tq[arm]{T} \right] = \left[ \tq[SI]{M}^*, \tq[arm]{N} \right] = 0 \,, \\
  \tq[SI]{M}^*\tq[arm]{T}^{\rm b.a.} = \tq[arm]{T}^{\rm b.a.} \,, \quad
  \tq[SI]{M}^*\tq[arm]{N}^{\rm b.a.} = \tq[arm]{N}^{\rm b.a.} \,, \quad
  \tq[SI]{M}^*\vs[arm]{t} = \vs[arm]{t} \,,
\end{gather*}
where $\tq[SI]{M}^*$ stands either for $\tq[SI]{M}^p$, or $\tq[SI]{M}^s$, or $\tq[SI]{M}$.

\subsubsection{Signal-recycled polarisation Sagnac interferometer.}

Now we can also write the exact expressions for the transfer matrices and response functions of the signal-recycled Sagnac interferometer with losses and imperfect PBS:
\begin{multline*}
  \vq{o}_{\mathrm{SR},\,P} = \tq[SI\,SR]{T}^{PP}\vq{i}_{\mathrm{SR},\,P} + \tq[SI\,SR]{T}^{P\bar{P}}\vq{i}_{\mathrm{SR},\,\bar{P}} + \\ +
                             \tq[SI\,SR]{N}^{PP}\vq{n}{}_P + \tq[SI\,SR]{N}^{P\bar{P}}\vq{n}{}_{\bar P} +
                             \vs[SI\,SR]{t}\frac{h}{\hSQL} \,,
\end{multline*}
where
\begin{align*}
  \tq[SI\,SR]{T}^{PP}       &= \tq[\alpha]{P} \bigl[ \left(1-\rho_{\rm SR}^2\right) \tq[SR]{P} \tq{C}^{PP} -
                                                     \rho_{\rm SR}\tq{I} \bigl] \tq[\alpha]{P} \,, \\
  \tq[SI\,SR]{T}^{P\bar{P}} &= \left(1-\rho_{\rm SR}^2\right) \tq[\alpha]{P} \tq[SR]{P} \tq{C}^{P\bar{P}} \tq[\alpha]{P} \,, \\
  \tq[SI\,SR]{N}^{PR}       &= \sqrt{1-\rho_{\rm SR}^2} \tq[\alpha]{P} \tq[SR]{P}\tq{\til{M}}_P \Bigl[ \tq{M}_P\tq[SI]{N}^{PR} + \\
                            & \hspace{10em} + \rho_{\rm SR}\tq{\til{T}}^{P\bar{P}}\tq[SR]{P}\tq{M}_{\bar{P}}\tq[SI]{N}^{\bar{P}R} \Bigr] \,, \\
  \vs[SI\,SR]{t}^P          &= \sqrt{1-\rho_{\rm SR}^2} \tq[\alpha]{P} \tq[SR]{P}\tq{\til{M}}_P \Bigl[ \tq{M}_P\vs[SI]{t}^P + \\
                            & \hspace{10em} + \rho_{\rm SR}\tq{\til{T}}^{P\bar{P}}\tq[SR]{P}\tq{M}_{\bar{P}}\vs[SI]{t}^{\bar{P}} \Bigr] \,, \\
  & \tq{C}^{PR}       = \tq{\til{M}}_P \left[ \tq{\til{T}}^{PR}  +  \rho_{rm SR}\tq{\til{T}}^{P\bar{P}}\tq[SR]{P}\tq{\til{T}}^{\bar{P}R} \right] \,, \\
  & \tq{\til{M}}_P    = \left[ \tq{I} - \rho_{\rm SR}\tq{\til{T}}^{P\bar{P}}\tq[SR]{P}\tq{\til{T}}^{\bar{P}P}\tq[SR]{P} \right]^{-1} \,, \\
  & \tq{\til{T}}^{PR} = \tq{M}_P\tq{T}^{PR}\tq[SR]{P} \,, \\
  & \tq{M}_P          = \left[ \tq{I} - \rho_{\rm SR}\tq{T}^{PP}\tq[SR]{P}^2 \right]^{-1} \,,
\end{align*}
and $\{P,R\} = \{p,s\}$ in all combinations.



\end{document}